\documentclass[twocolumn,showpacs,aps,prd]{revtex4}

\usepackage{graphicx}
\usepackage{dcolumn}
\usepackage{amsmath}
\usepackage{amssymb}
\usepackage{wasysym}
\usepackage{epsfig}
\usepackage{multirow}
\usepackage{maybemath}

\usepackage[italic]{hepparticles}

\usepackage{heppennames}

\def\BR                 {{\ensuremath{\cal B}\xspace}}

\newcommand{\forex}     {\mbox{\textsl{e.g.}}\xspace}
\newcommand{\ie}        {\mbox{\textsl{i.e.}}\xspace}
\newcommand{\vs}        {\mbox{\textsl{vs.}}\xspace}

\def\pizeta     {\Pgpz{}(\Pgh)\xspace}
\def\BB         {\ensuremath{\PB{}\PaB}\xspace}
\def\BzBzb      {\ensuremath{\PBz\PaBz}\xspace}
\def\BpBm       {\ensuremath{\PBp\PBm}\xspace}
\def\epem       {\ensuremath{\Pep\Pem}\xspace}
\def\ccbar      {\ensuremath{\Pqc\Paqc}\xspace}
\def\qqbar      {\ensuremath{\Pq\Paq}\xspace}
\def\tautau     {\ensuremath{\Pgtp\Pgtm}\xspace}

\def\mumu       {\ensuremath{\Pgmp\Pgmm}\xspace}
\def\mmg        {\ensuremath{\Pgm\Pgm\Pgg}\xspace}
\def\gg         {\ensuremath{\Pgg\Pgg}\xspace}

\mathchardef\Upsilon="7107
\def\Y#1S{\ensuremath{\Upsilon{(#1S)}}\xspace}

\def\FourS {\Y4S}

 \def\eg         {\ensuremath{E_{\gamma} }\xspace}

 \def\egcms      {\ensuremath{E^{*}_{\gamma}}\xspace}
 \def\egcmstrue  {\ensuremath{E^{* \: \mathrm{true}}_{\gamma}}\xspace}

 \def\egb        {\ensuremath{E^{B}_{\gamma}}\xspace}

 \def\mxs        {\ensuremath{m_{X_{s}} }\xspace}

 \def\mupsq      {\ensuremath{\mu_{\pi}^2}\xspace}
 \def\mugsq      {\ensuremath{\mu_{G}^2}\xspace}
 \def\mb         {\ensuremath{m_{b} }\xspace}

 \def\emiss      {\ensuremath{E_\mathrm{miss}^*}\xspace}
 \def\sig        {\ensuremath{\mathrm{S}}\xspace}
 \def\bkb        {\ensuremath{\mathrm{B}}\xspace}
 \def\bkc        {\ensuremath{\mathrm{C}}\xspace}

 \def\sigeff     {\ensuremath{\epsilon_\mathrm{sig}}\xspace}

 \def\emcfrac    {\ensuremath{x_\mathrm{EMC}}\xspace}
 \def\bsg        {\ensuremath{\HepProcess{\Pqb\to\Pqs\Pgg}}\xspace}
 \def\bdg        {\ensuremath{\HepProcess{\Pqb\to\Pqd\Pgg}}\xspace}
 \def\bbarsbarg  {\ensuremath{\HepProcess{\Paqb\to\Paqs\Pgg}}\xspace}
 \def\bbardbarg  {\ensuremath{\HepProcess{\Paqb\to\Paqd\Pgg}}\xspace}

 \def\bxsg       {\ensuremath{\PB \to X_{s} \Pgg}\xspace}
 \def\bxdg       {\ensuremath{\PB \to X_{d} \Pgg}\xspace}
 \def\bxsdg       {\ensuremath{\PB \to X_{s+d} \Pgg}\xspace}
 
 \def\bxclnu      {\ensuremath{\PB \to X_{c} \ell \nu }\xspace}
 \def\bxulnu      {\ensuremath{\PB \to X_{u} \ell \nu }\xspace}

\def\aveDelta#1 {\ensuremath{\langle \Delta_{total}#1 \rangle}\xspace}

\def\valerr#1#2#3 {\ensuremath{{#1}^{+#2}_{-#3}}\xspace}

\usepackage{relsize}
\def\babar{\mbox{\slshape B\kern-0.1em{\smaller A}\kern-0.1em
    B\kern-0.1em{\smaller A\kern-0.2em R}}}

\newcommand{\tev}{\ensuremath{\mathrm{\,Te\kern -0.1em V}}\xspace}
\newcommand{\gev}{\ensuremath{\mathrm{\,Ge\kern -0.1em V}}\xspace}
\newcommand{\mev}{\ensuremath{\mathrm{\,Me\kern -0.1em V}}\xspace}
\newcommand{\kev}{\ensuremath{\mathrm{\,ke\kern -0.1em V}}\xspace}
\newcommand{\ev}{\ensuremath{\mathrm{\,e\kern -0.1em V}}\xspace}
\newcommand{\gevc}{\ensuremath{{\mathrm{\,Ge\kern -0.1em V\!/}c}}\xspace}
\newcommand{\mevc}{\ensuremath{{\mathrm{\,Me\kern -0.1em V\!/}c}}\xspace}
\newcommand{\gevcc}{\ensuremath{{\mathrm{\,Ge\kern -0.1em V\!/}c^2}}\xspace}
\newcommand{\mevcc}{\ensuremath{{\mathrm{\,Me\kern -0.1em V\!/}c^2}}\xspace}

\def\cm   {\ensuremath{{\rm \,cm}}\xspace}

\def\invfb   {\ensuremath{\mbox{\,fb}^{-1}}\xspace}

\def\mus  {\ensuremath{\rm \,\mus}\xspace}

\def\mus        {\ensuremath{\,\mu{\rm s}}\xspace}    

\def\to                 {\ensuremath{\rightarrow}\xspace}

\def\pep2{PEP-II}

\newcommand{\chisq}{\ensuremath{\chi^2}\xspace}

\newcommand{\lum} {\ensuremath{\mathcal{L}}\xspace}

\def\gsim{{~\raise.15em\hbox{$>$}\kern-.85em
          \lower.35em\hbox{$\sim$}~}\xspace}
\def\lsim{{~\raise.15em\hbox{$<$}\kern-.85em
          \lower.35em\hbox{$\sim$}~}\xspace}

\def\CP                {\ensuremath{C\!P}\xspace}

\def\acp        {\ensuremath{A_{\CP}}\xspace}

\def\amcp       {\ensuremath{A^{\mathrm{meas}}_{\CP}}\xspace}

\def\Vtd  {\ensuremath{|V_{td}|}\xspace}

\def\Vts  {\ensuremath{|V_{ts}|}\xspace}
\def\Vub  {\ensuremath{|V_{ub}|}\xspace}
\def\Vcb  {\ensuremath{|V_{cb}|}\xspace}

\DeclareRobustCommand{\PsX}{\HepParticle{X}{\Pqs}{}\xspace}
\DeclareRobustCommand{\PdX}{\HepParticle{X}{\Pqd}{}\xspace}

\def\lumibb     {\ensuremath {382.8}\xspace}

\def\onlumi     {\ensuremath { 347.1 \invfb }\xspace}

\def\offlumi    {\ensuremath {36.4 \invfb }\xspace}

\newcommand{\BABARPubYear}    {12}
\newcommand{\BABARPubNumber}  {018}

\newcommand{\SLACPubNumber} {15181}
\newcommand{\LANLNumber} {xxx}

\def\figurebox#1#2#3{%
    \def\arg{#3}%
    \ifx\arg\empty
    {\hfill\vbox{\hsize#2\hrule\hbox to #2{\vrule\hfill\vbox to #1{\hsize#2\vfill}\vrule}\hrule}\hfill}%
    \else
    {\hfill\epsfbox{#3}\hfill}%
    \fi}

\begin{document}

\preprint{\babar-PUB-\BABARPubYear/\BABARPubNumber} 
\preprint{SLAC-PUB-\SLACPubNumber} 
\preprint{hep-ex/\LANLNumber}

\begin{flushleft}

\babar-PUB-\BABARPubYear/\BABARPubNumber\\
SLAC-PUB-\SLACPubNumber\\
\end{flushleft}

\title{
{\large \boldmath Measurement of \BR(\bxsg), the \bxsg photon energy spectrum, \\ and the direct \CP asymmetry in \bxsdg decays} 
}

%
\author{J.~P.~Lees}
\author{V.~Poireau}
\author{V.~Tisserand}
\affiliation{Laboratoire d'Annecy-le-Vieux de Physique des Particules (LAPP), Universit\'e de Savoie, CNRS/IN2P3,  F-74941 Annecy-Le-Vieux, France}
\author{J.~Garra~Tico}
\author{E.~Grauges}
\affiliation{Universitat de Barcelona, Facultat de Fisica, Departament ECM, E-08028 Barcelona, Spain }
\author{A.~Palano$^{ab}$ }
\affiliation{INFN Sezione di Bari$^{a}$; Dipartimento di Fisica, Universit\`a di Bari$^{b}$, I-70126 Bari, Italy }
\author{G.~Eigen}
\author{B.~Stugu}
\affiliation{University of Bergen, Institute of Physics, N-5007 Bergen, Norway }
\author{D.~N.~Brown}
\author{L.~T.~Kerth}
\author{Yu.~G.~Kolomensky}
\author{G.~Lynch}
\affiliation{Lawrence Berkeley National Laboratory and University of California, Berkeley, California 94720, USA }
% ...
\collaboration{The \babar\ Collaboration}
\noaffiliation

\date{November 16, 2012}
\begin{abstract}
The photon spectrum in $B\to X_s\gamma$ decay, where $X_s$ is any strange
hadronic state, is studied using a data sample of $(382.8\pm 4.2) \times 10^6$
$e^+ e^- \to \Upsilon(4S) \to B\overline{B}$ events collected by the 
\mbox{\slshape B\kern-0.1em{\smaller A}\kern-0.1em  B\kern-0.1em{\smaller A\kern-0.2em R}}\
experiment at the PEP-II collider.  The spectrum is used to measure
the branching fraction $\mathcal{B}(B\to X_s\gamma)=
(3.21 \pm 0.15 \pm 0.29 \pm 0.08)\times 10^{-4}$ and the first,
second, and third moments 
$\langle E_\gamma \rangle = 2.267 \pm 0.019 \pm 0.032 \pm 0.003\,\mathrm{GeV}$,
$\langle (E_\gamma -\langle E_\gamma \rangle)^2 \rangle = 
0.0484 \pm 0.0053 \pm 0.0077 \pm 0.0005\,\mathrm{GeV}^2$,
and $\langle ( E_\gamma -\langle E_\gamma \rangle )^3 \rangle =
-0.0048 \pm 0.0011 \pm 0.0011 \pm 0.0004\,\mathrm{GeV}^3$,
for the range $E_\gamma > 1.8\,\mathrm{GeV}$, where $E_\gamma$ is the
photon energy in the $B$-meson rest frame.  Results are also presented
for narrower $E_\gamma$ ranges. In addition, the direct $C\!P$ asymmetry
$A_{C\!P}(B \to X_{s+d}\gamma)$ is measured to be 
$0.057 \pm 0.063$. The spectrum itself
is also unfolded to the $B$-meson rest frame; that is the frame 
in which theoretical predictions for its shape are made.

\end{abstract}


\maketitle

\noindent
\section{Introduction}
\label{sec:intro}

\begin{figure}[tb]
\begin{center}   
  \includegraphics[width=.47\textwidth]{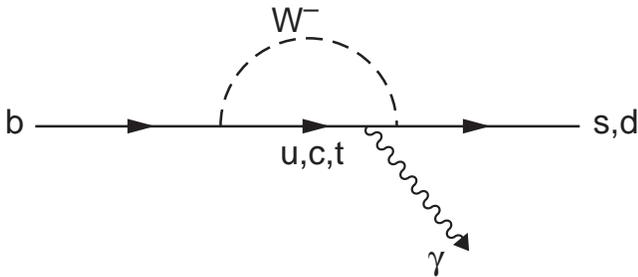}   
\end{center}
\vspace{-0.2in}
\caption{The leading order Feynman diagram for the electromagnetic radiative  decay of the \Pqb-quark in the SM.}
\label{fig:radiativedecay} 
\end{figure}

In the standard model (SM) the  electromagnetic radiative decay of the \Pqb quark,  \bsg or \bdg, proceeds 
at leading order via the loop diagram shown in Fig.~\ref{fig:radiativedecay} resulting in a photon
and a strange or down quark. The rate for \bdg relative to \bsg is suppressed by a factor $|V_{td}/V_{ts}|^2$ 
where $V_{td}$ and $V_{ts}$ are the Cabibbo-Kobayashi-Maskawa (CKM) matrix elements.
Interest in these decays is motivated by the possibility that new heavy particles might enter into 
the loop at leading order, causing significant 
deviations from the predicted SM decay rates.  
 There is an extensive theoretical literature evaluating the effects 
of new physics; some examples are given in references~\cite{Bertolini:1987pk,Hewett:1996ct,Carena:2000uj,
Baer:2003yh,Huo:2003vd,Buras:2002vd,Frank:2002nj,Agashe:2001xt}. New physics can also significantly enhance 
the direct \CP asymmetry for \bsg and \bdg decay~\cite{Kagan:1998bh, Ali:1998rr,Hurth:2003dk,Hurth:2001yb,Benzke:2010tq}.

The hadronic processeses corresponding to the underlying \bsg and \bdg decays are $\bxsg$ and 
$\bxdg$. Here $\PsX$ and $\PdX$ are any final state resulting from the 
hadronization of the $\Pqs\Paq$ or $\Pqd\Paq$ quark-level state,
respectively, where \Paq is the spectator from the \PB meson.
These  are predominantly resonances, including  $\PKsti$, $\PKi$
($\PsX$) or $\Pgr,\Pgo$ ($\PdX$) and higher-mass states,  but also  
nonresonant multihadron final states.
Theoretical predictions for the rates of such exclusive decays suffer 
from large uncertainties associated with the form factors of the  mesons. In contrast, the inclusive 
hadronic rates $\Gamma(\bxsg)$ and $\Gamma(\bxdg)$ can be equated with the 
precisely-calculable partonic rates 
$\Gamma(\bsg)$ and $\Gamma(\bdg)$  at the level of a few percent~\cite{Bigi:1992su} (quark-hadron duality), leading 
to significantly more accurate predictions. At next-to-next-to-leading order (up to four loops) the SM prediction
for the branching fraction is $\BR(\bxsg)=(3.15 \pm 0.23) \times 10^{-4} (\eg>1.6 \gev)$~\cite{Misiak:2006zs}. 
Measurements of the inclusive rates and asymmetries are therefore powerful probes of physics beyond the standard model.

The shape of the photon energy spectrum is determined by the strong
interaction of the \Pqb quark within the \PB meson and by the
hadronization process. The  Fermi motion of the quark within
the \PB meson and gluon radiation lead to an \eg  distribution, in the
\PB-meson rest frame, that is peaked in the range 2.2 to 2.5\gev, with 
a kinematic limit at $m_{\PB}/2 \approx 2.64\gev $ and a rapidly-falling 
low-energy tail.
The shape is insensitive to non-SM
physics~\cite{Kapustin:1995nr,Kagan:1998ym}, and can therefore 
provide information about the strong-interaction dynamics of the \Pqb quark.
Heavy Quark Effective
Theory (HQET)~\cite{Chay:1990da,Shifman:1984wx, Bigi:1992su,Bigi:1993fe,Manohar:1993qn,Manohar:2000dt} 
has been used most extensively to describe these dynamics. The shape of the photon spectrum provides
information on parameters of this theory related to the mass and
momentum of the \Pqb quark within the \PB meson; the definitions and
hence the values of these parameters differ
slightly between the ``kinetic scheme''~\cite{Benson:2004sg} and the
``shape function scheme''~\cite{Neubert:2005nt}.
The Heavy Flavor Averaging Group
(HFAG)~\cite{TheHeavyFlavorAveragingGroup:2010qj} has computed world-average
values of the parameters in the kinetic scheme based on previous
measurements of the inclusive semileptonic \PB-meson decay \bxclnu 
($\ell = e$ or $\mu$) and of \bxsg.
HFAG has also translated those values to the shape function scheme.
These parameters can be used to reduce the error in the extraction of the CKM matrix elements $\Vcb$ and $\Vub$ 
from the inclusive semileptonic decays, \bxclnu and \bxulnu~\cite{Bauer:2004ve,
Lange:2005yw,Bauer:2002sh,Gambino:2007rp}. The \bxsg spectral shape may also be compared to predictions in the
framework of Dressed Gluon Exponentiation~\cite{Andersen:2006hr}. 

The inclusive decay \bxsg was first measured by the CLEO collaboration ~\cite{Alam:1994aw,Coan:2000pu,Chen:2001fja} 
and  has been subsequently studied by the ALEPH~\cite{Barate:1998vz}, Belle~\cite{Abe:2001hk,Nishida:2003yw,
Koppenburg:2004fz,Abe:2008sxa,Schwanda:2008kw,Limosani:2009qg} and \babar~\cite{Aubert:2005cua,Aubert:2006gg,
Aubert:2007my} collaborations. All measurements have been made with 
\PB mesons produced in \epem collisions. The theoretical predictions,
which assume that the measurement is inclusive so that quark-hadron 
duality holds, are made in the \PB-meson rest frame for photons with  
$\eg>1.6$\gev. This means that ideally the measurement 
is made for all \PsX final states and for all photons $\eg>1.6 \gev$. The experimental challenge is to make the measurement as 
inclusive as possible while suppressing backgrounds from other processes producing photons or fake photons. 
The backgrounds arise from continuum events (\epem to $\Pq\Paq$ or $\Pgtp\Pgtm$
pairs, where $\Pq=\Pqu$, \Pqd, \Pqs or \Pqc),
with the photon coming from either
a \Pgpz or \Pgh decay or from initial state radiation, 
and from other \BB processes. The \BB background arises predominantly 
from \Pgpz or \Pgh decay  but also from decays of other light mesons, misreconstructed electrons and hadrons. It is
strongly dependent on photon energy and rises steeply at lower \eg. This places a practical lower limit
for \eg on the experimental measurements; 
measurements have been made to date with $\eg >1.7$, 1.8, and 1.9\gev.

Three experimental techniques have been pursued. They differ in the extent to which the final state is reconstructed. 
The first is the fully inclusive technique in which neither the \PsX from 
the signal \PB nor the recoiling \PaB meson is reconstructed.
(Charge conjugates are implied throughout this paper.)
The second is the semi-inclusive technique, in which as many exclusive
\PsX final states as possible are reconstructed and
combined. The recoiling \PaB meson is not reconstructed. The third is the 
reconstructed recoil-\PaB technique, in which inclusive \PB events are
tagged by fully reconstructing the recoiling \PaB mesons in as many final
states as possible, but \PsX is not reconstructed.
Each of the techniques has different strengths and weaknesses.

If the \PsX is not reconstructed, the sample includes all 
\PsX final states, but there are significant
backgrounds from other \BB decays that must be estimated. It also includes 
\PdX states from the Cabbibo-suppressed 
\bdg process. These can be subtracted by assuming the  \bdg photon spectrum to have a similar shape to the \bsg photon spectrum,
but scaled by the ratio of the CKM elements $(\Vtd/\Vts)^2=0.044 \pm 0.003$. 
This is believed to be a valid assumption.
Also, if the \PsX is not reconstructed then the signal \PB cannot be reconstructed. The \PB mesons have a small momentum 
in the \FourS rest frame.  As the \PB meson is not reconstructed, the direction 
of the momentum is not known. This leads to a Doppler 
smearing of the photon energy. This effect, along with the detector resolution, must be corrected for or unfolded in order 
to compare to predictions made in the \PB-meson rest frame. 
Quantities measured in the \FourS rest frame, \ie, the center-of-mass
(CM) frame, such as the 
photon energy \egcms, are denoted with an asterisk. 

No semi-inclusive measurement to date has reconstructed more than about
60\% of \PsX decays, due to the high combinatoric background for
higher multiplicity decays. Uncertainties in modeling the
mix of \PsX final states result in significant efficiency
uncertainties, as well as a large uncertainty in correcting for the
final states that are not reconstructed. However, the reconstruction
of the \PsX implies that the signal \PB can be fully reconstructed, providing
kinematic constraints to strongly suppress backgrounds, allowing the
measurement to be made directly in the \PB-meson rest frame.

In the reconstructed recoil-\PaB technique, only about 1\% of \PaB's
can be fully reconstructed, due to the presence of neutrinos in semileptonic decays and combinatoric backgrounds
to higher multiplicity decays. This severely limits the statistical precision, 
but does allow the measurement to be made in the \PB-meson rest frame.

This paper reports a fully inclusive analysis that supersedes the previous
\babar\ fully-inclusive result~\cite{Aubert:2006gg}, which is based on a
smaller data sample.   The \egcms photon spectrum is
measured in \bxsdg decays. It is used to measure the
branching fraction \BR(\bxsg) for $\eg > 1.8 \gev$ and for narrower energy
ranges. The effects of detector resolution and Doppler smearing are unfolded to provide
an \eg photon spectrum in the \PB-meson rest frame that can be used
to fit to theoretical predictions for the spectral shape. The unfolded
spectrum is also used to measure the first, second and 
third moments, given respectively by 
\begin{eqnarray}
E_{1} & = & \langle \eg \rangle  \nonumber \\
E_{2} & = & \langle (\eg - \langle \eg \rangle)^2 \rangle  \label{eq:momentdefs} 
\\
E_{3} & = & \langle (\eg - \langle \eg \rangle)^3 \rangle\ . \nonumber 
\end{eqnarray}

Although the SM predicts quite different asymmetries for \bxsg and \bxdg,
the \PsX and \PdX final states cannot be distinguished 
in the fully inclusive technique.  
Hence the \bxdg contribution to the fully-inclusive measurement cannot
be corrected for, and only the combination \acp(\bxsdg) can be measured:
\begin{equation*}
  \acp = 
  \frac{\Gamma(\bsg + \bdg)-\Gamma(\bbarsbarg + \bbardbarg)}
       {\Gamma(\bsg + \bdg)+\Gamma(\bbarsbarg + \bbardbarg)} \ .
\end{equation*}
This asymmetry is approximately $10^{-6}$ in the SM, with nearly exact cancellation
of opposite asymmetries for \bsg and \bdg. $\acp(\PB \to X_{s+d}\gamma)$ 
is sensitive to different new physics scenarios than \acp(\bxsg)~\cite{Hurth:2003dk}.
Thus measurements of this joint asymmetry complement those of
\acp in \bsg ~\cite{Coan:2000pu,Nishida:2003yw,Aubert:2004hq,Aubert:2008gvb} 
to constrain new physics models.

\section{Datasets, Detector, Simulation and Signal Models}
\label{sec:data}

The results presented are based on data samples of $\Pep\Pem \to \FourS\to \BB$ 
collisions collected with the \babar\ detector at the PEP-II
asymmetric-energy \epem collider. The on-resonance integrated
luminosity is \onlumi , corresponding to \lumibb million \BB
events. The continuum background is estimated with an off-resonance
data sample of \offlumi collected 40\mev below the \FourS resonance energy.

The \babar\ detector is described in detail in reference~\cite{Aubert:2001tu}.
Charged-particle momenta are measured with a 5-layer, double-sided
silicon vertex tracker (SVT) and a 40-layer drift chamber (DCH) inside
a 1.5-T superconducting solenoidal magnet.  A high resolution 
total-absorption electromagnetic calorimeter (EMC),
consisting of 6580 CsI(Tl) crystals, is used to measure localized 
electromagnetic energy deposits and hence to identify photons and electrons.
The EMC energy resolution for high-energy photons in the current
measurement is about 2.6\%.
A ring-imaging Cherenkov radiation
detector (DIRC), aided by measurements of ionization energy loss,
$dE/dx$, in the SVT and DCH, is used for particle identification (PID)
of charged particles.  Muons are identified in the instrumented flux
return (IFR), which consists of 18~layers of steel interleaved with
single-gap resistive-plate chambers.  For the last 38\% of the data
collected, 1/3 of these chambers in the central region of the detector
were replaced by 12~layers of limited-streamer tubes, interspersed
with 6~layers of brass (to increase absorption).

The \babar\ Monte Carlo (MC) simulation, based on 
\textsc{GEANT4}~\cite{Agostinelli:2002hh}, 
\textsc{EVTGEN}~\cite{Lange:2001uf} and 
\textsc{JETSET}~\cite{Sjostrand:1995iq}, is 
used to generate samples of \BpBm and \BzBzb, $\Pq\Paq$ (where \Pq is a 
\Pqu, \Pqd, \Pqs or \Pqc quark), $\Pgtp\Pgtm$, and signal events (\BB
events in which at least one \PB decays to $X_s \Pgg$).
To model beam backgrounds, each simulated event
is overlaid with one of a set of of random background data events collected
using a periodic trigger.

The signal models used to determine selection efficiencies are based on QCD
calculations of references~\cite{Benson:2004sg} (kinetic scheme) 
and~\cite{Lange:2005yw} (shape function scheme) and on an earlier
calculation by Kagan and Neubert~\cite{Kagan:1998bh} (``KN''). 
Each model uses an ansatz
for the shape that is constrained by calculations of the first and
second moments of the spectra. The models  approximate
the hadronic mass (\mxs) spectrum, which contains a number of overlapping 
resonances, as a smooth distribution. This
is reasonable, except at the lowest masses, where the \PKsti dominates
the spectrum.  Hence the portion of the \mxs spectrum below 1.1\gevcc
is replaced by a Breit-Wigner \PKsti distribution, normalized to yield
the same fraction of the integrated spectrum.  A particular signal model is
defined as the theoretical spectrum for specific HQET parameters,
with this \PKsti at low \mxs.  The photon energy in the \PB-meson rest frame
is related to \mxs via
\begin{equation}
  \mxs^2 = m_\PB^2 - 2 m_{\PB}\frac{\eg}{c^2} \ .
  \label{eq:kinematics}
\end{equation}
High-statistics MC signal samples for the non-\PKsti part of the spectrum
are generated uniformly in \eg -- separately for each of the two 
\PB-meson charge states -- and then weighted according to any particular model
of interest.

Monte Carlo samples of \BpBm and \BzBzb events are needed for background
evaluation.  They are produced, with nearly three times the effective
luminosity of the data sample, and include all known \PB decays,
except for events in which either \PB decays via \bxsdg.   
Monte Carlo samples
of continuum events (\qqbar, separately for \ccbar and for the light
quarks, and \tautau) are used to optimize the event selection
criteria, but are not otherwise relied upon.

\section{Analysis Overview}
\label{sec:anOverview}

The event selection is described in detail in Sec.~\ref{sec:evsel}.
The analysis begins by selecting hadronic events.  
A high-energy photon, characteristic
of \bxsg decays, is then required, while photons from \Pgpz and \Pgh 
decays are vetoed, reducing both the continuum and \BB backgrounds. 
The background from continuum events
is significantly suppressed by charged lepton tagging 
(requiring a high-momentum lepton, as would be expected
from the semileptonic decay of a \PB meson)
and by exploiting the more jet-like topology of the $\Pq\Paq$ or $\Pgtp\Pgtm$
events compared to the isotropic \BB decays.

The continuum MC simulation does not adequately model the
actual continuum background, primarily because it omits QED and
two-photon processes.  Hence the continuum background is estimated
with off-resonance data (Sec.~\ref{sec:contin}), which limits the
statistical precision of the signal-yield measurement.
However, the continuum simulation is used to optimize some of the
event-selection criteria (which must be done without reference to actual
data).  After preliminary event selection, which reduces the
unmodeled backgrounds, a simple scaling of the
continuum-MC predictions adequately models the event-yield
distributions relevant for optimization.

The lepton tagging and event topology criteria do not substantially 
reduce the \BB background relative to the signal, as these processes
have similar characteristics.  The remaining \BB background 
is estimated using MC simulation. There are several different \PB-meson decays that contribute. Section~\ref{sec:BB} describes
how each significant component is compared to an independent data control sample and weighted to replicate those data. 
The uncertainty in these weighting procedures is the dominant source of 
systematic uncertainty.

After the event selection, the continuum and reweighted \BB backgrounds 
are subtracted from the on-resonance data sample, resulting in the raw 
$\bxsdg$ photon spectrum (Sec.~\ref{sec:yields}).
The analysis was done ``blind'' in the range of reconstructed photon energy \egcms from
1.8 to 2.9\gev ; that is, the data were not looked at until all selection requirements
were set and the corrected backgrounds determined. The choice of signal range 
is limited by high \BB backgrounds at low \egcms. The regions 
$1.53 <\egcms < 1.8\gev$ and $2.9 < \egcms < 3.5\gev$ are dominated by
\BB and continuum backgrounds respectively. They provide control
regions to validate the background estimation for the signal region.

The raw spectrum is used to extract the direct \CP asymmetry 
(Sec.~\ref{sec:acp}) and the partial branching fraction for 
$1.8 < \eg < 2.8\gev$ (Sec.~\ref{sec:bf}).
Finally, in Sec.~\ref{sec:spectrum} 
the effects of detector resolution and Doppler smearing are unfolded 
in order to measure the shape of the photon energy spectrum in the \PB-meson
rest frame.

\section{Event Selection}
\label{sec:evsel}

The event selection was developed using MC samples of signal and
background events.  The model used for signal simulation, as defined in
Sec.~\ref{sec:data}, is based on a KN spectrum with $\mb=4.65\gevcc$.

\subsection{Selection of Hadronic Events}
\label{sec:evsel_hadB}

For each event, the analysis considers good-quality reconstructed tracks,
which have momenta transverse to the beam direction of at least 0.1\gevc and
originate from the vicinity of the interaction point (point of 
closest approach within 10\cm along the beam axis and 1.5\cm in the
transverse plane),
and EMC clusters of at
least 30\mev in the laboratory frame.
Hadronic events are selected by requiring at least three
reconstructed charged particles and the normalized second Fox-Wolfram
moment~\cite{Fox:1978vw} 
$R_2^*$ to be less than 0.90. To reduce radiative Bhabha and
two-photon backgrounds, the number of charged particles plus half the number 
of photons with laboratory-frame energy above 0.08\gev is required to be 
greater than 4.5.

\subsection{Requirements on the High-Energy Photon}
\label{sec:evsel_phot}

The photon selection requires at least one photon candidate with
$1.53<\egcms<3.5\gev$ in the event.  A photon candidate is a neutral
EMC energy cluster with a lateral moment consistent with that of a
single photon~\cite{Drescher:1984rt}. The latter requirement rejects
most background from neutral hadrons, which at these energies is
dominated by antineutrons that annihilate in the EMC.
The photon location is assigned
at a depth of 12.5\cm in the EMC, where 
it is required to be isolated by 25\cm from any other 
energy deposit (the lateral dimensions of the crystals are approximately
5\cm by 5\cm).  The cluster must also be well-contained in the
calorimeter ($-0.74 < \cos \theta_\gamma < 0.94$, where
$\theta_\gamma$ is the laboratory-frame polar angle with respect to
the direction of the electron beam).  A likelihood variable
$(L_{\Pgpz})$ based on the energy profile of the EMC cluster is used
to suppress the contribution of \Pgpz{}'s in which the two daughter
photons are not resolved.  The requirement on $L_{\Pgpz}$
retains essentially all isolated high-energy photons.  These photon
quality criteria are determined from studies of photons in \mmg events
and of \Pap's (\Pp's) from \PagL (\PgL) decays.  (Antiprotons are used
to estimate the detector response to background antineutrons.)

High-energy photons that are consistent with originating from $\Pgpz\to\gg$ 
or $\Pgh\to\gg$ decays are vetoed if the other \Pgpz or \Pgh daughter
is found. For the \pizeta veto, combinations are formed of the high-energy
photon with all other photon candidates 
that have laboratory-frame energy greater than 30 (230)\mev; it is required 
that the invariant mass not lie within a window around the nominal
\pizeta mass, $ 115 (508) < m_{\gg} < 155 (588)\mevcc$.  

\begin{figure*}[p]
 \begin{center}   
  \includegraphics[width=.49\textwidth]{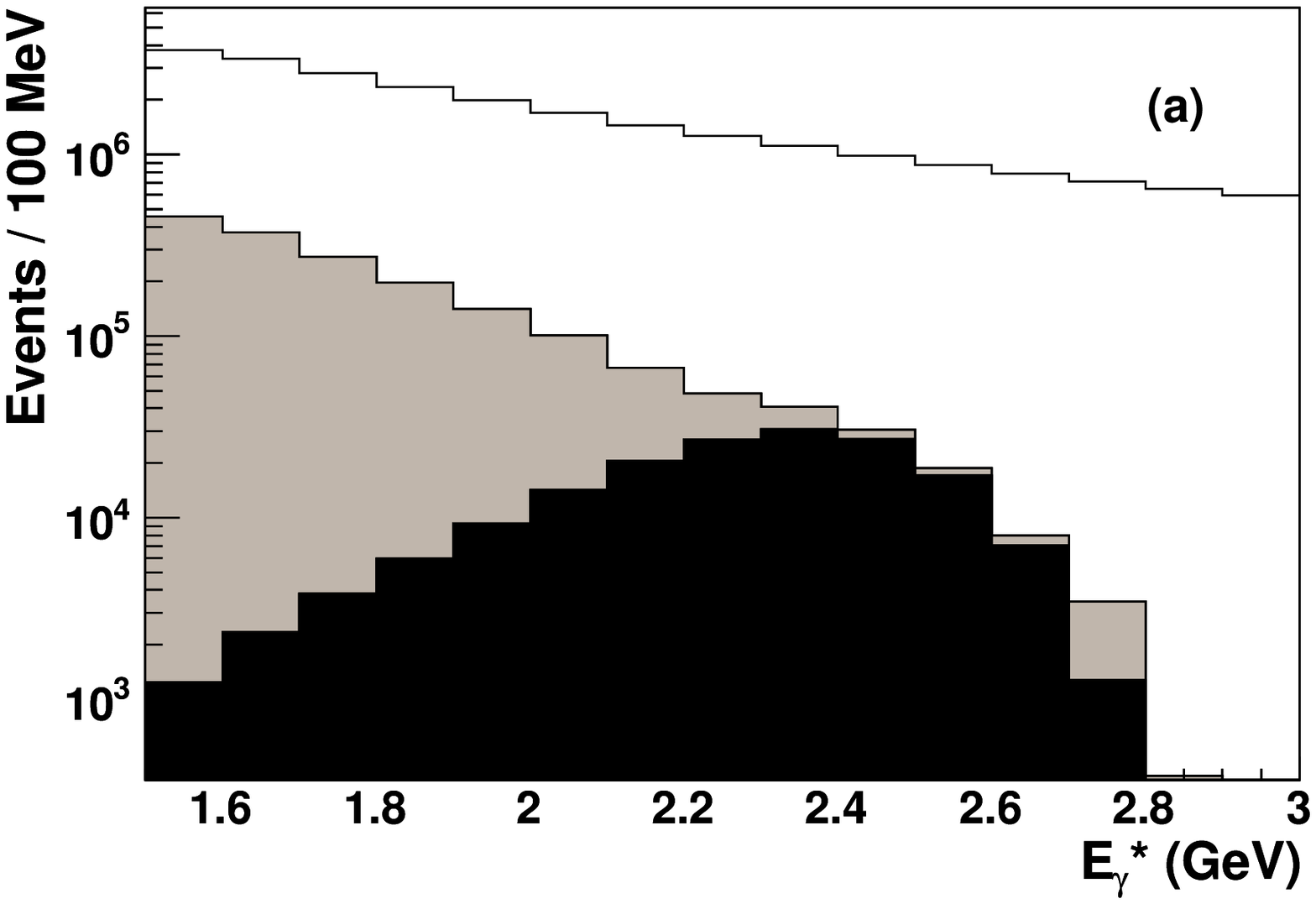}
  \includegraphics[width=.49\textwidth]{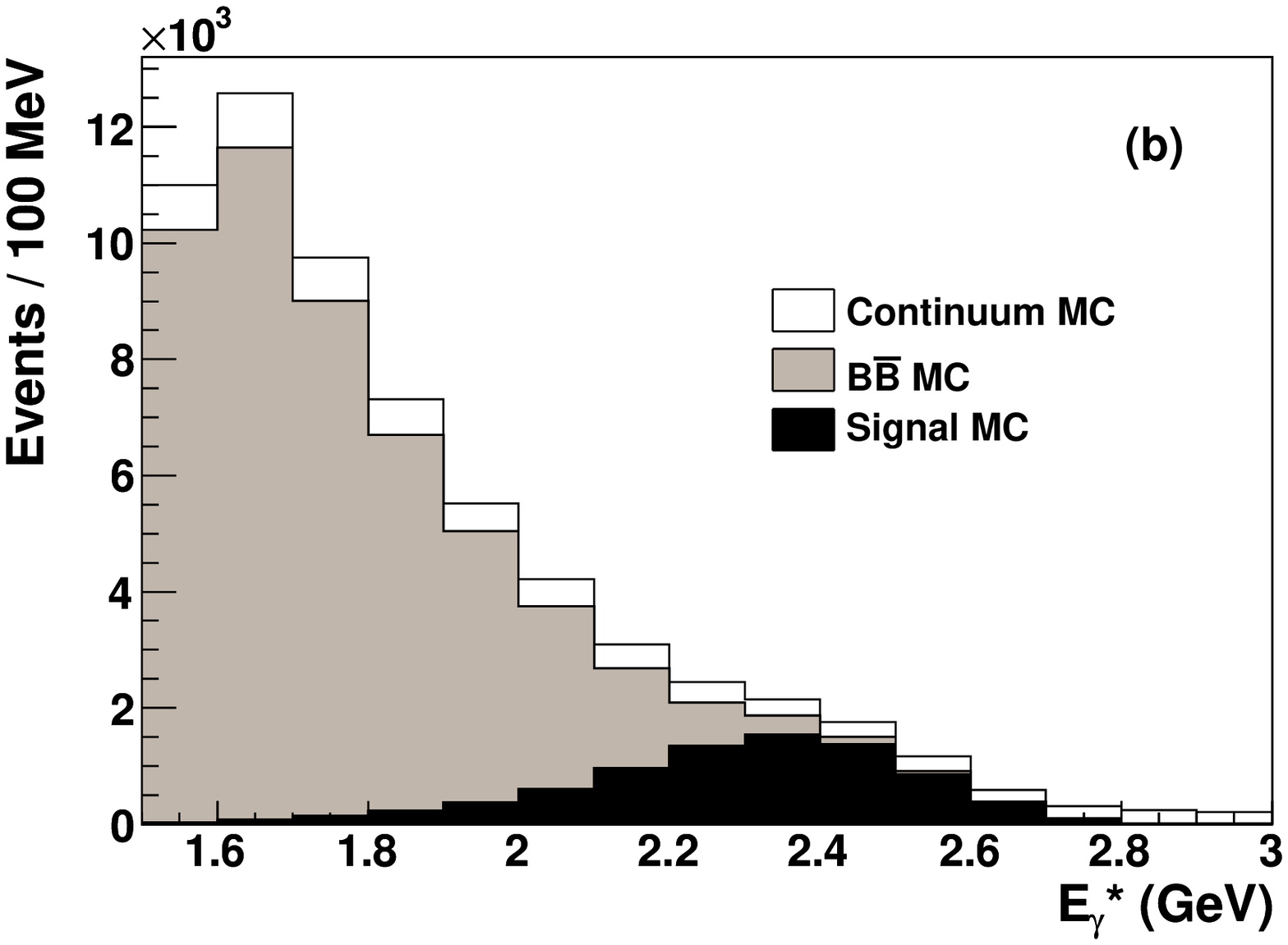}
 \end{center}
 \vspace{-0.2in}
 \caption{Estimated signal and background yields \vs photon energy in the
  CM frame based on MC simulation, at two stages of the event
  selection: (a) after requiring an unvetoed high-energy photon 
  (logarithmic scale); (b) after all selection requirements (linear scale).
  The three contributions are shown cumulatively.  The signal 
  distribution is for a KN model with $\mb=4.65\gevcc$, while the continuum 
  distribution has been scaled as described in Sec.~\ref{sec:anOverview}.}
 \label{fig:egamma_estimated}
\end{figure*}

The simulated distributions of signal and background at this stage of
the event selection are shown in
Fig.~\ref{fig:egamma_estimated}(a). The cumulative signal efficiency
up to this point is approximately 50\%, while 1.6\% of continuum and
0.4\% of \BB backgrounds are retained.  The remaining continuum
background arises predominantly from unvetoed \Pgpz and \Pgh decays, or
initial-state radiation in $\Pq\Paq$ events. The \BB background is
also dominated by unvetoed decays of \pizeta from $\PB\to X\pizeta$, but
also has a significant contribution from misidentified
electrons, and smaller components from antineutrons
and radiative \Pgo and \Pghpr decays.

\subsection{Lepton Tagging}
\label{sec:evsel_lept}

\begin{figure*}[p]
 \begin{center}   
  \includegraphics[width=.49\textwidth]{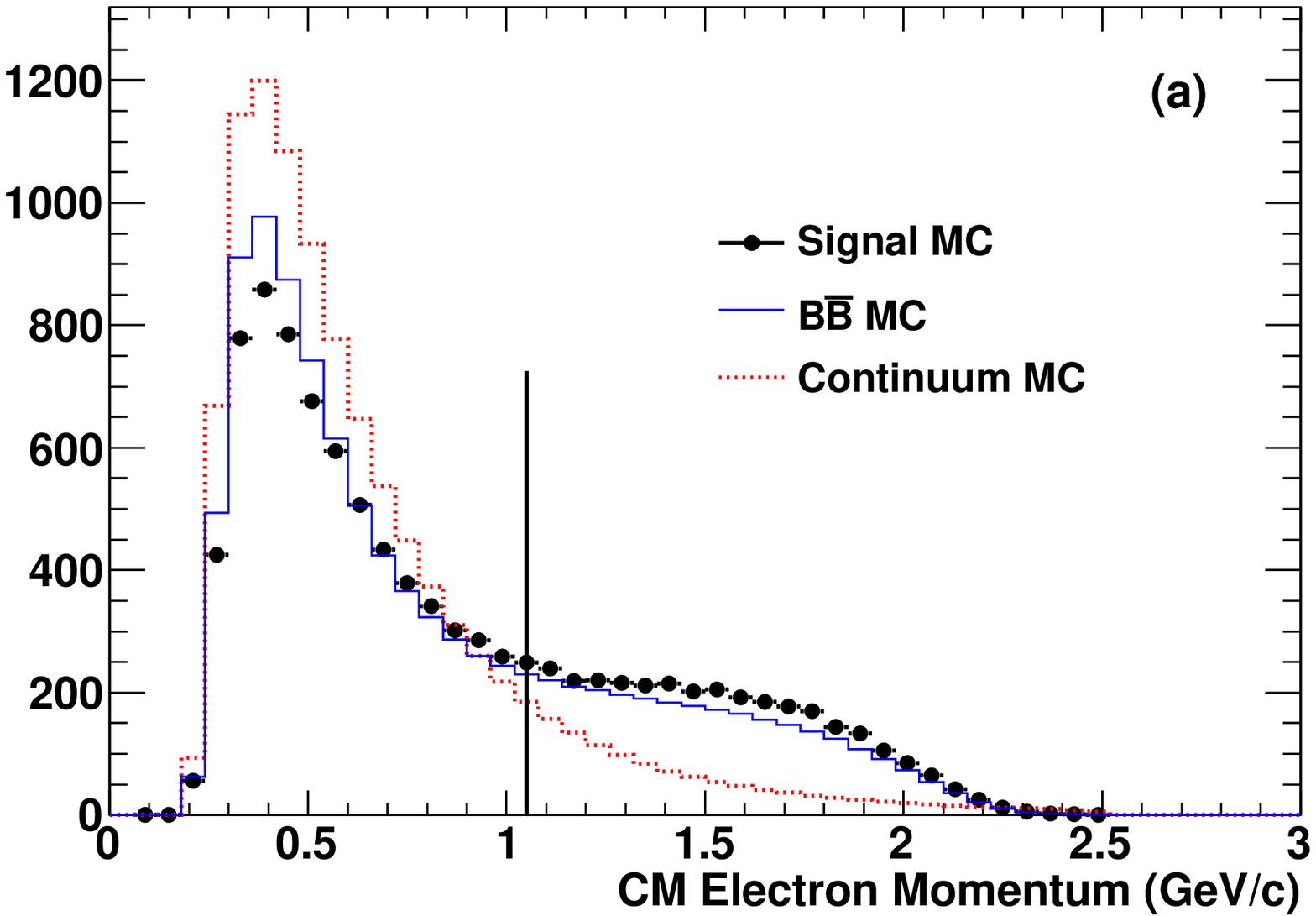}
  \includegraphics[width=.49\textwidth]{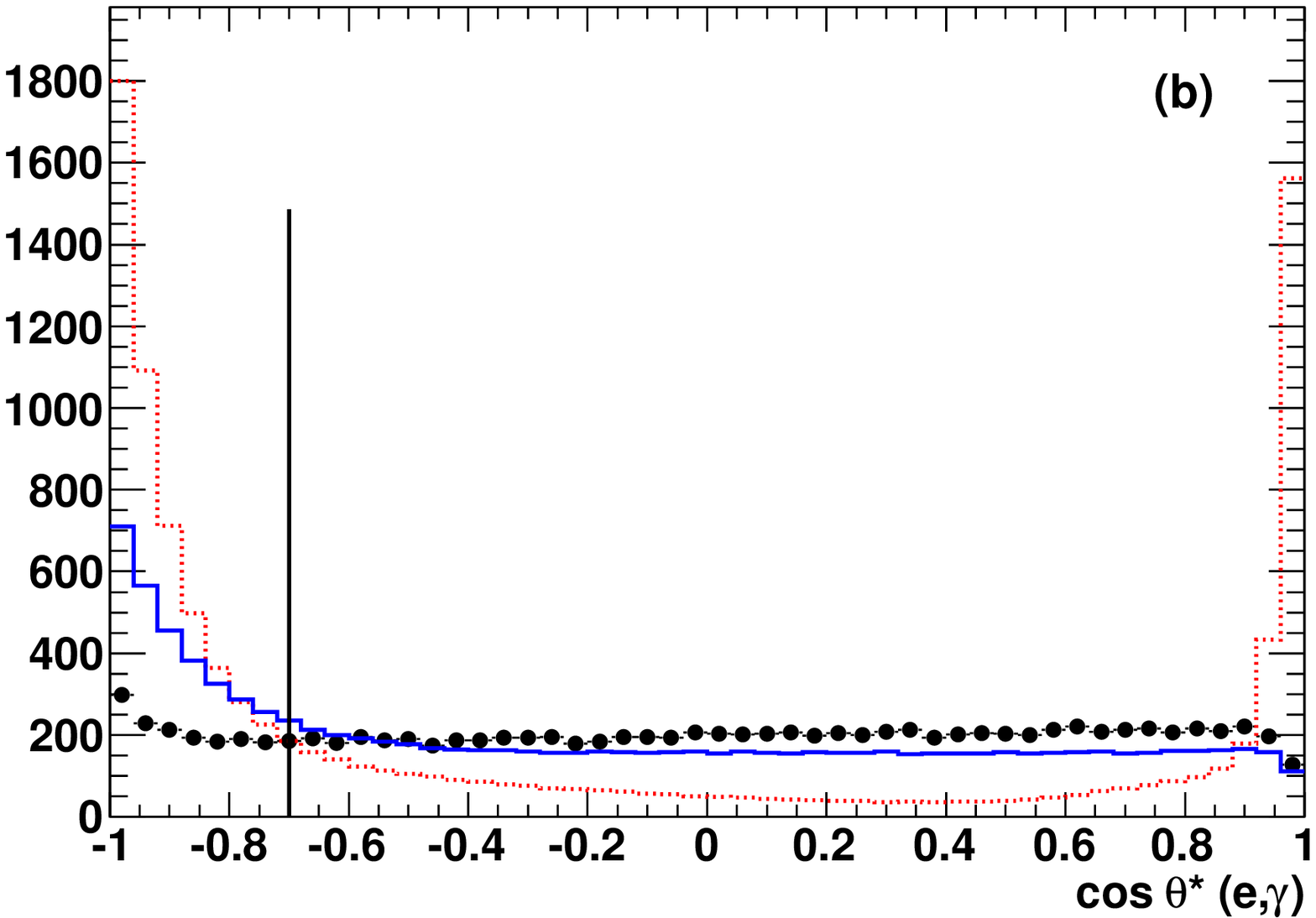} 
  \includegraphics[width=.49\textwidth]{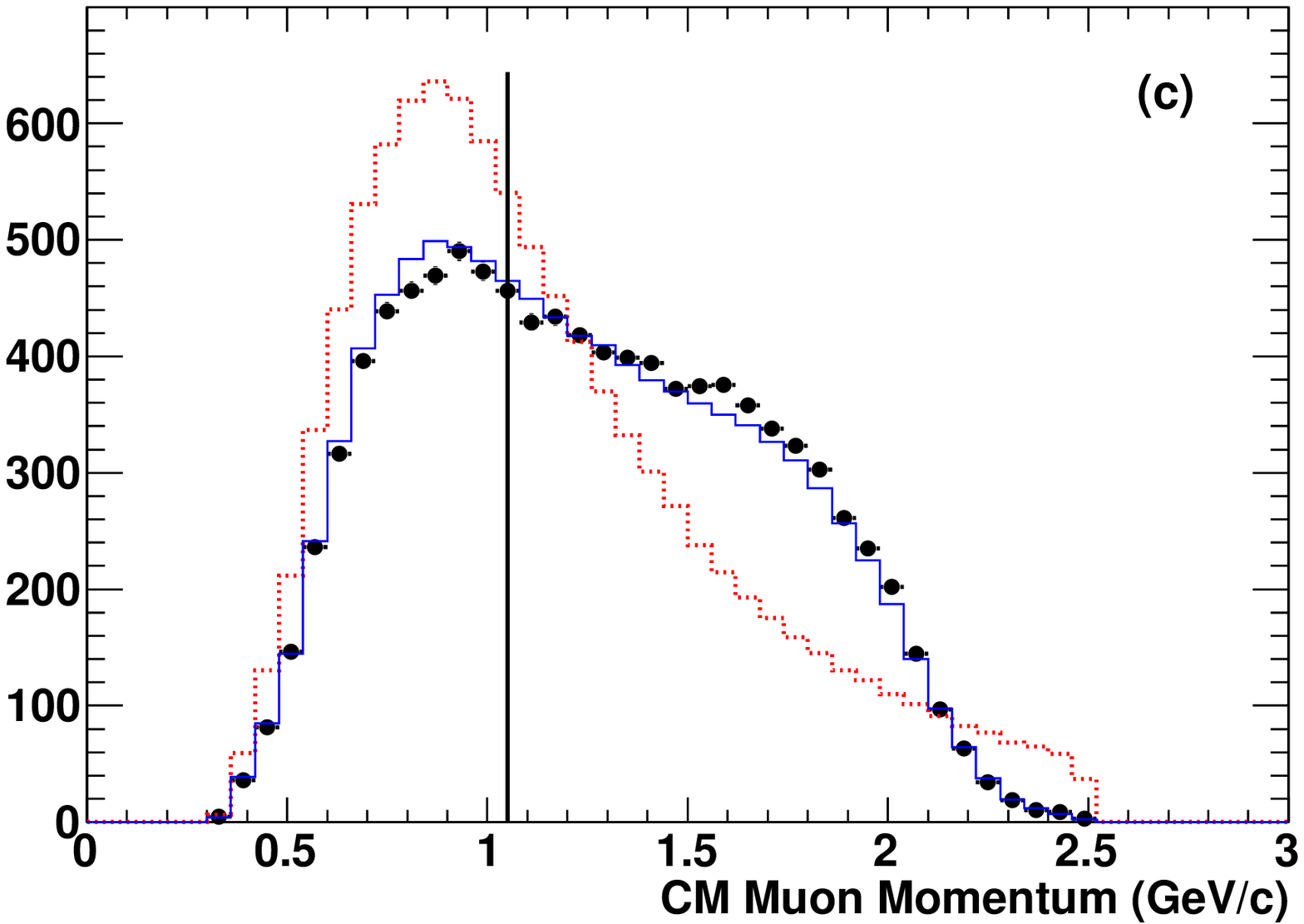}
  \includegraphics[width=.49\textwidth]{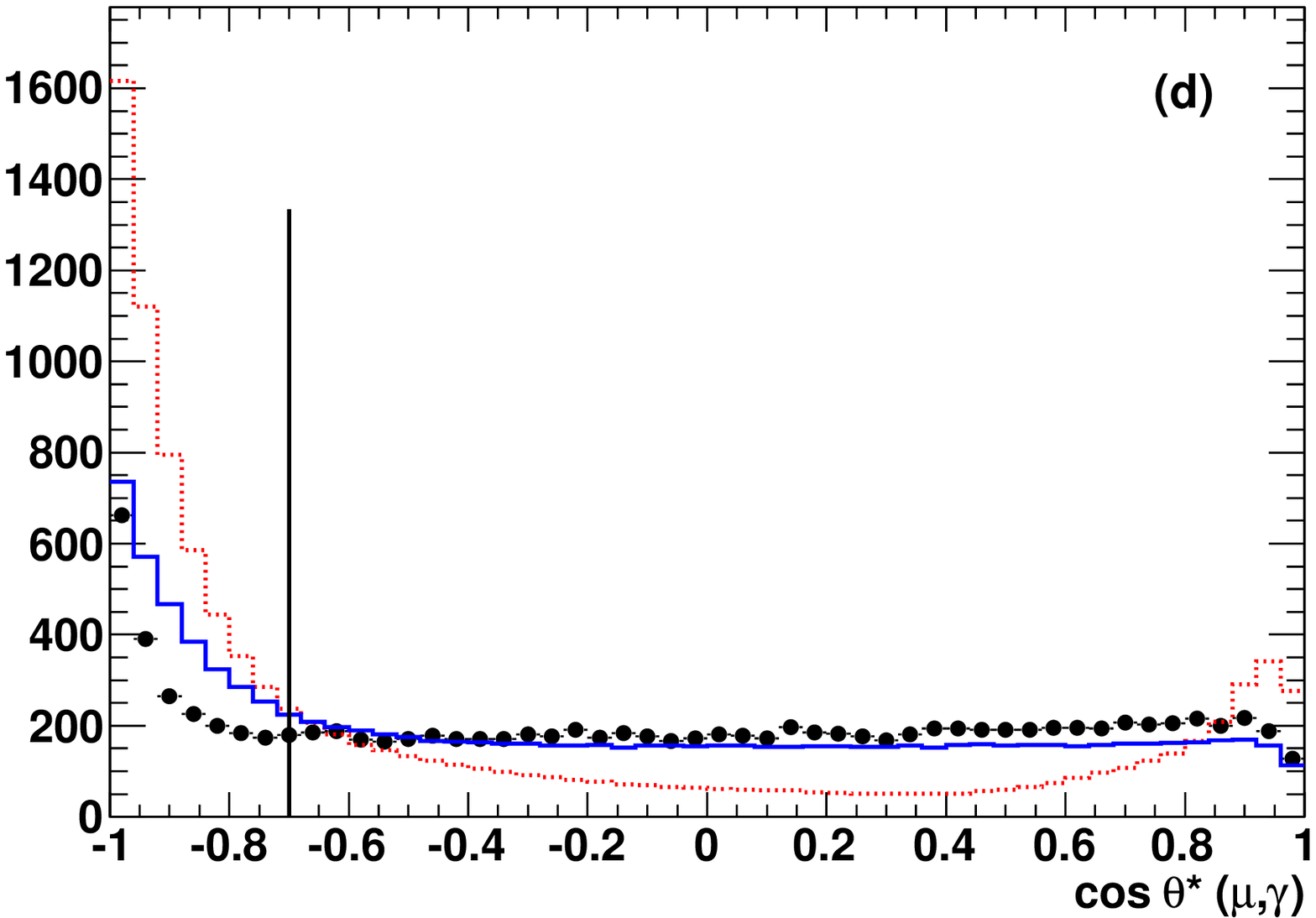} 
  \caption{Lepton distributions from MC simulation, after the photon
   selection requirements but before applying lepton tag and NN criteria. 
   Plots (a) and (b) are for electron tags, plots (c) and (d) for
   muon tags.  Plots (a) and (c) show the CM-frame momentum
   distributions, with vertical lines indicating the minimum selection
   requirements. Plots (b) and (d)  show the cosine of the CM angle
   between the lepton and the high-energy photon, after applying the
   momentum criteria; the vertical lines show the minimum requirement on
   this quantity.  The signal (black dots) is from a KN model with
   $\mb=4.65\gevcc$. The \BB background (solid blue histogram) and
   continuum background (dashed red) are from the MC
   simulations.  Each distribution is  separately normalized to best 
   illustrate its behavior. }
  \label{fig:leptontagplots} 
 \end{center}
\end{figure*}

About 20\% of \PB mesons decay semileptonically to either an electron
or muon, predominantly via $\PB\to X_c\ell\Pgn$. An additional 4\% of
\PB decays result in an electron or muon via $\PB\to
X_c\Pgt\Pgn$. Since the tagging lepton comes from the recoiling \PB
meson, this requirement does not compromise the inclusiveness of the
\bxsg\ selection.

Electrons are identified with a likelihood algorithm that incorporates
properties of the deposited energy and shower shapes of the EMC
clusters, the Cherenkov angles associated with the charged particle
passing through the DIRC, and the $dE/dx$ energy loss of the track.
Muons are identified using a neural-network selector containing
variables that discriminate between muons and electrons, primarily
through differences in EMC energy deposition, and those which
discriminate between muons and hadrons, mainly through differences in
IFR signatures.

The left plots of Fig.~\ref{fig:leptontagplots}
show that leptons from hadronic decays in continuum events tend to be
at lower momentum.  Hence the tagging lepton is required to have
momentum $p^{*}_{e,\mu}>1.05\gevc$. As seen in the right plots of
Fig.~\ref{fig:leptontagplots}, additionally requiring the
cosine of the CM-frame angle between the lepton and the high-energy 
photon $\cos\theta^{*}_{\Pgg\ell} > -0.7$ removes more
continuum background, in which the lepton and photon candidates tend
to be back-to-back. The peak at $\cos\theta^{*}_{\Pgg\ell} \approx 1.0$
for electrons in continuum events
arises predominantly from $\pizeta\to\gg$ decays in which one photon
satisfies the high-energy photon requirements and the other converts
to an \epem pair.  The peaks at $\cos\theta^{*}_{\Pgg\ell} \approx -1.0$
for the \BB background arise
from \PB decays in which the photon and lepton come from the same \PB.
A similar smaller peak for muon tags in signal events is due to pions
faking the muon signature.  These tag selection requirements are
designed as a loose preselection; a more stringent tag discrimination
is achieved by the multivariate selectors described in
Sec.~\ref{sec:evsel_event}.

The presence of a relatively high-energy neutrino in
semileptonic \PB decays is exploited by requiring the missing energy
of the event (\emiss) to be greater than 0.7\gev.  The lepton-tag 
requirements retain approximately 12\% of signal and \BB background events
after the photon selection, while 
retaining only 2.2\% of continuum backgrounds.

\subsection{Event Topology Requirement}
\label{sec:evsel_event}

As the continuum backgrounds are different for electron and muon tags,
each sample is divided according to the tag. For each lepton type the 
continuum backgrounds are then further suppressed by combining the
$p^{*}_{e,\mu}$ and $\cos\theta^{*}_{\gamma\ell}$ for the leptons with
event topology variables into a neural-network (NN)
discriminant.

\begin{figure*}[!bt]
 \begin{center}   
  \includegraphics[width=.49\textwidth]{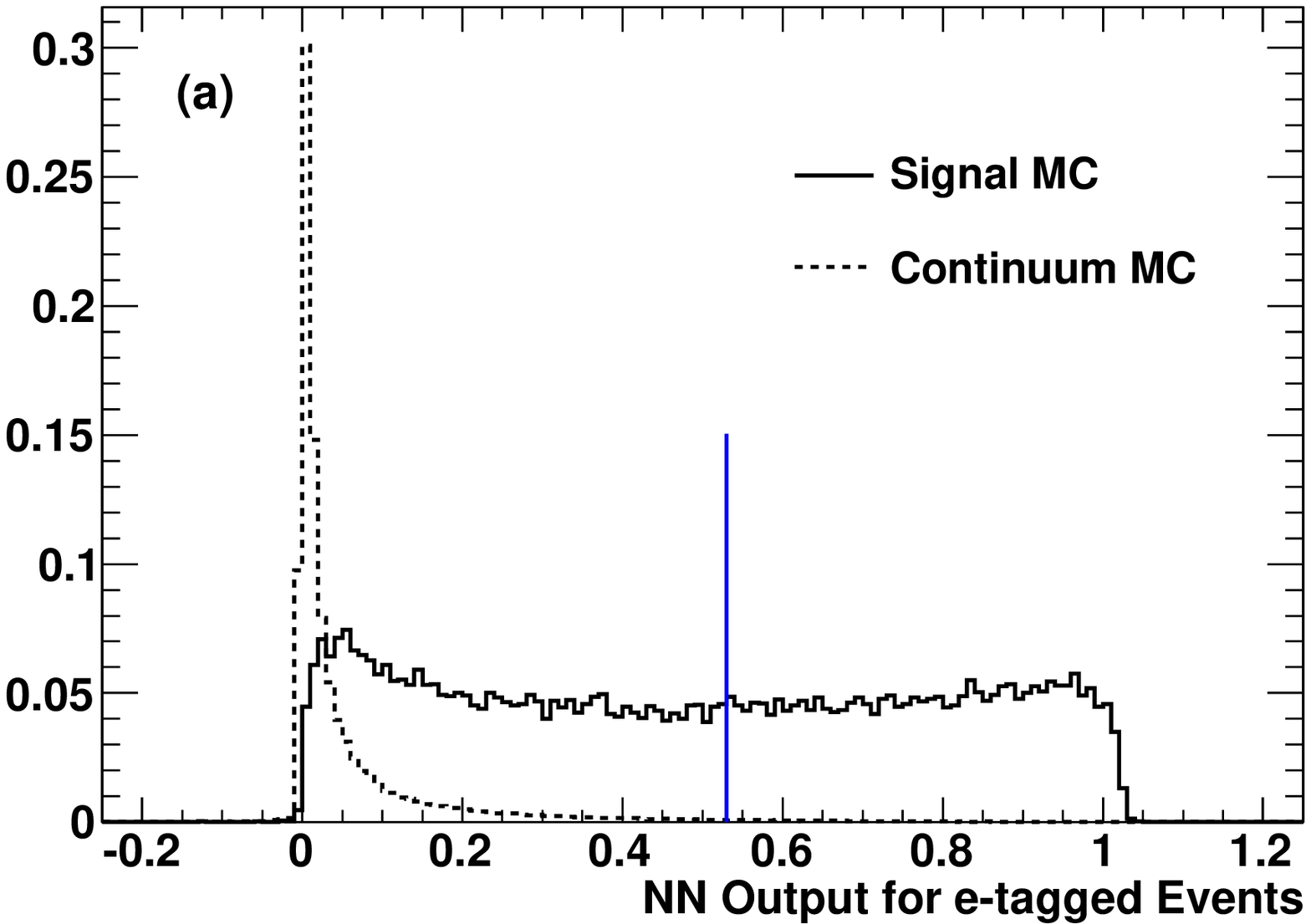}\hfill
  \includegraphics[width=.49\textwidth]{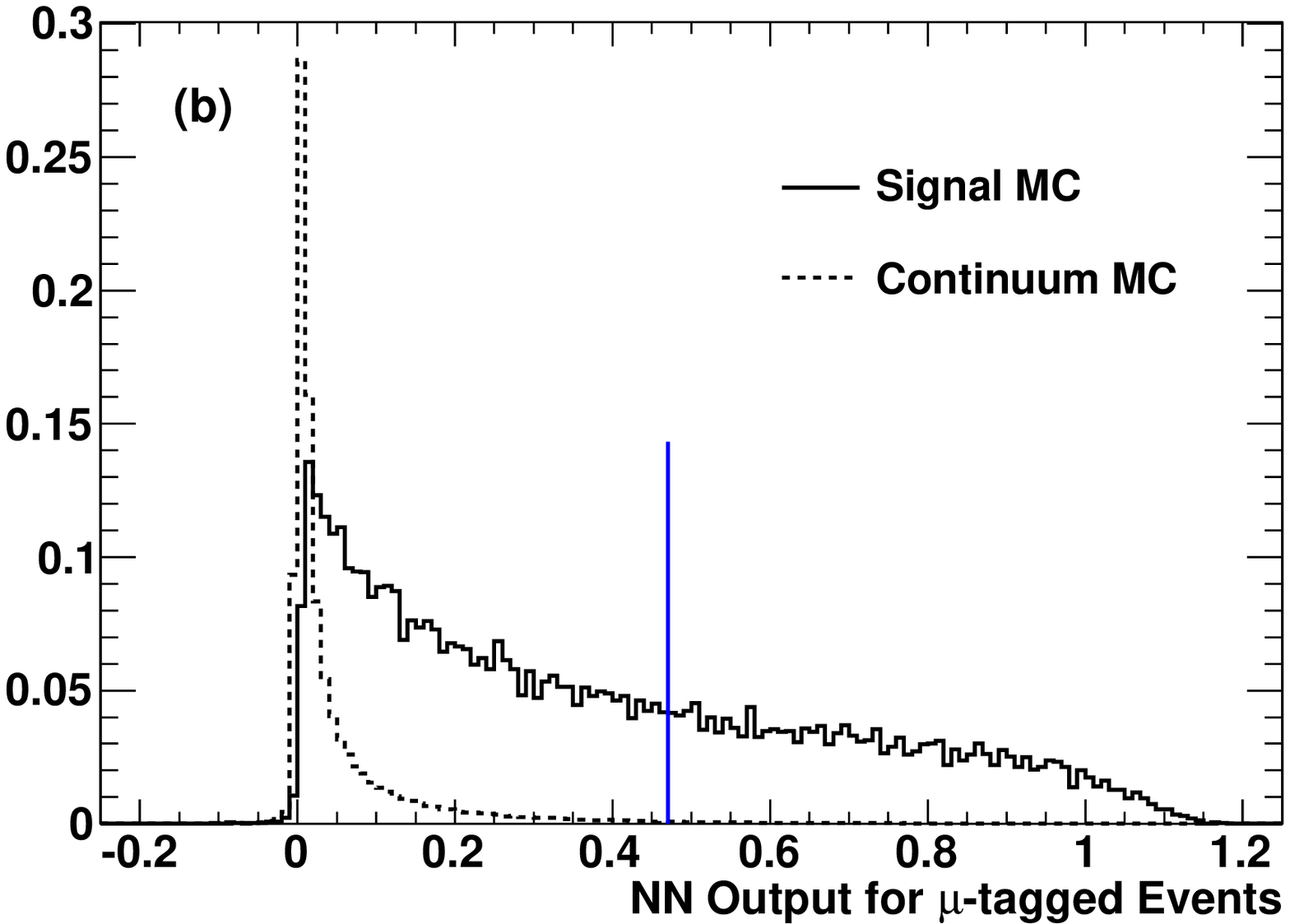} 
  \vspace{-0.1in}
  \caption{The distribution of the NN output, from MC simulation after the photon selection and lepton tag requirements,
   for (a) electron-tagged events and (b) muon-tagged events.
   The vertical lines show the minimum
   requirement on this quantity.  The signal is 
   from a KN model with $\mb=4.65\gevcc$. The continuum  is from the MC simulation.  Normalizations are arbitrary.
  }
  \label{fig:NNplots} 
 \end{center}
\end{figure*}

Several alternative choices of input variables were considered.
For each alternative, the electron and muon NN's are
trained, and the requirements on their output parameter optimized 
(see below).  
The choice of variables is designed to minimize the total error on the
branching fraction and spectral moment measurements, based on combining
in quadrature preliminary estimates
of statistical, systematic and model-dependence errors. 
The latter refers to a variation of the event-selection efficiency with
the choice of MC spectrum (``model'' in the sense of
Sec.~\ref{sec:data}) used to compute it. It arises primarily from the
increase
in efficiency as a function of \egcms; the stronger this trend,
the larger the model-dependence uncertainty.
The selection strategy aims for best signal precision,
while minimizing the dependence of efficiency
on \egcms. Since the backgrounds rise sharply as \egcms decreases, it is
impossible to completely eliminate the \egcms dependence.  Of
several multivariate discriminants (with different sets of
input variables) which were found to give approximately the same signal 
precision, the one resulting in the least \egcms dependence
was chosen.

The eight topology variables chosen for the NN
include $R_2^\prime/R_2^*$, where $R_2^\prime$ is the normalized
second Fox-Wolfram moment calculated in the frame recoiling against
the photon, which for ISR events is the \qqbar rest frame. Also included
are three momentum-weighted polar-angle moments,
$L_j/L_0,j=1,2,3$, where \begin{equation}
L_j = \sum_i |\mathbf{p}_i||\cos \theta_i|^j \ .
  \label{eq:legendre}
\end{equation}
Here $\mathbf{p}_i$ and $\cos \theta_i$ are the momentum and angle,
respectively, of the $i$th reconstructed particle with respect to the
high-energy photon axis in the recoil frame. Summation over $i$
includes every reconstructed charged and neutral particle except the
high-energy photon. The last four topology variables are derived from
the eigenvalues $\left(\lambda_1,\lambda_2,\lambda_3\right)$ and
eigenvectors of the momentum tensor~\cite{Ellis:1980nc}
\begin{equation}
  P^{nm}=\displaystyle{\frac{\sum_i p_i^n p_i^m /
  |\mathbf{p}_i|}{\sum_i |\mathbf{p}_i|}} \ ,
  \label{eq:tensor}
\end{equation}
where $p_i^n$ is the $n$th component of the $i$th reconstructed
particle's 3-momentum in the recoil frame.  The high-energy photon
candidate is excluded. The derived quantities used as NN inputs are:
\begin{eqnarray*}
  \lambda^{1d} & = & \max\left(\lambda_1,\lambda_2,\lambda_3\right) \\
  \lambda^{2d} & = & \lambda_1 \lambda_2 + \lambda_2 \lambda_3 +
  \lambda_3 \lambda_1 \\ \lambda^{3d} & = & \lambda_1 \lambda_2
  \lambda_3 \\ V_z^{1d} & = & z\mbox{-component of }\mathbf{V}_{\max}\, ,
\end{eqnarray*}
where $\mathbf{V}_{\max}$ is the eigenvector associated with the largest 
eigenvalue and $z$ is the electron beam direction.

The electron and muon NN's are trained with MC samples of
continuum and signal (KN model with $\mb=4.65\gevcc$) events that
contain a photon with energy in the range $1.9 < \egcms < 2.7\gev$.
The \BB background simulation sample is excluded from the training
because this sample is used for background subtraction and is
topologically very similar to the signal.  Training
with background and signal samples normalized to the expected event
yields at this stage of the event selection provides slightly better
statistical precision for signal (see Sec.~\ref{sec:evsel_opt})
than does training with background and signal samples with the same
normalization.  For an NN with equally-normalized training samples,
the NN output distributions would peak toward 0 and 1, respectively,
for background- and signal-like events.  Neural-network training based
on expected event yields, however, produces output distributions that
are qualitatively different, as demonstrated in
Fig.~\ref{fig:NNplots}, which shows the output distributions for
signal and continuum events, separated according to lepton tag.
Events with an electron (muon) tag are required to have an NN output
greater than 0.53 (0.47).  This selection accepts 42\% of signal
events ($1.8 < \egcms < 2.8 \gev$) that have passed the photon and
lepton selection requirements while retaining 1.7\% of continuum and
27\% of \BB background.  Events with more than one photon candidate
after the NN requirement are discarded (0.16\% of signal events).

\subsection{Optimization of the Event Selection}
\label{sec:evsel_opt}

The optimization for the selection criteria was performed iteratively 
on five variables:  the two NN outputs (Sec.~\ref{sec:evsel_event}), the 
minimum energy of the lower energy photon in the \Pgpz and \Pgh vetoes 
(Sec.~\ref{sec:evsel_phot}) and the missing energy
(Sec.~\ref{sec:evsel_lept}).  The figure
of merit (FOM) is the anticipated ratio of the signal yield to its statistical
uncertainty for \egcms between 1.8 and 2.8\gev, taking into account the
limited size of the off-resonance sample used for continuum subtraction:
\begin{equation}
 \mathrm{Statistical\ FOM} = \frac{\sig}{\sqrt{\sig+\bkb+(\bkc/f_\mathrm{off})}}
 \ .
\end{equation}
Here \sig, \bkb and \bkc are the estimated yields in the on-resonance
data of signal, \BB background and continuum background events,
respectively (after event selection), based on MC simulation, and  $f_\mathrm{off}$ is the fraction of total
luminosity accumulated off-resonance,
$\lum_\mathrm{off}/(\lum_\mathrm{on} + \lum_\mathrm{off}) = 0.0949$.

The selection criterion for each of the five variables was optimized in turn,
while holding the criteria for the others fixed, 
and the process repeated until a stable optimal selection was found.

\subsection{Overall Signal Efficiency}
\label{sec:evsel_effic}

The probability that a signal event is observed and survives the
event-selection process is approximately 2.5\%,
while only 0.0005\% of the continuum and 0.013\% of the \BB backgrounds
remain in the sample. Figure~\ref{fig:egamma_estimated}(b) shows
the expected signal and background distributions after all selection
criteria.

\begin{figure}[tb]
 \begin{center}
  \includegraphics[width=0.48\textwidth]{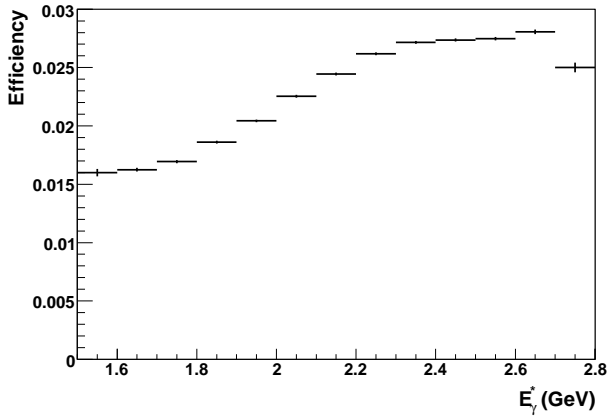}
  \caption{Combined acceptance and event-selection efficiency \vs
   measured \egcms for a KN model with $\mb=4.65\gevcc$.  Uncertainties
   are from MC statistics.  Corresponding efficiencies for
   a kinetic-scheme model with parameters set to HFAG world-average
   values are within 1\% (relative) of the values plotted.}
  \label{fig:efficKN465}
 \end{center}
\end{figure}

The photon spectrum is measured in bins of reconstructed \egcms.  Hence
the signal efficiency is presented here in terms of that quantity.
The selection efficiency for MC signal events, \ie, the fraction
of the events in a given range of \egcms that survive all the selection 
criteria described above,
is calculated in 100-MeV bins of reconstructed \egcms and also 
for wide ranges (such as 1.8 to 2.8\gev).  The overall
signal efficiency also includes an acceptance component, the probability
for the photon to enter the fiducial region of the EMC.  This is
available only as a function of true \egcms (the photon energy before
resolution smearing), since reconstructed
\egcms is defined only for accepted photons.  However, because the
variation of the acceptance efficiency is weak, 
it can be combined with the selection efficiency to
provide an overall efficiency in bins or ranges of reconstructed \egcms.
Figure~\ref{fig:efficKN465} shows the result.

\section{Continuum Backgrounds}
\label{sec:contin}

The continuum background is estimated using off-resonance data scaled
according to the ratio of the luminosity times 
the $\epem\to\qqbar$ cross section for the on- and off-resonance
data sets. Since continuum data are collected 40\mev below
the \FourS resonance, the center-of-mass energy is 0.4\% lower than the
center-of-mass energies for a typical \BB event.  In order to account
for this difference, the energy of a high-energy photon candidate in
off-resonance data is scaled by $m_{\FourS}/\sqrt{s_\mathrm{off}}$,
where $m_{\FourS}$ and $\sqrt{s_\mathrm{off}}$ are the mass of the
\FourS system and the center-of-mass energy of the off-resonance data
event, respectively.

\section{\BB Backgrounds}
\label{sec:BB}

\subsection{Overview}
\label{sec:BB_over}

\begin{table*}[!tb]
 \begin{center}
  \caption{The \BB background composition after all selection cuts, 
   according to the \babar\ Monte Carlo simulation, and the correction
   factors determined for each component.  Classification is according
   to the true MC particle associated with the high-energy photon,
   and to the true parent of that particle.  The ``{\PB}'' category under
   ``Parent'' corresponds to high-energy photons from final-state
   radiation.  The ``Other'' category consists of hadrons other than
   \Pan{}'s.  The ``None'' category consists of backgrounds unassociated
   with the primary event, mostly from out-of-time Bhabha-scattering
   events; such ``photons'' appear in the simulation via the beam-background
   mixing described in Sec.~\ref{sec:data}.  While all numbers are
   actually computed and applied in 100-MeV bins of \egcms, they are
   illustrated here for the overall signal region (1.80 to 2.80\gev) and \BB
   control region (1.53 to 1.80\gev).  The ``Subsection'' column refers to
   where each correction is discussed.   Note that the \Pgpz and \Pgh
   correction factors implicitly include the tagging efficiency correction
   described in Sec.~\ref{sec:BB_corr}; this tagging correction is 
   not included elsewhere in the table.
  }
  \label{tab:BBcomposition}
  \vspace{0.1in}
  \addtolength{\extrarowheight}{1.5pt}
  \begin{tabular*}{14cm}{@{\extracolsep{\fill}}lcccccc} \hline \hline
   \multicolumn{2}{c}{MC Category} & \multicolumn{2}{c}{1.53 to 1.8\gev} &
   \multicolumn{2}{c}{1.8 to 2.8\gev} & \multirow{2}{*}{Subsection} \\ \cline{1-2} \cline{3-4} \cline{5-6}
   Particle & Parent  & MC Fraction & Corr. Factor & MC Fraction & 
   Corr. Factor & \\ \hline
   Photon  & \Pgpz   & 0.5390 &$ 1.05 $& 0.6127 &$ 1.09 $&\ref{sec:BB_pizeta}\\
   	   & \Pgh    & 0.2062 &$ 0.79 $& 0.1919 &$ 0.75 $&\ref{sec:BB_pizeta}\\
   	   & \Pgo    & 0.0386 &$ 0.80 $& 0.0270 &$ 0.80 $&\ref{sec:BB_meson} \\
   	   & \Pghpr  & 0.0112 &$ 0.52 $& 0.0082 &$ 1.13 $&\ref{sec:BB_meson} \\
   	   & \PB     & 0.0362 &$ 1.00 $& 0.0194 &$ 1.00 $&\ref{sec:BB_fsr}   \\
   	   & \PJgy   & 0.0061 &$ 1.00 $& 0.0071 &$ 1.00 $&\ref{sec:BB_corr}  \\
   	   & \Pepm   & 0.0967 &$ 1.07 $& 0.0619 &$ 1.07 $&\ref{sec:BB_elec}  \\
   	   & Other   & 0.0035 &$ 1.00 $& 0.0032 &$ 1.00 $&\ref{sec:BB_corr}  \\
   	   & Total   & 0.9375 &  ---   & 0.9315 &  ---   &                   \\ \hline
   \Pepm   & Any     & 0.0411 &$ 1.65 $& 0.0333 &$ 1.68 $&\ref{sec:BB_elec}  \\
   \Pan    & Any     & 0.0170 &$ 0.35 $& 0.0243 &$ 0.15 $&\ref{sec:BB_nbar}  \\
   Other   & Any     & 0.0029 &$ 1.00 $& 0.0028 &$ 1.00 $&\ref{sec:BB_corr}  \\ \hline
   None    &         & 0.0015 &$ 1.00 $& 0.0079 &$ 1.00 $&                   \\ \hline
   \hline
  \end{tabular*}				  
 \end{center}
\end{table*}

The background from non-signal \BB events arises either from real
photons from the decays of low-mass mesons (with \Pgpz
and \Pgh responsible for most of the background) or from other
particles faking photons.  

The \BB background
remaining after event selection is estimated using the MC simulation
as an approximate starting point.  Various control samples are then
used to correct most of the significant components of this background 
according to data/MC yield ratios measured as a function of appropriate
kinematic variables.  The corrections are applied in 100-MeV bins
of \egcms.  The uncertainties of these factors (along with small
uncertainties from MC statistics) constitute the \BB systematic errors.
These can be highly correlated between \egcms bins.  The remainder
of Sec.~\ref{sec:BB} details the individual corrections, as well as a
more global correction to the lepton-tagging efficiency.  

The event simulation tells us the true (generated) particle that most
closely corresponds to the reconstructed high-energy photon candidate.
This allows the categorization of selected events according to the
origin of that candidate.
Table~\ref{tab:BBcomposition} lists the MC fractions by category and 
the corresponding correction factors averaged over two broad \egcms intervals,
covering the \BB control region and the signal region.

\subsection{\Pgpz and \Pgh Corrections}
\label{sec:BB_pizeta}

About 80\% of MC-predicted \BB background in the signal region arises from
$\PB\to X\pizeta$ with $\pizeta\to\Pgg\Pgg$.  This contribution is
dominated by highly-asymmetric \pizeta decays, in which a second photon
has much lower energy than the selected high-energy photon.
To correct MC predictions for these inclusive
\PB decays in the phase space region selected for the \bxsg analysis, 
inclusive \Pgpz and \Pgh samples are defined by applying the same selection 
criteria but omitting the \Pgpz and \Pgh vetoes.  To enhance statistics 
for these studies the minimum requirement on \egcms is relaxed from 1.53\gev to
1.03\gev, and for \Pgh{}'s the minimum laboratory-frame energy for
the low-energy photon is relaxed from 230\mev to 75\mev.  

\subsubsection{Scaling of MC \Pgpz and \Pgh yields to data}
\label{sec:BB_pizeta_corr}

The yields of \pizeta are measured in bins 
of $E^{*}_{\pizeta}$ by fitting the distributions of \gg mass ($m$) in
simulated \BB background, on-resonance data and off-resonance data.

The signal shape for \Pgpz is the sum of two Gaussian functions ($G_1$ and
$G_2$) with different means ($\mu_1$ and $\mu_2$) and rms widths
($\sigma_1$ and $\sigma_2$) plus a low-mass power-law tail (parameters
$p$ and $\lambda$):
\begin{equation}\label{eq:pi0shape}
 f(m) = \left\{
  \begin{array}{ll}
   A\left[f_1 G_1(m) + (1-f_1)G_2(m) \right]                 & m \ge m_0 \\
                                                             &       \\
   B{\left(\displaystyle\frac{p\sigma_1/\lambda}{m_0-m + p\sigma_1/\lambda}\right)}^p   & m <m_0 \ ,
  \end{array}
 \right.
\end{equation}
where $m_0 \equiv (\mu_1 - \lambda\sigma_1)$,
$A$ and $f_1$ govern the normalizations of the two Gaussian functions,
and $B$ is set by requiring continuity at $m = m_0$.
The signal shape for \Pgh is a single Gaussian with two such power-law
tails with separate parameters.

\begin{figure*}[ht]
 \begin{center}
  \includegraphics[width=0.95\textwidth]{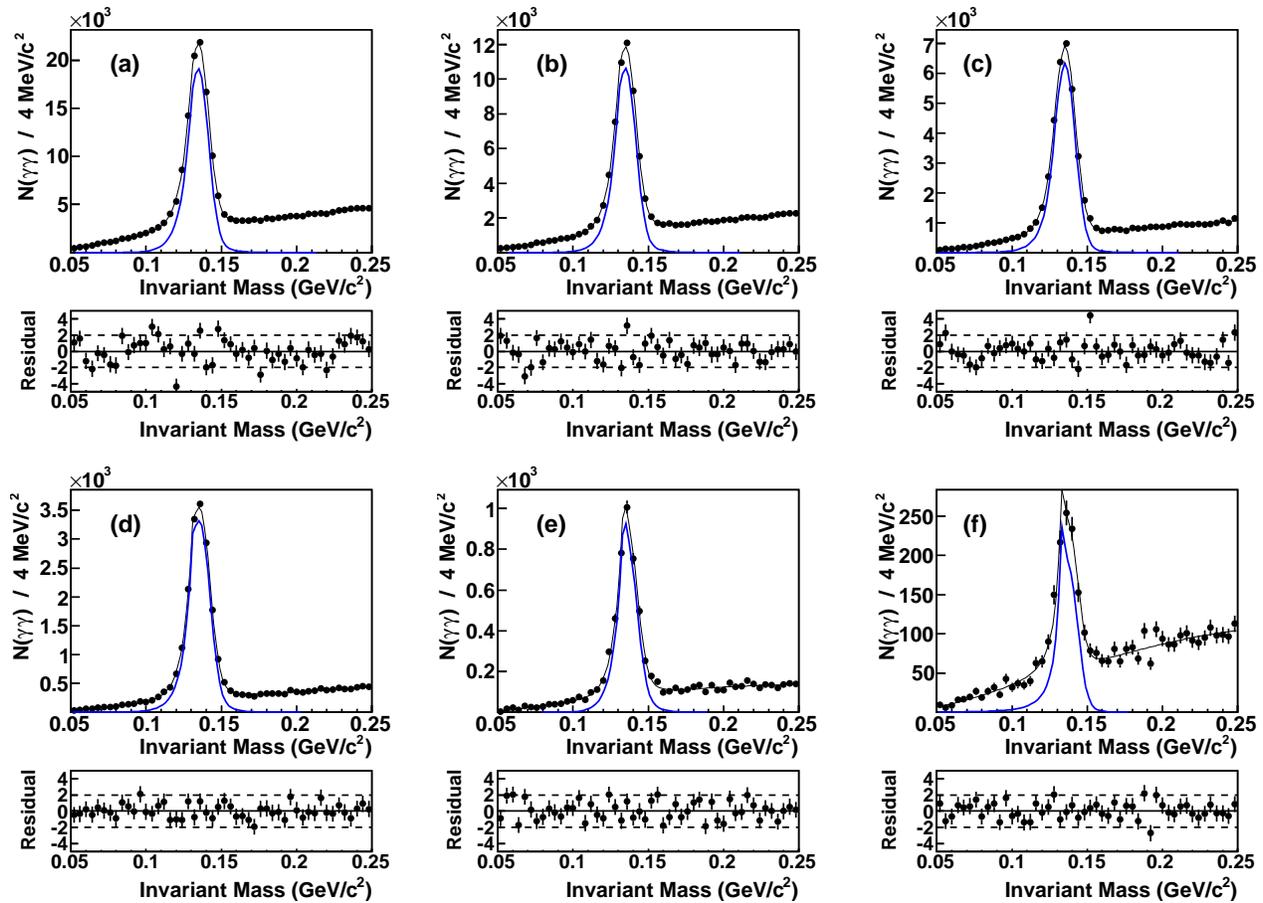}
  \vspace{-0.1in}
  \caption{Fits to spectra of \gg combinations per interval of \gg mass for
   \Pgpz on-resonance data, in bins of
   $E_\Pgpz^*$: (a) 1.4 to 1.6\gev, (b) 1.6 to 1.8\gev, (c) 1.8 to
   2.0\gev, (d) 2.0 to 2.2\gev, (e) 2.2 to 2.4\gev, and (f) 2.4 to
   3.0\gev.  For each bin, the top plot shows the data (points),
   the total fit (upper curve) and the signal component of the fit
   (lower curve); the bottom plot shows the residuals, defined as (data
   - fit)/(data uncertainty).  For the first few bins the signal shape
   doesn't precisely reproduce the center of the peak, an effect also
   seen in fits to MC \Pgpz signal only, but this does not affect the
   integrated signal yield.  }\label{fig:datafit_piz}
  \quad
 \end{center}
\end{figure*}

\begin{figure*}[ht]
  \begin{center}
    \includegraphics[width=0.95\textwidth]{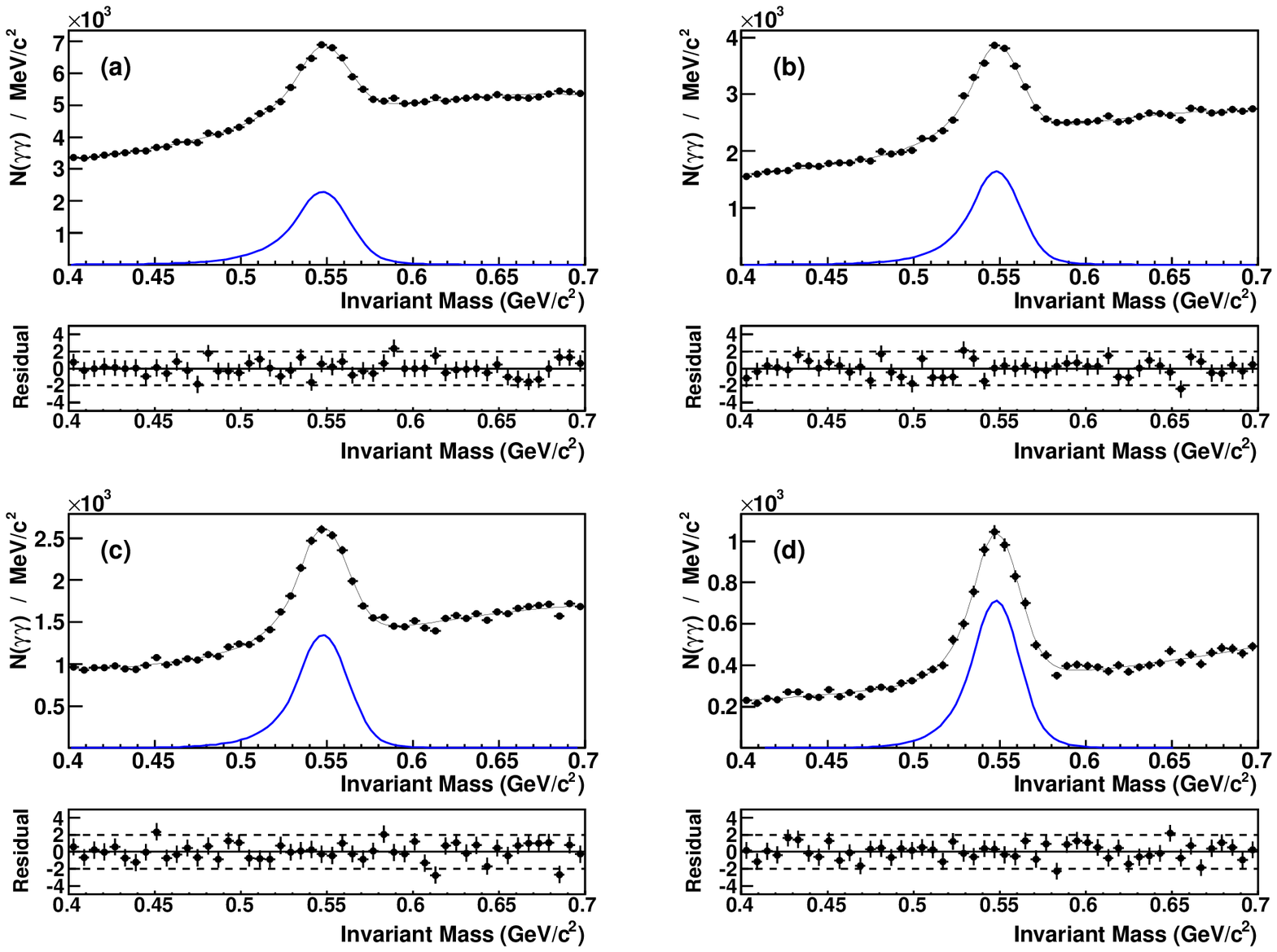}
    \vspace{-0.1in}
    \caption{Fits to spectra of \gg combinations per interval of \gg mass for
     \Pgh on-resonance data, in bins of
      $E_\Pgh^*$: (a) 1.5 to 1.7\gev, (b) 1.7 to 1.9\gev, (c) 1.9 to
      2.2\gev, and (d) 2.2 to 2.6\gev.  For each bin, the top plot shows
      the data (points), the total fit (upper curve) and the signal
      component of the fit (lower curve); the bottom plot shows the
      residuals, defined as (data - fit)/(data uncertainty).  }
    \label{fig:datafit_eta}
  \end{center}
\end{figure*}

The fit is carried out in several stages.  First,
a signal shape is determined for \BB MC events in which the reconstructed 
\gg pair derive from a true \Pgpz or \Pgh.  For purposes of this study,
these events are termed ``signal''.  Next, for both MC and on-resonance
data events, the mass spectrum of all \gg pairs which include the high-energy
photon is fit in the \pizeta mass region to 
signal plus a background shape, with some signal tail 
parameters fixed to their values from the signal-only fit.  
This procedure is validated by comparing the extracted
signal yields from this MC fit to those of true signal:  averaging
the absolute values of the differences over energy bins, 
the agreement is 1.3\% of the yield for \Pgpz and 2.1\% for \Pgh.  The fits to
on-resonance data are shown in Figs.~\ref{fig:datafit_piz}
and~\ref{fig:datafit_eta}.  Finally, the
off-resonance data are fit with all signal-shape parameters fixed to
their on-resonance fit values, with only the signal yield and background
parameters left free.  Then, in each $E^{*}_{\pizeta}$ bin, the
\pizeta correction factor is the ratio of the on-resonance minus
luminosity-weighted off-resonance \pizeta yield to the luminosity-weighted
MC \pizeta yield.  Systematic uncertainties from the fit are found
by individually varying the fixed parameters in the on-resonance data
fits, and also allowing for the MC fit-validation checks.
The resulting correction factors and their uncertainties are shown in
Tables~\ref{tab:piz_corrections} and~\ref{tab:eta_corrections}.  

\begin{table}[!b]
  \addtolength{\extrarowheight}{1.5pt}
 \begin{center}
  \caption{The \Pgpz correction factors from ratios of data-to-MC fitted yields.  
   The first and second sets of uncertainties are statistical and systematic,
   respectively.}
  \label{tab:piz_corrections}
  \vspace{0.1in}
  \begin{tabular*}{8.6cm}{@{\extracolsep{\fill}}cclc} \hline \hline
   & \Pgpz CM Energy (GeV) & \multicolumn{1}{c}{Correction Factor} & \\ \hline
   & 1.4 to 1.6               & $ 0.959 \pm 0.006 \pm 0.013 $  & \\
   & 1.6 to 1.8               & $ 0.933 \pm 0.009 \pm 0.012 $  & \\
   & 1.8 to 2.0               & $ 0.990 \pm 0.012 \pm 0.031 $  & \\
   & 2.0 to 2.2               & $ 0.992 \pm 0.016 \pm 0.013 $  & \\
   & 2.2 to 2.4               & $ 0.899 \pm 0.035 \pm 0.018 $  & \\
   & 2.4 to 3.0               & $ 1.489 \pm 0.259 \pm 0.076 $  & \\ \hline
   \hline
  \end{tabular*}
 \end{center}
\end{table}

Correction factors to the \BB MC predictions in 100\mev bins of \egcms,
along with their uncertainties and correlations, are determined by
applying the above factors event-by-event to MC events passing the \bxsg
selection criteria.

\begin{table}[!b]
  \addtolength{\extrarowheight}{1.5pt}
 \begin{center}
  \caption{The \Pgh correction factors from ratios of data-to-MC fitted yields.  
   The first and second sets of uncertainties are statistical and systematic,
   respectively.}
  \label{tab:eta_corrections}
  \vspace{0.1in}
  \begin{tabular*}{8.6cm}{@{\extracolsep{\fill}}cclc} \hline \hline
   & \Pgh CM Energy (GeV) & \multicolumn{1}{c}{Correction Factor} & \\ \hline
   & 1.5 to 1.7              & $ 0.948 \pm 0.029 \pm 0.034$ & \\
   & 1.7 to 1.9              & $ 0.744 \pm 0.026 \pm 0.029$ & \\
   & 1.9 to 2.2              & $ 0.654 \pm 0.024 \pm 0.017$ & \\
   & 2.2 to 2.6              & $ 0.864 \pm 0.049 \pm 0.027$ & \\ \hline
   \hline
  \end{tabular*}
 \end{center}
\end{table}

\subsubsection{Additional corrections for low-energy photon efficiency}
\label{BB_pizeta_lephot}

While the procedure described above accounts for data-MC differences in 
the produced \Pgpz and \Pgh yields after the full selection, including the 
efficiencies for lepton tagging and 
for detecting the high-energy photon, it does not properly account for
data-MC differences in the detection efficiency for the
low-energy photon from a \Pgpz or \Pgh decay.  This is because the fits to the
samples studied above count events in which that photon is detected and forms
a \gg pair in the \pizeta mass peak, whereas in the \bxsg analysis 
a $\PB\to X\pizeta$ background event is accepted if the low-energy
photon is \textsl{not} found (or forms a reconstructed \gg pair mass
outside the veto window).  Thus the procedure corrects in the
wrong direction for data-MC differences in low-energy photon detection
efficiency.  

Correcting for low-energy photon efficiency is another multi-step
process.  First, \babar\ measurements of \Pgpz detection efficiency
are taken from studies of the initial-state radiation (ISR) process 
$\epem\to\Pgo\Pgg$ with $\Pgo\to\Pgpp\Pgpm\Pgpz$.
Here the precise knowledge of the beam energies and the measured charged
pions and high-energy ISR photon allow the four-momentum of the
\Pgpz to be predicted.  The measured efficiency difference between
data and MC events is adjusted 
to match the \Pgpz CM-frame momentum distributions for $\PB\to X \Pgpz$
background in the \bxsg analysis and for the inclusive-\Pgpz studies
described in section~\ref{sec:BB_pizeta_corr}.  
The result is a data-MC fractional efficiency 
difference of $(-4.1\pm 0.7)\%$ for the \bxsg selection and
$(-3.5\pm 0.6)\%$ in the inclusive \Pgpz studies.  
Part of these data-MC efficiency differences are accounted for by a data-MC
difference of $(-1.15 \pm 0.65)\%$ measured for the high-energy
photon, as detailed in Sec.~\ref{sec:bf_syst_photon} below.
Subtracting this and combining the errors in quadrature leaves 
$(-2.95\pm 0.95)\%$ (\bxsg selection) and $(-2.35\pm 0.9)\%$ ($\PB\to X \Pgpz$ 
selection) as due to the low-energy photon.  

Finally, the $\PB \to X \Pgpz$ samples are used to separately study the
roughly 25\% of low-energy photons in the current measurements that
have laboratory-frame energies below 80\mev.  This is necessary because,
in order to suppress backgrounds, the ISR analysis effectively covers
cosines of the \Pgpz helicity angle (which equals the decay energy
asymmetry) only up to about 0.9.  Because of this, low-energy photons
below $\approx 80\mev$ are not adequately represented in the ISR analysis.
The data/MC ratios for $\PB \to X \Pgpz$ samples are sensitive to
branching fractions and detection efficiencies for high-energy and
low-energy photons.  These ratios in \Pgpz energy bins can be used to
determine the \textsl{relative} efficiency corrections for low-energy
photons below 80\mev compared to those above 80\mev, since both sets
of low-energy photons derive from \Pgpz mesons with the same kinematic
properties, and the accompanying high-energy photons hardly differ.
This is accomplished by separately applying the \Pgpz mass-spectrum
fitting technique for \Pgpz mass combinations involving low-energy
photons in these two regions.
An additional data-MC fractional efficiency correction 
of $(-3.6\pm 1.1)\%$ is derived for only those \Pgpz decays involving 
these photons below 80\mev.  There is no corresponding effect for \Pgh{}'s, 
where the minimum photon energy is always at least 75\mev.  

To determine the effect of these low-energy-photon efficiency differences
on the analysis, the \BB simulation is rerun with the specified fractions
of low-energy photons from \pizeta decays discarded.  The result
is an additional factor of
$1.105\pm 0.029$ for the \Pgpz component of \BB background in the \bxsg
analysis, and $1.041\pm 0.015$ for the \Pgh component.  The \Pgpz
and \Pgh errors are mostly correlated.  These factors multiply those
obtained from the inclusive \pizeta data/MC yield comparisons.

\subsection{Other Meson Decays}
\label{sec:BB_meson}

Radiative decays of inclusively produced \Pgo (in the $\Pgpz\Pgg$ mode)
and \Pghpr (in various decay modes) can lead to high-energy photons not
already accounted for among the inclusive \Pgpz{}'s.  
As seen in Table~\ref{tab:BBcomposition}, these contribute several
percent of the simulated \BB background.  
We have studied inclusive \Pgo and \Pghpr production in \FourS events.
Correction factors are determined in bins of CM-frame meson momentum 
($p_{\Pgo}^*$ or $p_{\Pghpr}^*$) as the ratios of measured inclusive
branching fractions to the values used in the MC simulation for the 
current analysis.

The \Pgo measurements cover the $p_{\Pgo}^*$ range from 0 to 2.25\gevc
in 0.25\gevc bins.  Correction factors range from 0.7 to 1.3, with
uncertainties averaging 0.17.  

\begin{figure*}[!bt]
 \begin{center}
  \includegraphics[width=0.35\textwidth]{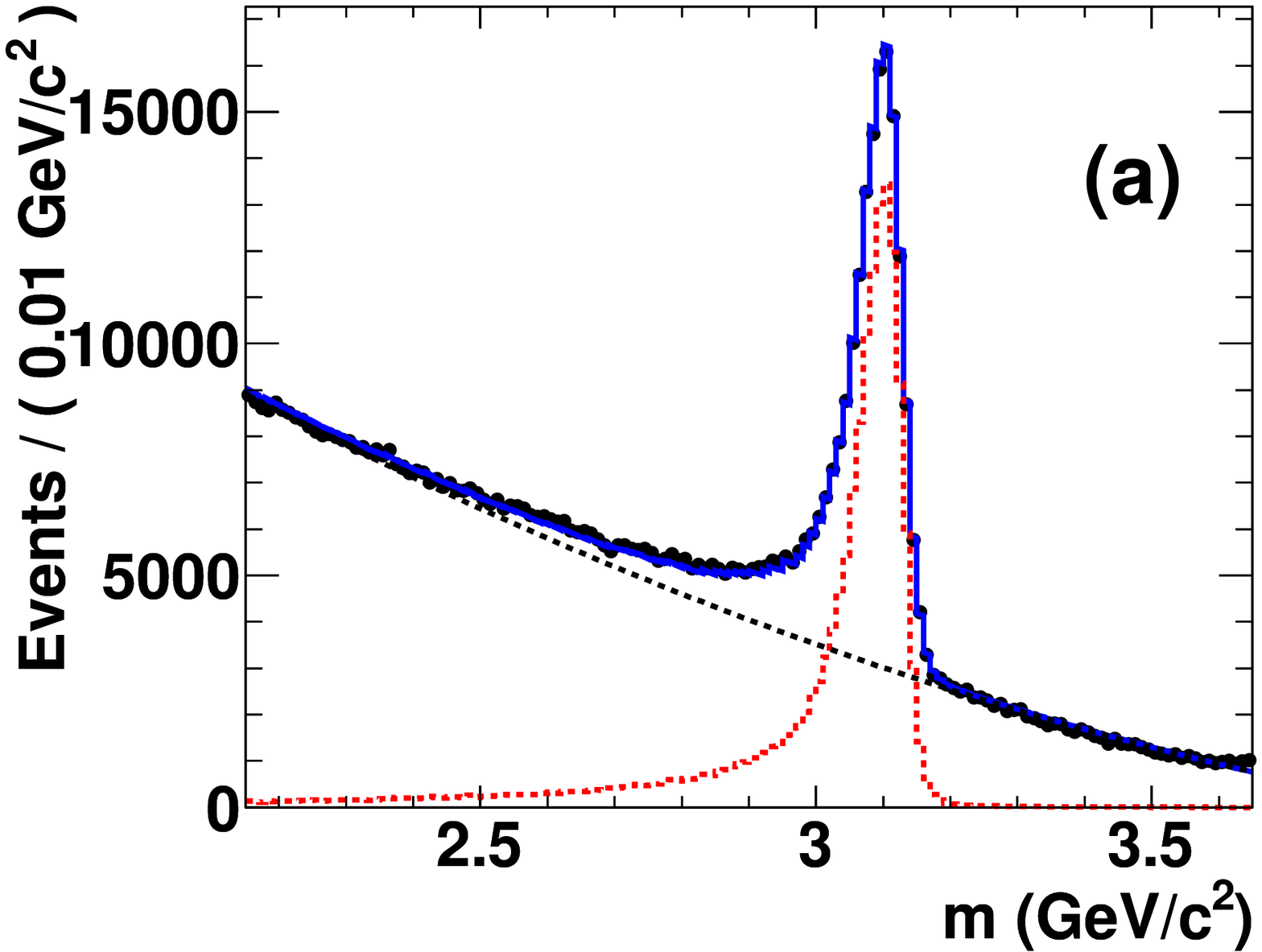}
  \includegraphics[width=0.35\textwidth]{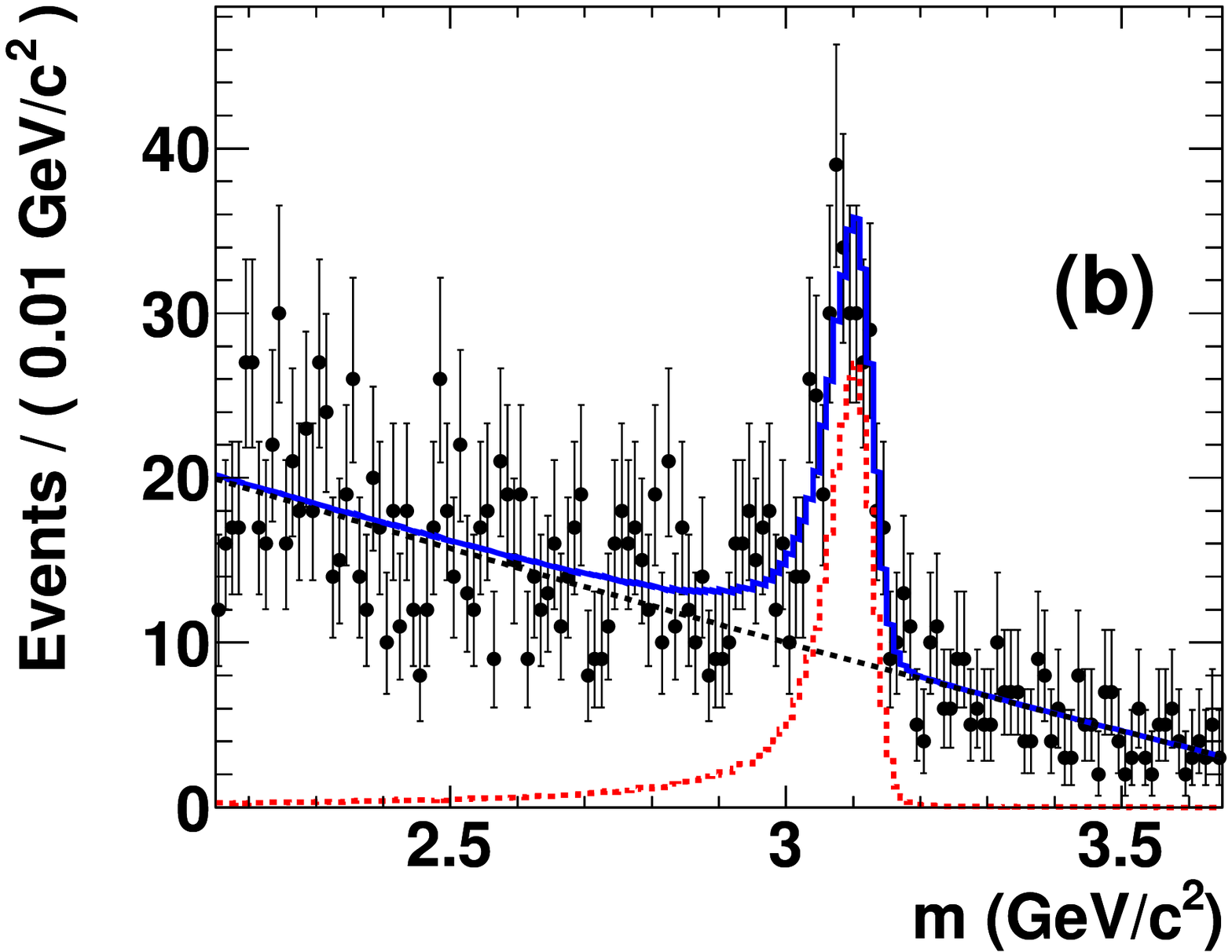}
  \includegraphics[width=0.35\textwidth]{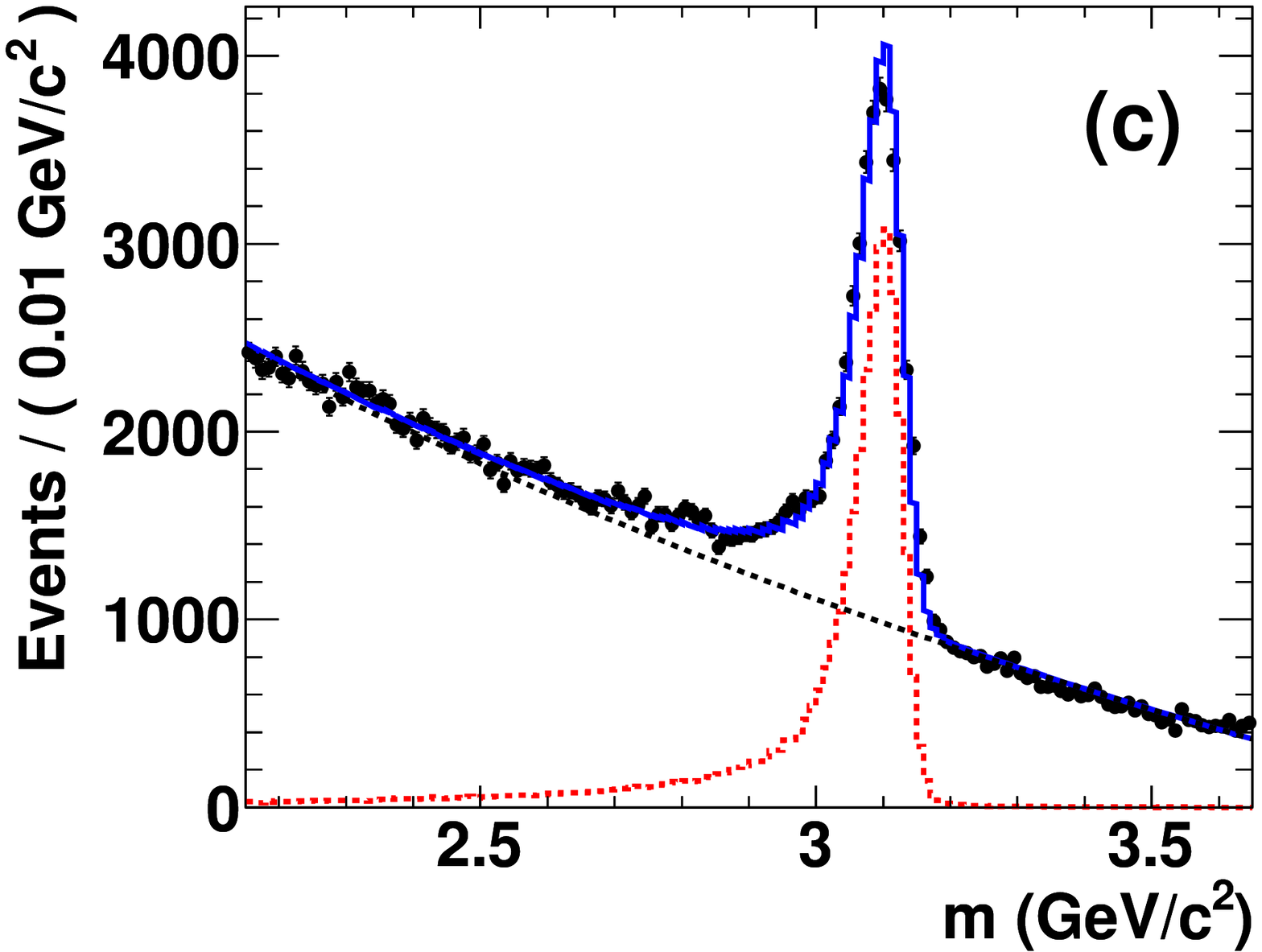}
  \includegraphics[width=0.35\textwidth]{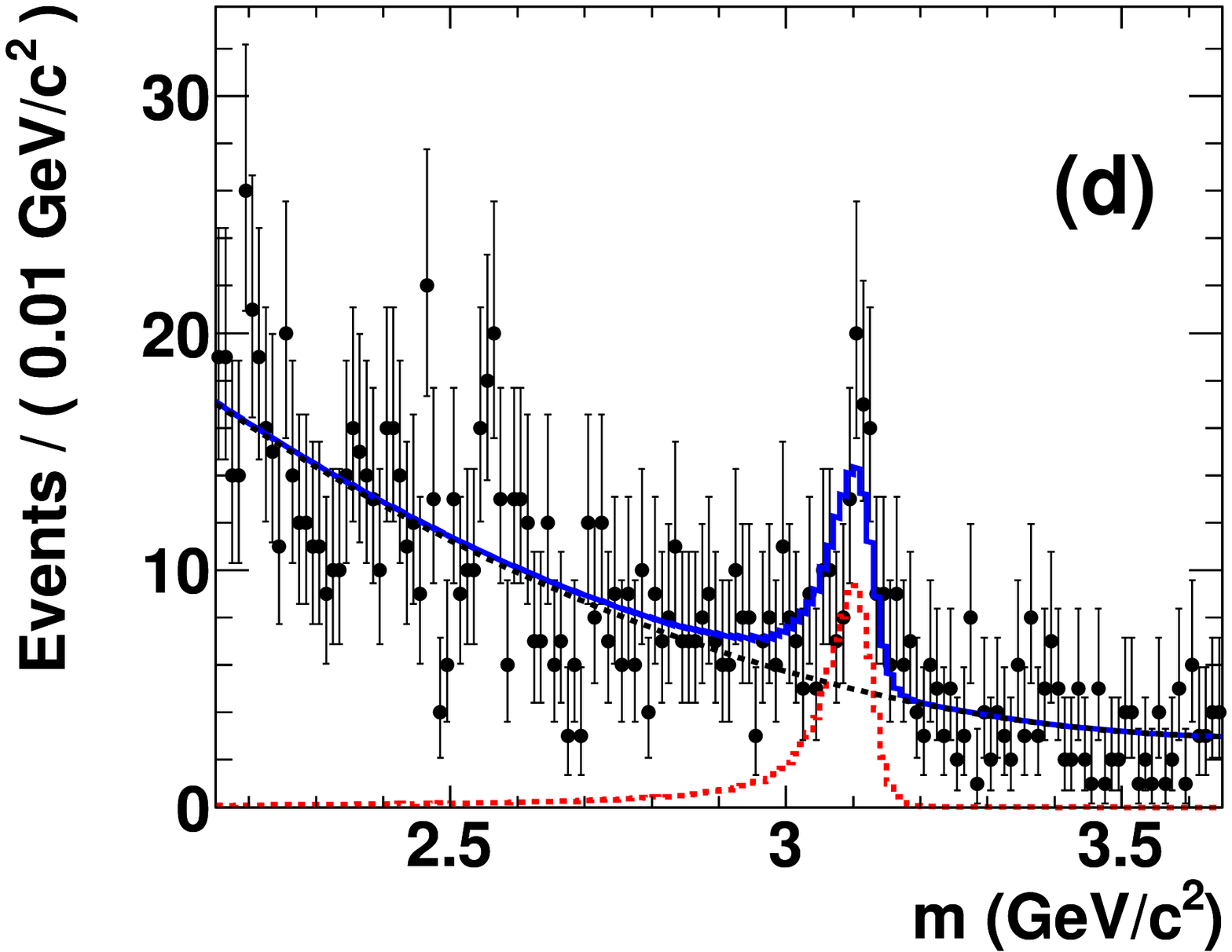}
  \caption{Invariant \epem mass distributions for $\PJgy\to\epem$ 
   electron-inefficiency studies.  Plots (a) and (b) are for MC samples;
   plots (c) and (d) are for data samples.  Plots (a) and (c)
   includes all (eClus, eTrk) and (eTrk, eTrk) pairs, while
   plots (b) and (d) include only the (eClus, eTrk) pairs.  
   Dotted, dashed and solid curves show,
   respectively, the \PJgy signal and background components of each fit,
   and their sum.  The numbers of fitted \PJgy signal events are
   $156327\pm550$, $313\pm27$, $35825\pm281$ and $109\pm18$
   for plots (a), (b), (c) and (d), respectively.
  }
  \label{fig:electron_psi_study}
 \end{center}
\end{figure*}

Results for \Pghpr are divided into two regions of reduced momentum,
$x_{\Pghpr}=p_{\Pghpr}^*/\sqrt{{E_{beam}^*}^2-m_{\Pghpr}^2}$.  
For $0.39 < x_{\Pghpr} < 0.52$, direct \HepProcess{\PB\to X \Pghpr}\xspace
decays are dominant, and the correction factor is $1.86\pm 0.61$. 
For $0.10 < x_{\Pghpr} < 0.39$,  decays via an intermediate charm-meson
state are dominant, and the correction factor is $0.35\pm 0.19$. The first
range is most important in the signal region for \bxsg, while the
second range is most important in the \BB control region.

Both \Pgo and \Pghpr corrections are applied event-by-event in the
\BB simulation in order to obtain correction factors in \egcms bins.

\subsection{Electron Backgrounds}
\label{sec:BB_elec}

Electrons and positrons contribute to the photon background in two ways
(see Table~\ref{tab:BBcomposition}).  First, there are events in which the 
reconstructed photon is from  hard bremsstrahlung from an \Pepm interacting 
with the material in the inner portion of the \babar\ detector (beam pipe,
SVT, and material between the SVT and the active area of the DCH).
Second, there are events in which the reconstructed 
photon is faked by an electron due to a failure to reconstruct a track or 
to match a track to the calorimeter energy deposit. The primary source of
the \Pepm in both of these categories is semileptonic \PB decay.

The bremsstrahlung process is reliably simulated by \textsc{GEANT4}, so
there is no correction to the simulation for this background.  But a 
3\% systematic error is assigned based on the precision with
which the amount of detector material has been measured.

The misreconstructed electron background is measured using a tag and probe
method with $\PB \to X \PJgy(\PJgy\to\Pep\Pem)$ data. This sample closely 
models the particle multiplicity in $\bxsg$ events.
The $\PJgy$ in this decay mode is normally
reconstructed by requiring two electrons with 
tracks associated with EMC clusters. If the
track is misreconstructed there will still be a cluster but without an
associated track.  In this case the \PJgy is reconstructed from
this unassociated cluster along with the other electron, which has a track
matched to a cluster.
Because either of the two leptons could have a misreconstructed track,
the track inefficiency may be measured as
\begin{equation}
 1-\epsilon =\frac{N(\PJgy(\mathrm{eClus,eTrk}))}
 {2N(\PJgy(\mathrm{eTrk,eTrk}))+N(\PJgy(\mathrm{eClus,eTrk}))}\ ,
\label{eq:trkIneff}
\end{equation}
where $N(\PJgy(\mathrm{eClus,eTrk}))$ and $N(\PJgy(\mathrm{eTrk,eTrk}))$
are the numbers of 
$\PJgy\to\epem$ events with one and two reconstructed tracks, respectively.
These yields are extracted from fits to distributions of \Pep\Pem invariant 
mass, computed from the four-momenta of the track found for one lepton
(the ``tag'') and the EMC cluster for the other.  
The value of $1-\epsilon$
is compared between data and MC samples to derive a correction
factor for the simulation. There is a  large combinatoric background
in the one-track (eClus, eTrk) sample due to actual photons. However,
when an electron track has been misreconstructed 
there are still a number of DCH hits around the trajectory from 
the vertex to the EMC cluster.  The background is significantly
reduced by requiring a minimum number of 20 hits in a road of 1-cm
radius around this trajectory. 

Figure~\ref{fig:electron_psi_study} shows an example of 
fits to the \epem mass combinations corresponding to the numerator
and denominator of Eq.~(\ref{eq:trkIneff}) for data and
MC simulation.  The mass is computed
from the track associated with a tag lepton and the EMC cluster 
associated with the other lepton; hence (eTrk, eTrk) combinations
are entered twice, with different masses, once for each tag.
The simulation underestimates the
fraction of misreconstructed tracks by a factor of 
$1.57 \pm 0.27(\mathrm{stat}) \pm 0.22 (\mathrm{syst})$,
where ``stat'' and ``syst'' denote the statistical and systematic
uncertainties, respectively.  The systematic error comes predominantly from 
uncertainties in the line shape assumed in the invariant mass fit and from
varying the requirements on the road width and the number of DCH hits.
Consequently, the MC estimate of the \BB background to \bxsg from 
misreconstructed electrons is increased by a factor of $1.57 \pm 0.35$.

\subsection{Antineutrons}
\label{sec:BB_nbar}

The only significant hadron background to high-energy photons is from
antineutrons, which have a neutral signature and can,
by annihilating in or just before the EMC, deposit
a large amount of energy.  A large fraction of such background is removed
by the requirement on maximum lateral moment (Sec.~\ref{sec:evsel_phot}).
There are two sources of potential bias in the predicted yield:  the 
inclusive $\PB\to X\Pan$ branching fraction and \Pan momentum spectrum
in the event simulation,
and the \textsc{GEANT4} simulation of the deposited energy
and its distribution in the EMC.  Because it is not possible to identify
or measure the four-momentum of an \Pan in the \babar\ detector, there
are no control samples of \Pan{}'s available to study these effects.
Hence estimates
of their size have been based on comparison of data to simulated events
involving \PB decays to \Pap{}'s.

The inclusive \Pan production spectrum in the \BB simulation is corrected
by the ratio of a measured inclusive \Pap spectrum to its corresponding
simulation.  Correction factors are applied as a function of CM-frame
antibaryon momentum.  They are close to 1.0 at momenta above 
about 0.9\gevc, increasing
to 2.0 for momenta from 0.5\gevc down to the lowest measured momentum
of about 0.3\gevc.  Uncertainties are typically 8\% to 12\%.
Below 0.3\gevc, a factor of 2.0 is assigned, with a larger uncertainty.
In addition, while the production of \Pap and \Pan from direct \PB
decays is related via isospin conservation, many of the antibaryons
arise from decays of \PgD{}'s or hyperons, which would require
separate correction factors. An additional uncertainty of 3\% accounts
for differences 
in fractions of direct \PB \vs \PgD \vs hyperon parentage of \Pan and \Pap.

Control samples of \Pap{}'s from the decay of
\PagL{}'s are used to compare data and MC
EMC response to \Pap{}'s as a function of laboratory-frame \Pap momentum.
Most \Pap{}'s are rejected by imposing the same upper limit on the
lateral moment of their EMC energy deposition pattern as used in
the \bxsg photon selection.
Correction factors are determined in bins of laboratory-frame momentum 
$p_\Pap$ \vs  \emcfrac, where 
\begin{equation}
 \emcfrac \equiv \frac{E_\mathrm{EMC}}{\sqrt{p_\Pap^2 + m_\Pp^2} + m_\Pp}
 \label{eq:emcfrac}
\end{equation}
is that fraction of the total energy from annihilation on a nucleon that 
is deposited in the EMC.  Corrections are computed as the ratio
of data to MC probability distribution functions (PDFs) for 
$\emcfrac > 0.5$, the only region that can yield an
apparent \egcms above 1.53\gev.  
The primary data-MC differences result from the larger average lateral
moment in data, an effect accentuated by restricting the lateral moment
to low (photon-like) values.  Hence the data have a considerably
smaller proportion of \Pap{}'s satisfying the selection than is the case for
MC events.  This inaccuracy of the simulation, and to a lesser extent an 
overestimate of the energy deposit itself, increases with increasing
\emcfrac, hence the data/MC correction factor becomes small as \emcfrac
increases.

\begin{figure}[!tb]
 \begin{center}
  \includegraphics[width=0.5\textwidth]{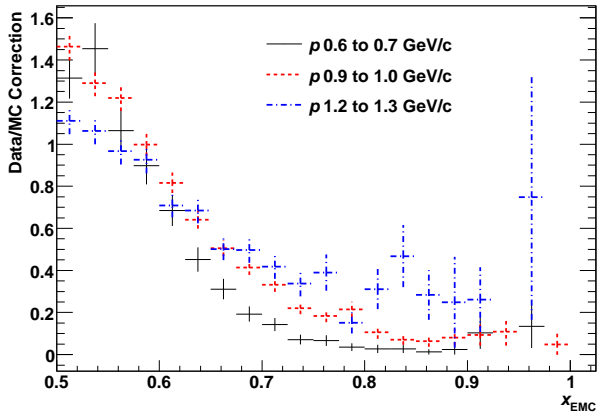}
  \caption{Data/MC correction factors for \Pap{}'s in 
   three illustrative intervals of laboratory
   momentum, \vs \emcfrac (see Eq.~\ref{eq:emcfrac}), from \PagL control
   samples, including the same maximum lateral moment criterion used for 
   \bxsg analysis.}
  \label{fig:pbar_ecal_low_corr}
 \end{center}
\end{figure}

Figure~\ref{fig:pbar_ecal_low_corr} illustrates the \Pap correction
factors in this two-dimensional space.  Note that in inclusive \PB
decays there are relatively few antibaryons with laboratory-frame 
momentum above 1.5\gevc.  

There are several difficulties in applying these results to antineutrons.  
First, because of energy loss in  the beam pipe and inner detector
components, \Pap{}'s
do not provide useful energy-deposit information
for laboratory-frame momenta below 0.5\gevc, whereas one-third of \Pan{}'s 
from \PB decay have momenta below 0.5\gevc.  A constant extrapolation of
correction factors to lower momenta is assumed, with a systematic uncertainty
set by including an additional factor of 1/2.  Second, \Pap{}'s enter the
EMC crystals at a larger angle of incidence than do \Pan{}'s,
because of the magnetic field, resulting in a larger lateral
moment for \Pap{}'s with $p_T < 0.7\gevc$ than for \Pan{}'s.  
A systematic uncertainty is assigned by increasing correction factors 
with laboratory $p_\Pap$ in this region to their values at just-higher,
unaffected, $p_\Pap$.  Third, because of a mistake in the version of
\textsc{GEANT4} implemented in \babar, simulated \Pap{}'s that stop before 
annihilating do not then annihilate.  This has been dealt with by increasing
the MC PDFs (decreasing the correction factors)
according to the fraction of \Pap{}'s that annihilate
for a given momentum.  Half of this correction is adopted as its
systematic uncertainty.

Correction factors to the simulated \Pan background in \egcms bins are
computed by applying event-by-event corrections for both the branching
fraction and the EMC response.  Systematic uncertainties are
obtained by redoing this for each of the systematic changes outlined above.
The resulting correction factors vary from about
0.4 to 0.04 as \egcms increases from 1.53 to 2.8\gev, with uncertainties
ranging from 1/4 to 1/2 of the correction factors. 

\subsection{Final-State Radiation}
\label{sec:BB_fsr}
Final-state radiation, most importantly from leptons, 
is incorporated into the \BB background 
simulation with PHOTOS~\cite{Golonka:2005pn}. The contribution is 
labeled as having \PB parentage in Table~\ref{tab:BBcomposition}. 
No correction is applied for this small component. Radiation from light 
quarks during the hadronization process is
not incorporated into the simulation. However, this contribution 
was computed for the previous \bxsg analysis~\cite{Aubert:2006gg},
where a photon spectrum based on the calculation in
Ref.~\cite{Ligeti:1999ea} was 
passed through the detector simulation and selection
criteria.  This contribution was found to be less than 0.3\%.

\subsection{Semileptonic Branching Fraction}
\label{sec:BB_semilept}

\begin{figure}[htb]
 \begin{center}
  \includegraphics[width=0.42\textwidth]{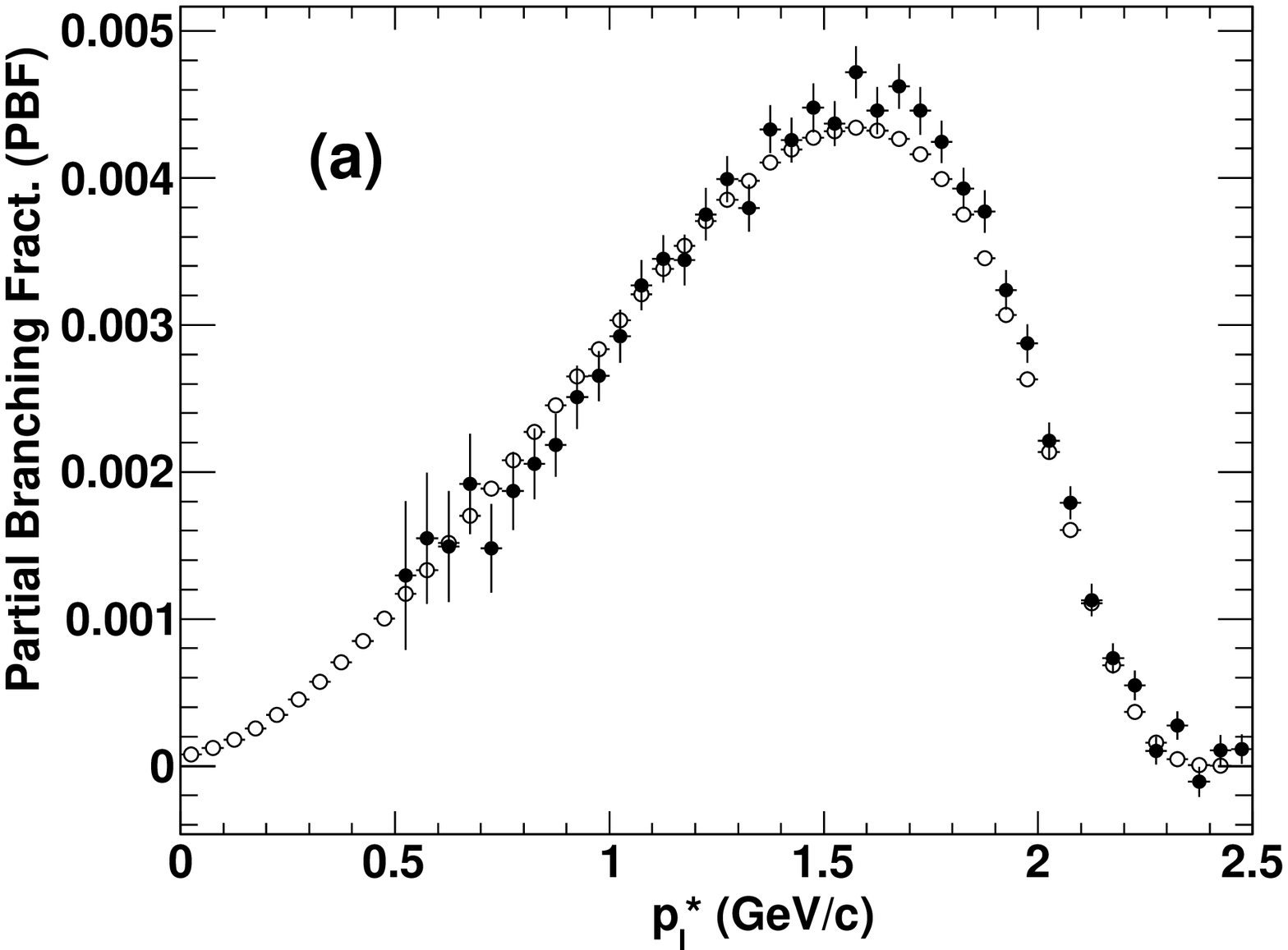}
  \includegraphics[width=0.42\textwidth]{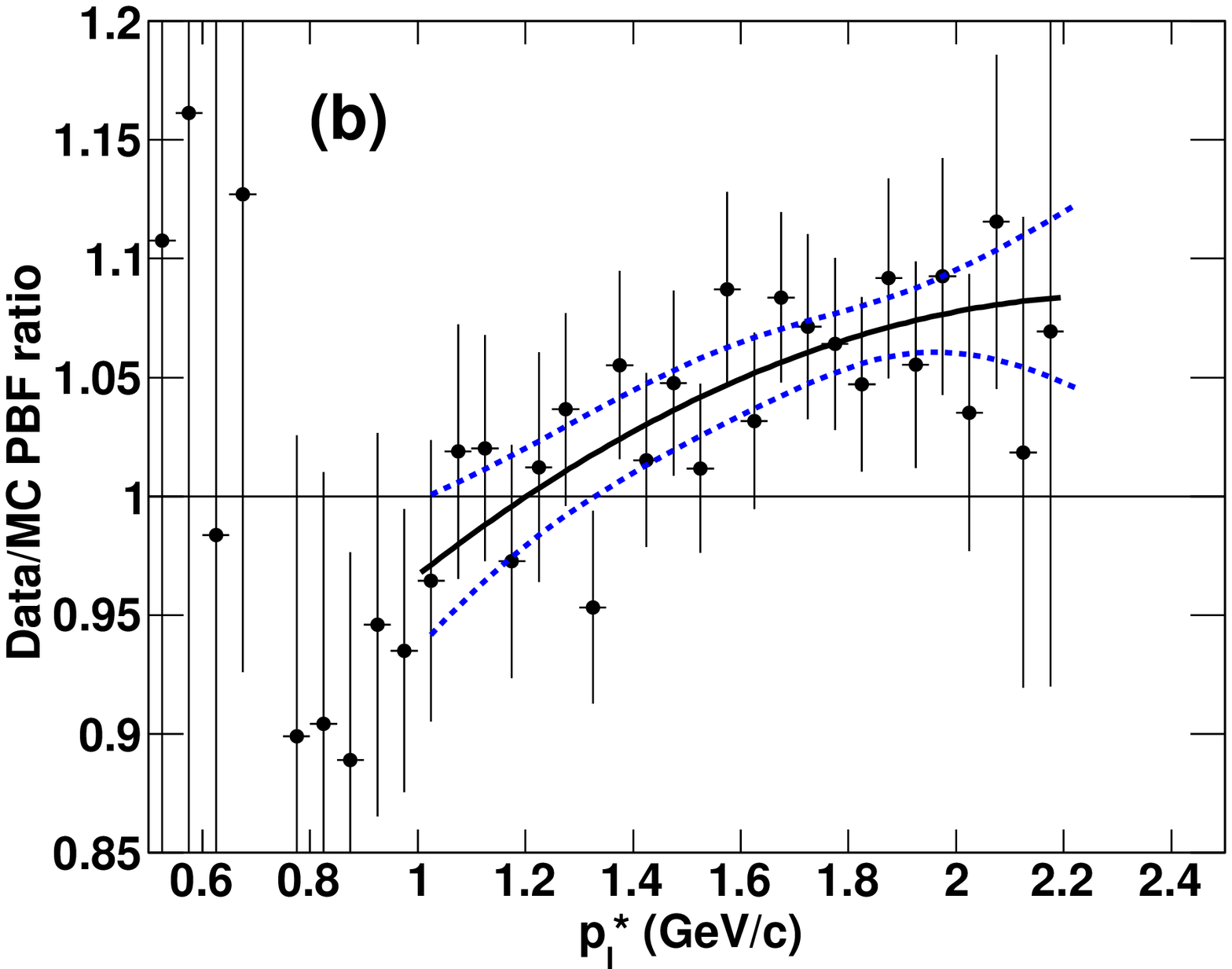}
  \caption{
   \PB-meson semileptonic partial branching fraction \vs 
   CM-frame lepton momentum, averaged over \PB charge states.
   Top:  \babar\ measurement~\cite{Aubert:2009qd} 
   (filled black circles) and values in the \BB MC simulation
   (open circles).  Bottom:  data/MC ratios, and results of a second-order
   polynomial fit from 1.0 to 2.2\gev.  
   The dashed curves show the $1\sigma$ error band.
  }
  \label{fig:SL_partial_BF}
 \end{center}
\end{figure}

The dominant source of tagging leptons above the minimum required momenta
(Sec.~\ref{sec:evsel_lept}) in both signal and \BB background events,
and also of electrons that fake high-energy photons 
(Table~\ref{tab:BBcomposition}) is the semileptonic decay of \PB mesons.
The MC simulation models \PB semileptonic decays as a sum of
exclusive processes.  But this sum does not accurately
reproduce inclusive measurements of semileptonic 
decays~\cite{Nakamura:2010zzi}.  A \babar\
inclusive electron measurement~\cite{Aubert:2004td, Aubert:2009qd} is used to
renormalize the simulated branching fractions as a function of CM-frame
lepton momentum $p_{\ell}^*$.  Figure~\ref{fig:SL_partial_BF} shows the
data and MC points and their ratio.  Correction factors are applied based
on the polynomial fit.  For most leptons relevant to this analysis
the correction is larger than unity.

This correction enters in two places in the analysis.  First, it affects
tagging efficiency.  By integrating over all lepton tags in events passing
selection criteria, a correction factor of $1.047 \pm 0.013$ is obtained, 
for \bxsg signal events, while for the
\BB MC sample the factor is $1.051 \pm 0.013$.  This correction is
independent of \egcms.  However, the procedure for normalizing the \Pgpz
and \Pgh background components to data implicitly takes this into account.
Hence the correction is applied only to \textsl{other} \BB components. 
(However, the corrections given in  Table~\ref{tab:BBcomposition}
for these components are derived before applying this additional semileptonic
correction.)

In addition to its effect on lepton tagging, the semileptonic
correction affects the two backgrounds in which an \Pepm fakes
a high-energy photon.  The corrections (which \textsl{are} included in 
Table~\ref{tab:BBcomposition} along with the \Pepm corrections
described in Sec.~\ref{sec:BB_elec}) depend upon \egcms; 
their average value for both backgrounds is $1.058 \pm 0.013$.

The two effects are taken to be fully correlated in computing their
contribution to the overall \BB yield uncertainty.

\subsection{Overall \BB corrections}
\label{sec:BB_corr}

The above subsections describe corrections by \BB component for all but a
few percent of the predicted makeup of the \BB background, as summarized in
Table~\ref{tab:BBcomposition}.  Several other small categories 
(\forex, ``\PJgy'' and ``Other'') are left as-predicted.  Finally,
several small corrections computed in the context of signal
efficiency are also applicable to \BB backgrounds:  a
high-energy photon efficiency correction of $0.9885 \pm 0.0065$
(Sec.~\ref{sec:bf_syst_photon}),  and a correction of
$0.989 \pm 0.004$ for lepton identification efficiency in
a multiparticle environment (Sec.~\ref{sec:bf_syst_lepton}).
Like the semileptonic tag
correction, these need only be applied to the 20 to 25\% of \PB
backgrounds other than \Pgpz or \Pgh, and hence are small effects.  More
significant is a global factor of $0.991 \pm 0.004$ from different 
probabilities
between MC and data events of the \Pgpz veto being activated by a
background photon.  Uncertainties also include a small contribution
from \BB MC statistics.

The \BB corrections described above are applied to each component and
for each 100\mev bin of \egcms.  Correlations between bins, due both
to \egcms-independent corrections and to corrections dependent upon
parent energies, are tracked, resulting in a table of corrected \BB
yields and a correlation matrix.  This information is used to compute
the results presented in the next section.  The largest systematic
uncertainties on the \BB yields are those due to the low-energy
photon correction to the \Pgpz and \Pgh components, 
with uncertainties in the no-track electron component and the 
\Pgpz inclusive spectrum next most significant.

\section{Signal Yields and Validation of Background Estimation}
\label{sec:yields}

\begin{figure}[!tb]
\begin{center}   
  \includegraphics[width=.5\textwidth]{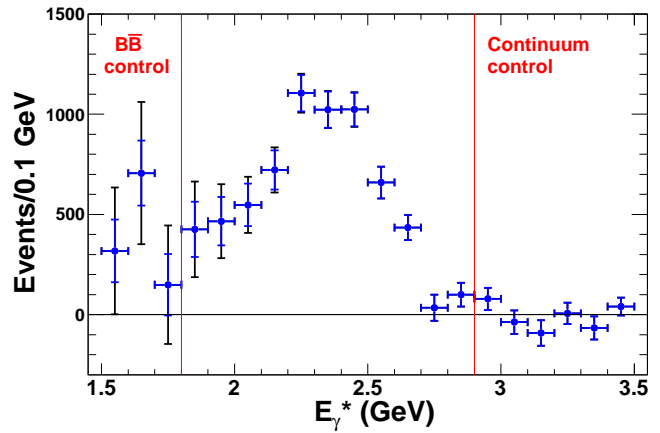}   
\end{center}
\vspace{-0.2in}
\caption{The photon spectrum in \onlumi of data after background subtraction.
 The inner error bars are statistical only, while the outer include both
 statistical and systematic errors in quadrature.}
\label{fig:egamma_dataminusbackground} 
\end{figure}

Figure~\ref{fig:egamma_dataminusbackground} shows the
photon energy spectrum in data after background subtraction. 
Table~\ref{tab:unblindedYields} gives the signal yields and background
estimations in bins of \egcms. 
The bin-to-bin correlations between the errors on
the signal yields are given in Table~\ref{tab:EGammaStarMeasCorrelation}. 
The continuum background is estimated with off-resonance data, while
the \BB background is estimated from MC simulation, with all
the corrections described in Sec.~\ref{sec:BB} applied.

\begin{table*}[!tb]
 \addtolength{\extrarowheight}{1.5pt}
 \begin{center}
  \caption
    {The event yields in bins of \egcms.  The continuum background is estimated
     from off-resonance data normalized to on-resonance
     luminosity. The \BB background is estimated using Monte Carlo
     simulation, corrected as described in Sec.~\ref{sec:BB}.  The
     extracted signal yield is computed by subtracting the continuum
     and \BB backgrounds from the on-resonance data yield. It is
     quoted with statistical uncertainties (from on-resonance minus
     off-resonance subtraction) and \BB systematics.
     The last set
     of rows show yields in wide \egcms bins, taking into account the
     correlations of \BB backgrounds between 100-MeV bins.
    } 
  \vspace{0.1in}
  \begin{tabular*}{14cm}{@{\extracolsep{\fill}}ccccc} \hline \hline
   \multirow{2}{*}{\egcms (GeV)} & On-resonance  & Continuum  & \BB & Signal \\
                                 & Data     & Background & Background & Yield \\
   \hline 
   1.53 to 1.60 &$11869\pm 109$ & $1319\pm 112$ &  $10232\pm 275$ &    $ 318\pm 156 \pm  275$   \\
   1.60 to 1.70 &$13531\pm 116$ & $1327\pm 113$ &  $11497\pm 316$ &    $ 706\pm 162 \pm  316$   \\
   1.70 to 1.80 &$10366\pm 102$ & $1371\pm 115$ &  $\; 8846\pm 252$ &  $ 150\pm 153 \pm  252$   \\ 
   \hline 
   1.80 to 1.90 & $8054\pm  90$ & $1118\pm 105$ &  $6511\pm 195$  &  $\;\; 426\pm 138 \pm  195$   \\ 
   1.90 to 2.00 & $6083\pm  78$ & $ 885\pm  93$ &  $4732\pm 139$  &  $\;\; 466\pm 121 \pm  139$   \\ 
   2.00 to 2.10 & $4429\pm  67$ & $ 717\pm  82$ &  $3165\pm 91\;$ &  $\;\; 548\pm 106 \pm   91\;$   \\ 
   2.10 to 2.20 & $3124\pm  56$ & $ 659\pm  80$ &  $1743\pm 56\;$ &  $\;\; 722\pm 98\;\;\pm  56\;$   \\ 
   2.20 to 2.30 & $2465\pm  50$ & $ 603\pm  77$ &  $ 757\pm  33$  &  $1105\pm  91\;\;\pm   33\;$   \\ 
   2.30 to 2.40 & $1977\pm  45$ & $ 639\pm  79$ &  $ 314\pm  20$  &  $1024\pm  90\;\;\pm   20$   \\ 
   2.40 to 2.50 & $1712\pm  41$ & $ 537\pm  73$ &  $ 152\pm  19$  &  $1024\pm  84\;\;\pm   19$   \\ 
   2.50 to 2.60 & $1225\pm  35$ & $ 499\pm  71$ &  $  67\pm  9$   &  $ 659\pm  79\;\;\pm    9$   \\ 
   2.60 to 2.70 & $ 795\pm  28$ & $ 328\pm  55$ &  $  32\pm  7$   &  $ 435\pm  62\;\; \pm    7$   \\ 
   2.70 to 2.80 & $ 457\pm  21$ & $ 404\pm  62$ &  $  18\pm  3$   &  $ \; 35\pm  66\;\; \pm    3$   \\ 
   \hline
   2.80 to 2.90 & $ 410\pm  20$ & $310 \pm  55$ &  $  9 \pm   4$ &  $ \;\; 91 \pm  59  \pm  4$ \\ 
   2.90 to 3.00 & $ 370\pm  19$ & $292 \pm  52$ &  $  8 \pm   4$ &  $ \;\; 71 \pm  55  \pm  4$ \\
   3.00 to 3.10 & $ 298\pm  17$ & $335 \pm  56$ &  $  6 \pm   3$ &  $ -44 \pm  59  \pm  3$ \\
   3.10 to 3.20 & $ 305\pm  18$ & $396 \pm  61$ &  $  5 \pm   3$ &  $ -96 \pm  64  \pm  3$ \\
   3.20 to 3.30 & $ 279\pm  17$ & $273 \pm  51$ &  $  6 \pm   2$ &  $ \;\;\;\; 0  \pm  54  \pm  2$ \\
   3.30 to 3.40 & $ 252\pm  16$ & $318 \pm  56$ &  $  3 \pm   2$ &  $ -69 \pm  58  \pm  1$ \\
   3.40 to 3.50 & $ 222\pm  15$ & $182 \pm  42$ &  $  3 \pm   1$ &  $ \;\; 38 \pm  44  \pm  1$ \\
   \hline
   \hline
   1.80 to 2.80 &$30321\pm 174$ & $6387\pm 249$ & $17490\pm 496$ &  $6444\pm 304 \pm  496$   \\
   1.90 to 2.80 &$22267\pm 149$ & $5270\pm 226$ & $10980\pm 313$ &  $6018\pm 271 \pm  313$   \\ 
   2.00 to 2.80 &$16184\pm 127$ & $4385\pm 206$ &  $6248\pm 187$ &  $5551\pm 242 \pm  187$   \\ 
   2.10 to 2.80 &$11755\pm 108$ & $3669\pm 189$ &  $3083\pm 110$ &  $5004\pm 218 \pm  110$   \\ 
   \hline \hline
  \end{tabular*}
 \label{tab:unblindedYields}
 \end{center} 
\end{table*}

\begin{table*}[htbp]
\addtolength{\extrarowheight}{1.5pt}
\begin{center}  
\caption{ The correlation matrix for the signal yield errors from 
 Table~\ref{tab:unblindedYields} in 100-MeV bins of \egcms.
 Systematic (\BB background) and statistical contributions are
 included.  Rows and columns are labeled by the value 
 of \egcms at the lower edge of the bin.}
 \label{tab:EGammaStarMeasCorrelation}
\vspace{0.1in}
\begin{tabular*}{14cm}{@{\extracolsep{\fill}}cc|cccccccccccccc} \hline \hline 
   \quad & \egcms (GeV) &1.53&1.6&1.7&1.8&1.9&2.0&2.1&2.2&2.3&2.4&2.5&2.6&2.7 & \quad \\ \hline 
   \quad & $1.53$& 1.00 &  0.75 &  0.71 &  0.65 &  0.58 &  0.46 &  0.32 &  0.16 &  0.07 &  0.03 &  0.01 &  0.01 &  0.00 & \quad \\ 
   \quad & $1.6$ &      &  1.00 &  0.74 &  0.68 &  0.61 &  0.48 &  0.33 &  0.17 &  0.07 &  0.03 &  0.02 &  0.01 &  0.00 & \quad \\ 
   \quad & $1.7$ &      &       &  1.00 &  0.67 &  0.60 &  0.47 &  0.33 &  0.17 &  0.08 &  0.03 &  0.02 &  0.01 &  0.00 & \quad \\ 
   \quad & $1.8$ &      &       &       &  1.00 &  0.58 &  0.46 &  0.32 &  0.17 &  0.08 &  0.04 &  0.02 &  0.01 &  0.00 & \quad \\ 
   \quad & $1.9$ &      &       &       &       &  1.00 &  0.44 &  0.31 &  0.16 &  0.08 &  0.03 &  0.02 &  0.01 &  0.00 & \quad \\ 
   \quad & $2.0$ &      &       &       &       &       &  1.00 &  0.28 &  0.15 &  0.07 &  0.03 &  0.02 &  0.01 &  0.00 & \quad \\ 
   \quad & $2.1$ &      &       &       &       &       &       &  1.00 &  0.14 &  0.07 &  0.03 &  0.02 &  0.01 &  0.00 & \quad \\ 
   \quad & $2.2$ &      &       &       &       &       &       &       &  1.00 &  0.05 &  0.03 &  0.02 &  0.01 &  0.00 & \quad \\ 
   \quad & $2.3$ &      &       &       &       &       &       &       &       &  1.00 &  0.02 &  0.01 &  0.01 &  0.00 & \quad \\ 
   \quad & $2.4$ &      &       &       &       &       &       &       &       &       &  1.00 &  0.01 &  0.00 &  0.00 & \quad \\ 
   \quad & $2.5$ &      &       &       &       &       &       &       &       &       &       &  1.00 &  0.00 &  0.00 & \quad \\ 
   \quad & $2.6$ &      &       &       &       &       &       &       &       &       &       &       &  1.00 &  0.00 & \quad \\ 
   \quad & $2.7$ &      &       &       &       &       &       &       &       &       &       &       &       &  1.00 & \quad \\ 
\hline\hline  
\end{tabular*}  
\end{center}
\end{table*}

To validate the background estimation, two control regions are set
aside in the photon spectrum.  In the upper control region ($2.9 < \egcms <
3.5\gev$), the event yield after subtracting continuum and \BB
backgrounds is $-100 \pm 138 \,(\mathrm{stat}) \pm 14
\,(\mathrm{syst})$ events, where the statistical uncertainty results
from off-resonance subtraction.  The systematic error is from the
uncertainty of out-of-time Bhabha-scattering events in \BB background
(see Table~\ref{tab:BBcomposition} caption).  This subtracted yield is 
consistent with the expectation of zero events.

In the lower control region ($1.53 < \egcms < 1.8\gev $), there remain
$1174 \pm 272 \,(\mathrm{stat}) \pm 828 \,(\mathrm{syst})$ events
after background subtraction.
The errors in the \BB estimates in these \egcms bins are
highly correlated; these correlations have been included when
computing the control-region systematic error.  The
agreement with zero in this region is at the $1.4\sigma$ level,
assuming no signal events. However this energy region contains a few
hundred signal events, with the exact number depending on the assumed
signal model.  For example, using predictions based on the kinetic and
shape function schemes with parameters close to HFAG's world-average
values~\cite{TheHeavyFlavorAveragingGroup:2010qj}, on average about
275 signal events would be expected in the lower control region.
Allowing for this, the data-background difference is reduced to the
$1.0\sigma$ level.

\section{Obtaining Physics Results:  an Overview}
\label{sec:resOverview}

Three physics results are extracted from the measured signal yield:
\begin{itemize}
 \item the \CP asymmetry, \acp(\bxsdg), 
 \item the inclusive branching fraction, \BR(\bxsg) (for several wide ranges 
  of true \eg in the \PB-meson rest frame), and 
 \item the true spectral shape and energy moments for \bxsg (in both the 
  CM frame and the \PB frame).  
\end{itemize}
The presence of new physics beyond the SM can affect the branching fraction
and \acp. The spectral 
shape, however, depends only on the dynamics of the \Pqb quark within the 
\PB meson; it is independent of any new physics contributions.
Three different approaches are optimal for the three physics results.

The branching fraction and spectral shape measurements require corrections 
for efficiency.
The partial branching fraction for signal in any range of measured
photon energy \egcms is obtained from the signal yield \sig in
that same range by
\begin{equation}
  \mathcal{B}(\bxsdg) = \frac{1}{2N_{\BB}}\frac{\sig}{\sigeff}\ ,
  \label{eq:bfrac}
\end{equation}
where \sigeff is the signal efficiency for that range and
$N_{\BB}$ is the number of \BB events in the on-resonance data set
before event selection.
$\mathcal{B}(\bxsg)$ is obtained from this by removing the small
constant fraction contributed by \bxdg.  Applying Eq.~(\ref{eq:bfrac})
brings in additional systematic uncertainties related to the efficiency
and to $N_{\BB}$.  The inclusive branching fraction and spectral shape
measurements are made in terms of reconstructed \egcms in the CM frame,
while theoretical predictions are made for true photon energy \eg
in the \PB frame.  These differ due to resolution and Doppler smearing.
The measurements must
be converted to corresponding measurements in terms of true \egcms
or \eg, in order to allow for detector-independent
comparisons.

\begin{figure}[!th]
 \begin{center}
  \includegraphics[width=0.47\textwidth]{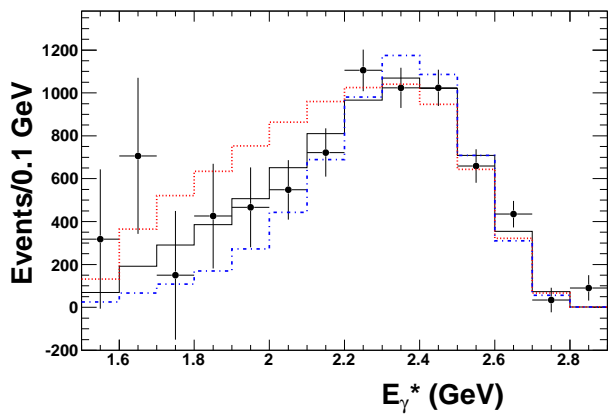}
  \caption{Comparison of data spectrum in reconstructed \egcms to the
   predictions of three models, each normalized for best agreement with
   the data above 1.8\gev, based on \chisq (including bin-to-bin
   correlations).  The
   solid histogram is for a shape-function-scheme model with $\mb = 4.51\gev$
   and $\mupsq = 0.46\,\mathrm{GeV}^2$   which resembles the data in this
   range.  The dot-dash histogram is for a kinetic scheme model with parameters
   $\mb = 4.60\gev$ and $\mupsq = 0.45\,\mathrm{GeV}^2$,
   close to the HFAG world average.  The dotted histogram is for a 
   shape-function-scheme model with $\mb = 4.40\gev$ and
   $\mupsq = 0.52\,\mathrm{GeV}^2$.  The minimum values of \chisq are
   6.7, 13.4 and 19.6, respectively.}
  \label{fig:dataVsModels}
 \end{center}
\end{figure}

Efficiency factors 
and also the transformation from one 
definition of photon energy to another depend upon the choice
of signal model, \ie, on the values of the HQET parameters in the
kinetic or shape function scheme.  
HFAG~\cite{TheHeavyFlavorAveragingGroup:2010qj} has extracted world-average
values of these parameters by combined fits to measurements of \bxclnu
decays and previous \bxsg measurements.  For the present
inclusive branching fraction measurement the range of models considered
is based on these HFAG central values and errors.
On the other hand, for the spectrum measurement such a restriction would
prejudice the results; the range of models considered must instead 
be driven by the data.  Put another way, for the branching
fraction measurement the MC model plays a subsidiary role, used only
to estimate the efficiency and transformation factors, so it makes
sense to use the best available information to constrain
the model, while for the spectrum the model is itself the object
of the measurement.  With these procedures, the model-dependence
uncertainties for both measurements are small compared to the combined
statistical and systematic uncertainties. 

In Fig.~\ref{fig:dataVsModels} predictions of three models are
superimposes on the measured data.  The first 
resembles the data for measured \egcms above 1.8\gev.  The second, 
which has HQET parameters very close to the HFAG world-average values in
the kinetic scheme, is about one standard deviation (``$1\sigma$'')
below the data in the first few energy bins above 1.8\gev, where 
\BB background is large.
The third is somewhat more than $1\sigma$ above the data in this region.
Differences between data and a particular model may be due either to the 
model being an incorrect description or to systematic fluctuations in the 
\BB background contribution.  This recognition is a key element
of the approaches used to measure both the branching
fraction (see below) and the shape of the true energy 
spectra (Sec.~\ref{sec:unfolding_overview}).

Branching fraction results are determined for $\egcms > 1.8\gev$.
The branching fraction is computed
by applying Eq.~(\ref{eq:bfrac}) to a single wide bin, \forex, 
$1.8 < \egcms < 2.8\gev$, using the average efficiency \sigeff computed
for an HFAG-based model.  
If 1/\sigeff factors were instead applied in 100-MeV
bins, the smaller values of \sigeff at low energies
(Fig.~\ref{fig:efficKN465}) would amplify the larger systematic uncertainties
on the event yield in this region as well as any data-model differences, 
in effect translating possible background fluctuations into a larger 
branching fraction bias.  Because of the energy-dependent \sigeff, 
statistical precision also improves with fewer bins.  Note also that 
the model dependence of the branching fraction computed using 100-MeV 
bins is comparable to that for a wide bin.  Thus, overall,
the wide-bin approach is both more accurate and more precise.
Full details are in Sec.~\ref{sec:bf}.

In contrast, the spectral shape must be determined by applying
Eq.~\ref{eq:bfrac} in each 100-MeV bin of reconstructed \egcms.  This is
the first step of a four-step unfolding procedure, 
detailed in Sec.~\ref{sec:spectrum}, leading to the true photon
energy spectra in the CM and \PB-meson rest frames.
Each model shown in Fig.~\ref{fig:dataVsModels} is used in all four
steps, to obtain the measured spectrum and its model dependence.
Energy moments and their correlations are computed from the
unfolded spectra.  This information is a needed input to the
HFAG fitting procedure, and may facilitate other potential comparisons
with theory.

The effects of efficiency and smearing cancel in the
extraction of \acp.  A raw asymmetry is thus directly computed from the 
measured yields \vs \egcms, using the lepton charge to tag \PB \vs \PaB
mesons.  Systematic corrections and
uncertainties arise only from possible 
charge dependence of the efficiencies (which would be a bias),
as well as from mistagging (which dilutes the asymmetry).
The full \acp analysis procedure is described in Sec.~\ref{sec:acp}.

\section{\boldmath Measurement of Direct \CP Asymmetry}
\label{sec:acp}

\begin{figure}[!tb]
\begin{center}
  \includegraphics[width=.5\textwidth]{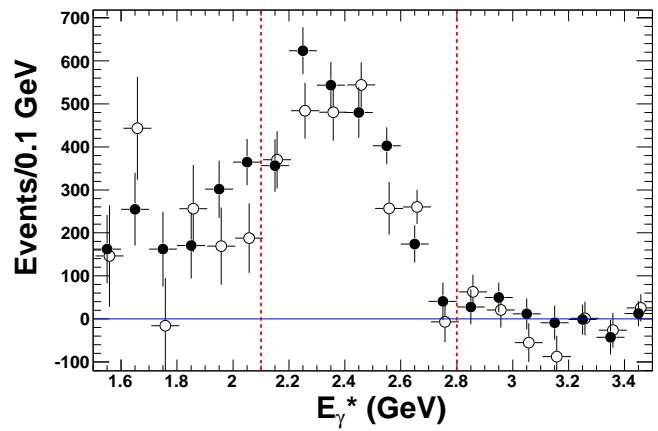}
\end{center}
\vspace{-0.2in}
\caption{The background-subtracted photon spectrum of 
 Fig.~\ref{fig:egamma_dataminusbackground} separated into yields for
 positive (filled circles) and negative (open circles) tagging lepton 
 charges.  Errors are statistical only.  The dashed vertical lines show
 the range utilized for the \acp measurement.}
\label{fig:egamma_databycharge}
\end{figure}

The direct \CP asymmetry, \acp(\bxsdg) 
is measured by dividing the signal sample into $\PB$ and $\PaB$ decays 
according to the charge of the lepton tag and computing
\begin{equation}
 \amcp(\bxsdg) = \frac{N^{+}-N^{-}}{N^{+}+N^{-}} \ ,
 \label{eq:acpmeas}
\end{equation}
 where $N^{+(-)}$ are the positively (negatively) tagged signal yields. 
Figure~\ref{fig:egamma_databycharge} shows these yields \vs \egcms.
The asymmetry must be corrected for
the dilution due to the mistag fraction  $\omega$:
\begin{equation}
\acp(\PB \to X_{s+d}\Pgg) = \frac{1}{1-2\omega}\amcp(\PB \to X_{s+d}\Pgg) \ .
\end{equation}
As can be seen in
Fig.~\ref{fig:egamma_estimated}(b) and Table~\ref{tab:unblindedYields},
the \BB background decreases at higher photon energies.
It was determined (prior to looking at the data) that restricting the
\acp signal region to $2.1<\egcms<2.8\gev$ optimizes the statistical
precision, and also the total precision including the uncertainty on the 
\BB background asymmetry described below.  Other systematic uncertainties
on \acp have negligible variation with \egcms.
The theoretical SM prediction of a near-zero asymmetry is not
affected for a minimum-energy requirement of 
2.1\gev~\cite{Kagan:1998bh,bib:thurth}.
All of the other selection requirements (Sec.~\ref{sec:evsel})
were found to be optimal also for the \acp measurement. 

The tagged signal
yields are $N^{+}=2620 \pm 158 (\mathrm{stat})$ and 
$N^{-}=2389 \pm 151(\mathrm{stat})$, giving an asymmetry of 
\begin{equation}
\amcp(\bxsdg)=0.046 \pm 0.044\ .
\end{equation}
To correct for dilution we compute the mistag fraction
\begin{equation}
   \omega = \frac{\chi_d}{2} + \omega_{\mathrm{cascade}} + 
   \omega_{\mathrm{misID}}\ .
   \label{eq:mistag}
\end{equation}
The largest contribution is from $\PBz-\PaBz$ oscillation, with mixing 
probability $\chi_d = 0.1863 \pm 0.0024$~\cite{Nakamura:2010zzi}; the factor 
of 1/2 accounts for the \PBpm mesons, which do not oscillate.  Smaller
contributions are $\omega_{\mathrm{cascade}}=0.0328 \pm 0.0035$, the fraction of events with 
wrong-sign leptons from the \PB decay chain, and $\omega_{misID}=0.0073\pm0.0037$, the 
mistag fraction due to misidentification of hadrons as leptons (almost
entirely in muon rather than electron tags).  Their
values are taken from the MC simulation averaging over electron and muon
tags.  An additional uncertainty in $\omega$ arises because our MC simulation assumes
 $\BR(\FourS \to \BzBzb)=0.50$ which leads to the factor of $1/2$ in the first term of Eq.~(\ref{eq:mistag}). The 
measured value is $\BR(\FourS \to \BzBzb)=0.484\pm 0.006$~\cite{Nakamura:2010zzi} 
so we take as a systematic the difference between the measured and assumed
values, $\Delta \omega = 0.016\chi_d$. This and the errors on $\chi_d$, $\omega_{\mathrm{cascade}}$ and
 $\omega_{\mathrm{misID}}$ are added in quadrature to give $\omega = 0.131 \pm 0.006$.   

The uncertainty in the \BB background estimation described in
Sec.~\ref{sec:BB} cancels in the numerator of
Eq.~(\ref{eq:acpmeas}) but not in the denominator, leading to an
uncertainty in \amcp of 0.022.  This uncertainty is combined with the
uncertainty in $\omega$ to give a multiplicative systematic
uncertainty on $\acp(\bxsdg)$ of $0.029\acp$.  
Table~\ref{tab:AcpErrorsMultiplicative} summarizes all of the
contributions to $\omega$ and to this uncertainty.

The measured asymmetry could be biased if there were (a) an asymmetry in
the \BB background not modeled in the simulation or (b) a
charge asymmetry in the lepton tag efficiency. To assess the potential
bias due to \BB subtraction, we use the data in the control region
$1.53 < \egcms < 1.8 \gev$. where the signal yield is much smaller than
\BB background.  After continuum subtraction, 
$\amcp(\PB\ \mathrm{control})=0.006 \pm 0.009 (\mathrm{stat})$.  Interpreting
this as a bias, it translates to a correction for the \acp signal region of
$\Delta\amcp(\bxsdg)=-0.004 \pm 0.006$.  The $ \PB \to \Pgpz X$ 
background sample described in Sec.~\ref{sec:BB_pizeta} is
used to confirm that there is no \egcms dependence to this
correction in the signal region. Lepton charge tag asymmetries have
been measured in $\epem\to\epem\Pgg$, $\epem\to\mmg$ and 
$\PB \to K^{(*)}\PJgy(\ell^{+}\ell^{-})$ events. No significant 
asymmetries are observed to a precision of 0.011, which is assigned as 
a systematic error on $\amcp(\bxsdg)$.  Table~\ref{tab:AcpErrorsAdditive} 
summarizes these additive systematic effects, showing a combined
error in quadrature of 0.013.

\begin{table}[!bt]
 \begin{center}
  \caption
   { Contributions to \acp multiplicative systematic correction and error.
   }
  \label{tab:AcpErrorsMultiplicative}
  \vspace{0.1in}
  \addtolength{\extrarowheight}{2pt}
  \begin{tabular*}{8.6cm}{lcc} \hline \hline
   Source  &  $ \omega \pm \Delta\omega $ & $\Delta\acp/\acp$ \\ 
   \hline
   \BzBzb oscillation         &  $ (0.1863 \pm 0.0023)/2 $ &       \\
   Fake lepton ID             &  $ 0.0073 \pm 0.0037 $     &       \\
   Cascade decays of \PB{}'s  &  $ 0.0328 \pm 0.0035 $     &       \\ 
   $\BzBzb:\BpBm=1:1$         &  $ 0.0000 \pm 0.0030 $            &       \\ \hline
   Total $\omega$             &  $ 0.133  \pm 0.0064 $      & 0.018 \\ 
   \BB yield                  &                            & 0.022 \\ \hline
   Total uncertainty          &                            & 0.029 \\ \hline
   \hline
  \end{tabular*}
 \end{center}
\end{table}

\begin{table}[!bt]
 \begin{center}
  \caption{\amcp additive systematic corrections and errors}
  \label{tab:AcpErrorsAdditive}
  \vspace{0.1in}
  \addtolength{\extrarowheight}{2pt}
  \begin{tabular*}{8.6cm}{@{\extracolsep{\fill}}clcc} \hline \hline
    & Source               &  Correction ($10^{-2}$)  & \\ \hline
    & \BB background       &  $ -0.4 \pm 0.6$         & \\
    & Detection asymmetry  &  $ 0.0\pm 1.1$           & \\ \hline
    & Total                &  $ -0.4 \pm 1.3$         & \\ \hline 
    \hline
  \end{tabular*}
 \end{center}
\end{table}

Since the SM prediction of $\acp\approx 0$ depends upon cancellation of
\bxsg and \bxdg asymmetries, a difference in their selection efficiencies
could also cause a bias.  We have used MC simulations with 
the same underlying model (KN with $\mb=4.65\gevcc$) to compare selection
efficiencies following $\Pqs\Paq$ \vs $\Pqd\Paq$ hadronization.
For $\egcms > 2.1\gev$, the \bxdg efficiency is
larger by a factor of $1.028 \pm 0.014$, so we conservatively assign
a 4.2\% uncertainty (MC central value plus one standard deviation)
in the yield of \bxdg events.  Given the SM-predicted yields
and asymmetries~\cite{Hurth:2001yb}, that would
change \amcp by less than 0.0002, which is negligible.

Finally the  $\amcp(\bxsdg)$ is corrected for mistags and bias to give
\begin{eqnarray*}
  \acp &  =  &\frac{(0.046 \pm 0.044(\mathrm{stat})) 
   - (0.004 \pm 0.013 )}{0.734 (1 \pm 0.029)} \\
   &  =  & 0.057 \pm 0.060 (\mathrm{stat}) \pm 0.018 (\mathrm{syst}) \ ,
  \label{eq:ACP_result}
\end{eqnarray*}
where the two systematic errors have been combined in quadrature. The result is
consistent with no asymmetry.

\section{\boldmath Measurement of $\mathcal{B}(\bxsg)$ }
\label{sec:bf}

As discussed in Sec.~\ref{sec:resOverview},
\BR(\bxsdg) is measured by applying Eq.~(\ref{eq:bfrac}) 
to a single wide bin in measured \egcms.  Results are computed for three 
choices of energy range:  1.8 to 2.8\gev, 1.9 to 2.8\gev and 2.0 to 2.8\gev.
Note that \sigeff here means the overall signal efficiency, including
both acceptance and event-selection, as discussed in
Sec.~\ref{sec:evsel_effic}.
A small adjustment, by a factor $\alpha$ which is close to 1.0, converts
each result to a branching fraction in the same range of the true \eg in
the \PB-meson rest frame.  This corrects for the effects of EMC resolution
and Doppler smearing.  Finally, the factor $1/(1+(\Vtd/\Vts)^2)$ is applied to
account for the contribution of \bxdg events, yielding a branching fraction
for \bxsg only.

Section~\ref{sec:bf_syst} describes systematic corrections and 
uncertainties affecting the efficiency \sigeff in Eq.~(\ref{eq:bfrac}),
and computes the total fractional systematic uncertainty on the branching
fraction.  The choice of the central values for \sigeff and $\alpha$
depend upon the choice of the signal model used in MC simulation.
Section~\ref{sec:bf_modeldep} addresses this choice and determines the
model-dependence uncertainty of the branching fraction.  
Section~\ref{sec:bf_results} presents branching fraction results 
first in terms of measured \egcms, then presents the conversion to the
branching fraction in the \PB-meson rest frame, along with associated
uncertainties.

\subsection{Systematic Corrections and Uncertainties}
\label{sec:bf_syst}

Each of the factors in Eq.~(\ref{eq:bfrac}) can contribute to the
uncertainty in the branching fraction.  The signal yield has contributions
from statistics (of the on-peak and off-peak data yields) and from the
systematics of the \BB background subtraction.  The number of
produced \FourS events, $N_{\BB}$, has a systematic uncertainty of 1.1\%.
The focus here is on the systematic uncertainty of the remaining factor, 
the signal efficiency \sigeff.  For each event-selection criterion,
an efficiency is computed using MC simulation.  But
the actual efficiency in data may differ from that in the simulation.
Systematic corrections are determined by comparing data to MC
events for various control samples; the precision of each comparison
provides a systematic uncertainty.  A summary of these corrections
and uncertainties is presented in Table~\ref{tab:efficSyst}.  

\subsubsection{Systematics of the High-Energy Photon Selection}
\label{sec:bf_syst_photon}

Two dedicated studies of high-energy photon detection efficiency
have been done using \mmg ISR events.  These events are overconstrained,
so the measured \Pgmp and \Pgmm tracks, along with known beam kinematics,
can be used in a one-constraint fit to predict the three-momentum of the 
photon.  Naively
one would look for a detected photon ``close'' to this predicted photon,
and the data/MC correction would be the ratio of the probability of
finding such a photon in \mmg data events to the probability of finding
one in \mmg MC events.  However, this is complicated by the effects of
EMC resolution and by the possibility that the likelihood of photon
conversion (in detector material) isn't accurately simulated.
The earlier and more recent of the 
\mmg studies took rather different approaches to these issues.
The first applied acceptance criteria (particularly a minimum energy)
to detected photons, and folded EMC resolution into the predicted photon
properties before making the same cuts.  That study also in effect
measured the conversion fraction, separately for data and MC samples.
The second study
did not use acceptance cuts for the detected photon, instead loosely
matching its parameters to those of the predicted photon, and used an
electron veto to suppress photon conversions.  
Results of the two studies are in good agreement, with the data/MC 
efficiency corrections differing by 0.3\% when weighted by the \bxsg
photon polar angle distribution.  Systematic uncertainties 
on these corrections are 0.65\% and 0.55\%.  The two correction factors are
averaged, giving 0.9885, and an uncertainty of 0.65\% is assigned.
The assigned correction and uncertainty are independent of the photon
energy \egcms.  Hence they affect the branching fraction, but not
the spectral shape or energy moments.

The \mmg samples from data and MC simulation are also used to
assess the photon energy scale and resolution, by comparing the distributions
for data and MC events of the ratio of detected to predicted photon
energy.  For the energy scale, the energy balance in the decay
$\PBz\to\PKsti(\to\PKp\Pgpm)\Pgg$ is also used.  After small energy scale
adjustments already included in event reconstruction, both processes show
no remaining bias for either MC or data events, with a conservative
uncertainty of 0.3\%.  For photon energy resolution, inclusion of an
additional 1\% energy smearing of MC photons brings the \mmg ratio
distribution into good agreement with that for \mmg data.  This is taken
as a systematic uncertainty.  The energy scale and resolution effects
translate into the small uncertainties on the inclusive branching
fraction shown in Table~\ref{tab:efficSyst}.  

Lastly, the \mmg samples are used to assess shower shape, in 
particular the efficiency of the selection cut on lateral moment.
After a small adjustment of the simulated lateral moment, there is
good agreement between MC and data efficiencies of this selection,
with the uncertainty given in Table~\ref{tab:efficSyst}.

The high-energy photon efficiency is calibrated using the low-multiplicity
\mmg events, but could also be affected by the hadronic-event environment
in \BB events (including signal).  The requirement that the high-energy
photon be isolated from any other EMC energy deposition by at least 25\cm 
is meant to reduce data-MC efficiency differences.  The systematic
uncertainty of 2\% is estimated by embedding high-energy photon signatures 
into hadronic events, separately for data and MC samples, and determining
the fractions of events passing the isolation requirement.

\subsubsection{Systematics of the \Pgpz and \Pgh Vetoes}
\label{sec:bf_syst_veto}

The \Pgpz and \Pgh vetoes can remove events not only if the high-energy
photon originates from an actual \Pgpz or \Pgh, but also if there is a
random (``background'') photon with which the high-energy photon forms
a \gg invariant mass combination lying inside one of the veto windows.  
The efficiencies of the vetoes for simulated events can
differ from those for data if the number of background photons
in simulation differs from data.  
Off-resonance-subtracted data and \BB MC events are compared
for high-energy photons in the control region below 1.8\gev, with
all selection criteria except the vetoes applied.  Sidebands of the 
\gg mass windows are used to estimate the numbers of low-energy 
background photons that result in masses inside the windows.  It is
found that there are more such low-energy photons in the data than in 
the simulation (as much as 8\% more at the lowest energies, below 80\mev, 
decreasing monotonically with photon energy to approximately 
2.5\% above 250\mev).

Monte Carlo studies are used to correct for the effects of these
differences on event-selection efficiency when the vetoes are imposed:
$-0.4\%$ for signal events, and $-0.9\%$ for nonsignal generic \BB events.
Uncertainties are taken to be half of the corrections.  Differences
between \Pgpz and \Pgh line shapes in data and simulated events could
also potentially affect the \BB efficiencies, but such differences proved
to be negligible.

\subsubsection{Systematics of the Lepton Tag Efficiency}
\label{sec:bf_syst_lepton}

There are two contributions to signal efficiency systematics from
the lepton tagging.  The first is the uncertainty in the semileptonic
branching fraction (for the nonsignal \PB in the event), averaged
over the lepton acceptance for the current analysis.  This is
addressed in Sec.~\ref{sec:BB_semilept}, and results in a
systematic correction and uncertainty of $1.047 \pm 0.013$.

The second contribution arises from possible differences between data
and MC samples in the lepton identification efficiencies.
These identification efficiencies in the simulation are calibrated 
as a function of lepton momentum to those in data using control samples
of low-multiplicity (Bhabha and \mmg) events.  To measure the additional
effect of the high-multiplicity environment in signal events, fitted
\PJgy yields are compared in data and MC samples of reconstructed
$\PB \to \PJgy\ (\PJgy \to \Plp\Plm)$ events, both with and without particle 
identification requirements applied to the leptons.  This is done separately 
for \epem and \mumu decays.  The resulting systematic
correction factor for a single lepton, averaged over the mix of electron 
and muon tags in the current analysis, is $0.989 \pm 0.004$.
This result is also included for the \BB background systematics
in Sec.~\ref{sec:BB_corr}.

\subsubsection{Other Uncertainties in Event-selection Efficiency}
\label{sec:bf_syst_event}

A systematic uncertainty is assigned to the MC computation
of the efficiency of the neural-network selection criteria.  The
control samples used to compare data and MC efficiencies
are inclusive \Pgpz samples, created by applying the standard
event selection criteria to data and to \BB background events, but with the 
\Pgpz veto inverted, \ie, an event is accepted if it has a \gg mass
combination inside the veto window.  The \gg mass spectra confirm that
most of these events are due to actual \Pgpz production.
Off-resonance-subtracted data are compared
to the simulated \BB sample.  The efficiencies of the neural-network criteria
for signal MC and \BB background MC events show similar
increases with \egcms.  To validate use of the \Pgpz control sample, 
neural-network
output distributions for signal and control samples were compared
in a narrow range of $1.8 < \egcms < 2.0\gev$ and found to be quite similar. 
Data-MC efficiency comparisons 
for the control samples are made separately for the electron and muon
neural networks, and differences are weighted by the fractions
of electron and muon tags in the standard event selection.
This average difference of 1.2\% is taken as a systematic uncertainty.

Lastly, the signal efficiency has some small variation with the 
specific final hadronic \PsX state.  The overall efficiency is
thus sensitive to whether the \textsc{JETSET} model 
implemented in the simulation properly describes the hadronization process.
Measured data-MC differences from the \babar\ sum-of-exclusives 
\bxsg analysis~\cite{Aubert:2005cua} are used to reweight the hadronic
multiplicity distribution of the simulated \PsX final state, and,
separately, the fraction of final states which contain at least one
\Pgpz.  Each efficiency change is taken as a systematic uncertainty.  
Combining the two effects in quadrature, the total systematic uncertainty
due to modeling of the hadronization process is 1.1\%.

\subsubsection{Overall Efficiency Systematics}
\label{sec:bf_syst_effic}

Table~\ref{tab:efficSyst} summarizes the efficiency corrections and
their estimated uncertainties.  Nearly
all of these effects are independent of photon energy \egcms, so the
tabulated values apply both to wide bins and 100-MeV bins.  The only
exceptions are the small energy-scale and resolution uncertainties, which are
folded into the yield spectrum (Fig.~\ref{fig:egamma_dataminusbackground});
the Table presents the values for a bin from 1.8 to 2.8\gev.  The correction
factors are included in all values of efficiency quoted subsequently
in this paper.

\begin{table}[!tb]
  \addtolength{\extrarowheight}{1.5pt}
 \begin{center}
  \caption{Systematic correction factors and uncertainties on the signal 
   efficiency in \bxsg
   branching fraction measurements.  Corrections are relative to the signal
   Monte Carlo simulation.  ``HE\Pgg'' stands for the high-energy photon.}
  \label{tab:efficSyst}
  \vspace{0.1in}
  \begin{tabular*}{8.6cm}{@{\extracolsep{\fill}}clcclclc} \hline \hline
   & \multicolumn{1}{c}{Effect} &&&  \multicolumn{3}{c}{Value} & \\ 
   \hline 
   & HE\Pgg detection efficiency          &&& 0.9885 & $\pm$ & 0.0065  & \\
   & HE\Pgg energy scale        &&&  1.0   & $\pm$ & 0.0025  & \\
   & HE\Pgg resolution          &&&  1.0   & $\pm$ & 0.001   & \\
   & HE\Pgg lateral moment requirement  &&&  1.0   & $\pm$ & 0.003   & \\
   & HE\Pgg isolation requirement       &&&  1.0   & $\pm$ & 0.020   & \\
   & \Pgpz and \Pgh vetoes      &&& 0.996  & $\pm$ & 0.002   & \\
   & Lepton PID                 &&& 0.989  & $\pm$ & 0.004   & \\
   & \PB semileptonic BF        &&& 1.047  & $\pm$ & 0.013   & \\
   & Neural network             &&&  1.0   & $\pm$ & 0.012   & \\
   & Hadronization model        &&&  1.0   & $\pm$ & 0.011   & \\
   \hline		     						  
   & Combined                   &&& 1.019  & $\pm$ & 0.030   & \\
   \hline 
   \hline
  \end{tabular*}				  
 \end{center}
\end{table}

\subsubsection{Combining Yield and Efficiency Uncertainties}
\label{sec:bf_syst_total}

Table~\ref{tab:totalSyst} summarizes all systematic uncertainties for
the branching fraction measurement.

\begin{table}[!tb]
 \begin{center}
  \caption
   {Summary of relative systematic uncertainties on the signal branching
   fraction.  In addition to the contributions from the three factors in
   Eq.~(\ref{eq:bfrac}) (the systematic uncertainty on signal yield is due
   to that on \BB background), there is a cross-term arising from
   correlations between background-yield and signal-efficiency uncertainties.}
  \label{tab:totalSyst}
  \vspace{0.1in}
  \addtolength{\extrarowheight}{2pt}
  \begin{tabular*}{8.6cm}{@{\extracolsep{\fill}}lccc} \hline \hline
   \egcms Range (GeV)     & 1.8 to 2.8 & 1.9 to 2.8 & 2.0 to 2.8 \\  \hline
   Signal efficiency      & 0.031      & 0.031      & 0.031      \\
   \BB background         & 0.078      & 0.051      & 0.032      \\
   Cross-terms            & 0.029      & 0.024      & 0.019      \\
   Count of \PgUc events  & 0.011      & 0.011      & 0.011      \\  \hline
   Total (quadrature sum) & 0.090      & 0.065      & 0.050      \\  \hline
   \hline
  \end{tabular*}
 \end{center}
\end{table}

The fractional branching fraction uncertainty due to \BB background
is energy-dependent primarily because the ratio \bkb/\sig of background 
yield to signal yield decreases sharply with increasing \egcms.

Similar contributions to efficiency affect the MC computations of
both the \BB background yield \bkb and the signal efficiency \sigeff, so
some systematic uncertainties are common to both 
and hence are treated
as correlated in evaluating Eq.~(\ref{eq:bfrac}).  Because of the direct
calibration of \Pgpz and \Pgh contributions to the \BB background yield 
against data, some correlated effects are reduced to an insignificant
level.  We consider these remaining correlated effects:
\begin{itemize}
 \item the systematic uncertainty due to high-energy photon efficiency,
  which enters the \BB yield predominantly via the low-energy photon 
  efficiency correction to the \Pgpz and \Pgh components; 
 \item the \pizeta veto efficiency, which affects all background
  components; 
 \item the semileptonic branching fraction, which for \BB backgrounds
  affects lepton tags for non-\Pgpz/\Pgh components, and also
  those events in which an electron fakes the high-energy photon
  signature.
\end{itemize}
These correlated effects result in a cross-term between
the uncertainties in \sigeff and \bkb of $0.0178\sqrt{\bkb/\sig}$ 
for the energy ranges considered in Table~\ref{tab:totalSyst}.  Like
the \BB yield contribution itself, this decreases with increasing
\egcms.  
For the 100-MeV bins used in the spectrum measurements
(Sec.~\ref{sec:spectrum}), an additional energy dependence is
allowed for in the semileptonic branching fraction cross-term.  It
arises because the variation of the uncertainty with lepton energy
given in Fig.~\ref{fig:SL_partial_BF} directly applies to the
electron backgrounds.  For each of the three cross-term contributions,
the uncertainty is treated as fully correlated between energy bins,
and the three corresponding error matrices are then summed. 

The total systematic uncertainty on the branching fraction, also given
in Table~\ref{tab:totalSyst}, is the sum in 
quadrature of the contributions from yield, efficiency, cross-terms and
$N_{\BB}$.  

\subsection{Model-dependence Uncertainties of the Signal Efficiency}
\label{sec:bf_modeldep}

The signal efficiency \sigeff is estimated with MC 
simulated spectra.  The central value depends on the \bxsg model chosen, and
thus has an associated model-dependent uncertainty.  
HFAG~\cite{TheHeavyFlavorAveragingGroup:2010qj} 
has provided world average values of the HQET parameters \mb and \mupsq 
(and others) in the kinetic scheme, obtained from
a combined fits to measurements of \bxclnu moments and previous
measurements of \bxsg moments.  (The small samples of earlier \babar\
\bxsg data used by HFAG do not lead to significant correlations
between the fit results and the data presented here.)
The central values of the efficiency
in each of three energy ranges for the current analysis are
determined by computing efficiencies for several kinetic-scheme models
with \mb and \mupsq close to the values found in the global HFAG fit, and
interpolating to the HFAG values.  For an energy range 
$1.8 < \egcms < 2.8\gev$, the corresponding signal efficiency is 0.02573.

Three considerations enter the estimate of model dependence.  First,
the error ellipse associated with the HFAG fit is used to estimate an
efficiency uncertainty.  Second, the central values for \mb and \mupsq 
from the HFAG fit to \bxclnu moments only, and from a similar 
fit~\cite{bib:schwanda} using \bxsg moments only but constraining
other HQET parameters based on the combined fit, are considered.  The
largest efficiency deviation is
that from the \bxsg-only fit, so that is assigned as the kinetic-scheme
uncertainty.  Third, a procedure which translates HQET parameters from the 
kinetic scheme to the shape function scheme is applied to the combined-fit
results to provided \mb and \mupsq
values~\cite{TheHeavyFlavorAveragingGroup:2010qj,bib:golubev} 
for an efficiency estimate in the shape function scheme.  The difference
between that estimate and the central value in the kinetic scheme is added
to the kinetic-scheme uncertainty (linearly because both effects are
systematic shifts rather than random variations), and taken as a
symmetric uncertainty.  Lastly, this is combined in quadrature with an
uncertainty due to the choice of scale factor in the scheme translation.
For the range 1.8 to 2.8\gev, the three effects together yield
$\Delta\sigeff = (0.00025 + 0.00019) \oplus 0.00024 = 0.00051$.

Another possible source of model-dependence is the choice of the \mxs cutoff
used to define the \PKsti region (Sec.~\ref{sec:data}).  But changing
that cutoff from 1.1\gevcc to 1.0 or 1.2\gevcc results in an
efficiency change small compared to the other effects computed here.

The signal efficiency, and associated model errors, for
three photon energy ranges is given in Table~\ref{tab:efficModel}.

\begin{table}[!tb]
 \begin{center}
  \caption{Signal efficiency central values and model-dependence
   errors, for various ranges of measured \egcms.}
  \label{tab:efficModel}
  \vspace{0.1in}
  \addtolength{\extrarowheight}{2pt}
  \begin{tabular*}{8.6cm}{@{\extracolsep{\fill}}cccc} \hline \hline
   & \egcms Range (GeV)& $\epsilon_\mathrm{sig} (\%)$ & \\
   \hline 
   & 1.8 to 2.8 & $2.573 \pm 0.051$ & \\
   & 1.9 to 2.8 & $2.603 \pm 0.038$ & \\
   & 2.0 to 2.8 & $2.641 \pm 0.029$ & \\
   \hline \hline
  \end{tabular*}				  
 \end{center}
\end{table}

\subsection{Branching Fraction Results}
\label{sec:bf_results}

\begin{table*}[!tb]
 \begin{center}
  \caption{Branching fractions in several photon energy ranges of both
    measured \egcms (CM frame) and true \eg (\PB rest frame), along with
    the adjustment factor $\alpha$ between them.  Uncertainties on
    branching fractions are statistical, systematic and model-dependence,
    respectively.  The error on the adjustment factor $\alpha$ is a 
    model-dependence uncertainty, treated as fully correlated with that on 
    the initial \BR.}
  \label{tab:bfResultsTable}
  \addtolength{\extrarowheight}{2pt}
  \vspace{0.15in}
  \begin{tabular*}{17.8cm}{@{\extracolsep{\fill}}ccccc} \hline \hline
                  & $\BR(\bxsdg)\ (10^{-4})$              & Factor $\alpha$     & $\BR(\bxsdg)\ (10^{-4})$              & $\BR(\bxsg)\ (10^{-4})$              \\
   Energy Range   & in Measured \egcms Range              & to True \eg         & in True \eg Range                     & in True \eg Range                    \\ \hline
   1.8 to 2.8\gev & $3.271 \pm 0.154 \pm 0.294 \pm 0.065$ & $1.0233 \pm 0.0042$ & $3.347 \pm 0.158 \pm 0.301 \pm 0.080$ & $3.207 \pm 0.151 \pm 0.288 \pm 0.077$\\ 
   1.9 to 2.8\gev & $3.019 \pm 0.136 \pm 0.196 \pm 0.044$ & $1.0356 \pm 0.0045$ & $3.126 \pm 0.141 \pm 0.203 \pm 0.059$ & $2.995 \pm 0.135 \pm 0.194 \pm 0.057$\\ 
   2.0 to 2.8\gev & $2.745 \pm 0.120 \pm 0.137 \pm 0.030$ & $1.0657 \pm 0.0045$ & $2.925 \pm 0.128 \pm 0.146 \pm 0.045$ & $2.802 \pm 0.122 \pm 0.140 \pm 0.043$\\ \hline \hline
  \end{tabular*}
 \end{center} 
\end{table*}

\begin{table*}[!t]
 \caption[Correlation matrix between measured branching fractions in three
  energy ranges.]
  {The correlation matrix for the measured branching fractions in three
  energy ranges, including all statistical and systematic but not
  model-dependence uncertainties.}
 \label{tab:bfCorrelation}
 \addtolength{\extrarowheight}{1.5pt}
 \begin{center}  
  \begin{tabular*}{14cm}{@{\extracolsep{\fill}}c|ccc} \hline \hline 
     \quad \quad \eg Range       \quad \quad & 1.8 to 2.8\gev & 1.9 to 2.8\gev & 2.0 to 2.8\gev    \quad \quad\\   \hline
     \quad \quad 1.8 to 2.8\gev  \quad \quad  &  1.00         &  0.94         &   0.84  \quad \quad\\   
     \quad \quad 1.9 to 2.8\gev  \quad \quad  &               &  1.00         &   0.92  \quad \quad\\   
     \quad \quad 2.0 to 2.8\gev  \quad \quad  &               &               &   1.00  \quad \quad\\     \hline \hline
  \end{tabular*}  
 \end{center}
\end{table*}

Table~\ref{tab:bfResultsTable} shows the branching fractions 
\BR(\bxsdg) for three ranges of measured \egcms, from applying
Eq.~(\ref{eq:bfrac}) with the efficiencies obtained in 
Sec.~\ref{sec:bf_modeldep}.

In order to compare directly to theoretical predictions, the measurement for
each energy range in the CM frame is converted to a branching fraction 
in the corresponding range of true energy in the \PB frame.  The
factor $\alpha$ needed to accomplish this is determined from
MC simulation using the same methods for choosing a central
value (based on the HFAG world-average HQET parameters in the kinetic
scheme) and for estimating model dependence as are used for \sigeff
(Sec.~\ref{sec:bf_modeldep}).  Values of $\alpha$ and the resulting values
of \BR(\bxsdg) are also presented in Table~\ref{tab:bfResultsTable}.
The model-dependence uncertainties on $\alpha$ and 1/\sigeff are positively
correlated:  models with a larger fraction of the spectrum at 
low energy have larger average 1/\sigeff and usually larger $\alpha$.
Hence the fractional model-dependence errors on \sigeff and
$\alpha$ are linearly added.  Should the HFAG values for the
kinetic-scheme parameters change in the future, the Appendix provides
a prescription for adjusting both \sigeff and $\alpha$, and hence 
the central branching fraction values, for such a change.

Finally, the contribution of \BR(\bxdg) is accounted for by multiplying
\BR(\bxsdg) by $1/(1+(\Vtd/\Vts)^2) = 0.958 \pm 0.003$.  
This leads to the results, also presented in Table~\ref{tab:bfResultsTable},
for \BR(\bxsg) in true-\eg ranges, with the small additional uncertainty
from this factor included in the systematic error.

Because the events
in the three energy ranges are mostly in common, even the statistical
uncertainties on the three branching fractions are highly correlated.
The overall correlation matrix for all statistical and systematic effects
(including yield-efficiency cross-terms, but excluding
model dependence) are given in Table~\ref{tab:bfCorrelation}.

\section{Unfolded Spectrum}
\label{sec:spectrum}
The theoretical predictions of the photon energy spectrum are made in the \PB-meson rest frame in terms of the 
true photon energy \egb. However the measured spectrum in Fig.~\ref{fig:egamma_dataminusbackground} is measured in
the \FourS frame in terms of the reconstructed \egcms after the event selection requirements. To 
convert the measured spectrum to one that can be directly compared to predictions requires correcting for 
selection efficiency and detector acceptance, and unfolding two resolution effects. 
These are detector resolution and Doppler smearing of the photon energy.  The transformation of the measured \egcms spectrum to an \egb spectrum thus requires four steps:
\begin{enumerate}
\item Correcting for the event selection efficiency.
\item Unfolding the effects of detector resolution.
\item Correcting for the detector acceptance. 
\item Unfolding the Doppler smearing.
\end{enumerate}
   
Each of these steps requires the use of the MC simulation to either
estimate the efficiency and acceptance or model the resolution and
smearing. The effects of calorimeter resolution and Doppler smearing
on the photon spectrum are unfolded using a simplified version of an
iterative method~\cite{Malaescu:2009dm}. This simplified method has
been used previously by the \babar\ collaboration in a measurement of
the $\epem \to \pi^+ \pi^- (\gamma)$ cross
section~\cite{bib:2009fg}. An introduction to this method is followed
by a description of the implementation used here and then the results
and systematic uncertainties. The notation used for the photon energy is:
\begin{itemize}
\item  $\egcms$ is the energy  measured in the \FourS rest frame after event selection.
\item $\egcmstrue$ is the true photon energy in the \FourS rest frame. Its spectrum is
      obtained after steps 1-3 above.
\item $\egb$ is the true photon energy in the \PB-meson rest frame.  Its spectrum is obtained 
       after steps 1-4 above.
\end{itemize}

\subsection{Overview of the Unfolding Technique}
\label{sec:unfolding_overview}
The effects of detector resolution and Doppler smearing each require a separate unfolding but the procedure for each
is identical. In this overview  the unfolding of the detector resolution is described. The unfolding of
the Doppler smearing uses the same procedure. First some general considerations for the unfolding are given 
before describing the features of implementation used.

The spectrum is measured in twelve 100-MeV bins between 1.6 and 2.8\gev and one 70-MeV bin
between 1.53 and 1.6\gev. The detector resolution can cause a migration between bins  which is
described by a transfer matrix $A$, whose elements $A_{ij}$ are the number of 
events generated in bin $j$ that are reconstructed in bin $i$. Identical binning is
used for the generated and smeared spectra so that $A_{ij}$ is a square matrix. The transfer matrix is derived
from MC simulation using an assumed model for the spectrum. It is then used to construct a folding
matrix ${P}_{ij}$ and an  unfolding
matrix $\tilde{P}_{ij}$
\begin{eqnarray*}
 P_{ij} & = & \frac{A_{ij}}{\sum_{k=1}^{N} A_{kj} } 
 \label{equ:foldingmatrix}  \\
 \tilde{P}_{ij} & = & \frac{A_{ij}}{\sum_{k=1}^{N} A_{ik}} \ ,
 \label{equ:unfoldingmatrix}
\end{eqnarray*}
where $P_{ij}$ is the probability for an event generated in the bin $j$ 
to be reconstructed in the bin $i$ and $\tilde{P}_{ij}$ is the probability 
of the reconstructed event in the bin $i$  coming from the 
generated  bin $j$. $N$ is the number of bins. In principle the unfolding 
matrix can now be directly applied to the reconstructed spectrum to unfold the
resolution effects. There are, however, two significant problems with this 
approach. The first is that it assumes
the simulated model perfectly describes the data. The second is that any significant statistical fluctuations
in the reconstructed spectrum can be unfolded into several bins, causing unstable and unreliable results.

The technique adopted mitigates these problems. It begins by simulating an approximate  model of the spectrum that 
is normalized to data in the range $1.8 < \egcms < 2.8 \gev$. This model is referred to as  the initial model.
The difference between this  model and the data is then divided into two parts. The first part is attributed to
a genuine difference between the model and the true data spectrum and is used to modify the 
transfer matrix, equivalent to changing the initial model. The second is attributed to statistical and systematic
fluctuations
and is not unfolded using the unfolding matrix, but rather used to correct the model spectrum so that significant 
fluctuations in the  reconstructed spectrum are propagated to the unfolded true data spectrum. The division
of the difference into these two parts is accomplished using a bin-dependent regularization function $f$
with a tunable parameter $\lambda$. The value of $f$  
varies from 0 to 1 according to the value of $\lambda$ so that a fraction 
$f$ comprises the true model-to-data difference and a fraction $1-f$ the 
statistical and systematic fluctuation. \textsl{A priori} the value of 
$\lambda$ is unknown, but can be estimated using an MC technique described in Sec.~\ref{sec:unfolding_implementation}. 

The technique has been tested extensively in simulated data and found to give reliable and stable results.

\subsection{Implementation of the Unfolding}
\label{sec:unfolding_implementation}

This nominal initial model is found by comparing the data to a set of models in the kinetic, shape function, 
and KN  schemes using different values of HQET parameters. Each model is passed through the full simulation
and event selection. Figure~\ref{fig:dataVsModels} shows the comparison of the data to a range of models that
describe the data at the one-$\sigma$ level. The closest match is chosen by
constructing a  $\chi^{2}$ function formed from the bin-by-bin differences of the data and the generated  spectrum using 100-Mev bins in the signal range $1.8 < \egcms < 2.8 \gev$ and the full covariance matrix. 
It is found that a model in  the shape function scheme with ($\mb=4.51 \gev ,\mupsq =0.46 \gev^{2}$) best
describes the data. The other models shown in Fig.~\ref{fig:dataVsModels} are used to optimize the $\lambda$ 
parameters for the two unfolding steps and to estimate model-dependence systematic uncertainties.

The unfolding method begins by correcting the measured data spectrum for selection efficiency in each bin.
It is then compared  with the  reconstructed simulated spectrum of the initial model by computing the difference
\begin{displaymath} \label{equ:dataMCdifference}
  \Delta d_i= d_{i}-Cr_{i} \ .
\end{displaymath}  
Here $d_i$ is the number of efficiency-corrected reconstructed data events 
in the $i^{th}$ bin, $r_i$ is the number of efficiency-corrected reconstructed 
simulated events and $C$ normalizes the initial
model spectrum to the data in the signal range $ 1.8 < \egcms < 2.8\gev$. 
A fraction $f$ of $\Delta d_i$ comes from a true difference between the model
and the data spectrum, while the remaining fraction $1-f$ is due to a  fluctuation in either the signal 
or in the background subtraction. The function $f$ is a regularization 
function with a tunable parameter $\lambda$:
\begin{eqnarray*}
f(\Delta d_{i}, \sigma_{i}, \lambda ) & = & 1-e^{-(\frac{\Delta d_{i}}{\lambda \sigma_{i}})^2} \ ,
\label{eqn:regularizationfunction}
\end{eqnarray*}
where $\sigma_{i}$ is the error in $d_{i}$.
There are several choices of regularization functions suggested in reference~\cite{Malaescu:2009dm}.
Each function has the property that it varies monotonically between  0 and 1 as
the combination $\Delta d_{i}/(\lambda \sigma_{i})$ changes from 0 to $\infty$.
The procedure is found to be insensitive to the particular choice, so the simplest is chosen.  The value of
the regularization parameter $\lambda$ thus determines the fraction
$f$ of the difference that is unfolded in each bin.

An ensemble of 40,000 simulated model spectra is used both to optimize
$\lambda$ and to derive the error matrix for the unfolded spectrum.
These spectra have been generated using
an error matrix that is constructed from the errors in Table~\ref{tab:unblindedYields},  
the bin-to-bin correlations in Table~\ref{tab:EGammaStarMeasCorrelation} and the correlations between the background
subtraction and the efficiency systematics described in Sec.~\ref{sec:bf_syst_total}.
Each spectrum  is unfolded with a range of values of $\lambda$ and then corrected for acceptance.
The error matrix of the unfolded spectra, $O$, is computed from the ensemble using the output
distributions of energy in each bin and the correlations between these distributions. This is then repeated using 
different models to construct the unfolding matrix. A $\chi^{2}$ function is formed 
using a vector of the unfolded yields, the inverse of the error matrix $O^{-1}$ and a vector of the 
true value of the original generated MC spectra $\vec{t}$:
\begin{eqnarray*}
  \chi^2 = (\vec{u}-C\vec{t})^T O^{-1} (\vec{u} - C\vec{t}) \ .
\label{eqn:optimizationchisq}
\end{eqnarray*}
 Only bins in the signal region $1.8 < \egcms <  2.8\gev$ are used for the optimization. 
The $\chi^2$ function is then used to find the value of $\lambda$ that most closely reproduces $\vec{t}$ for all models.
The optimal value of $\lambda$ is then used to unfold the data. The detector resolution
unfolding is performed using  $\lambda=0.5$.

The unfolding matrix used to unfold the data $\tilde{P}_{ij}^{\prime}$ is constructed from a modified
transfer matrix $A_{ij}^{\prime}$. It is modified by adding the folded difference between the initial
model and the data:
\begin{eqnarray*} \label{eqn:unfoldingMatrixImprove}
 A_{ij}^{\prime} &=& C A_{ij} + \Delta d_{j}^{1} \cdot P_{ij}  \nonumber \\
 \tilde{P}_{ij}^{\prime} & = & \frac{A_{ij}^{\prime}}{\sum_{k=1}^{N} A_{ik}^{\prime}} \ .
\end{eqnarray*}

The unfolded data spectrum $u_j$ is then obtained from  
\begin{eqnarray*} 
u_j &=& C t_{j} +  \sum_{i=1}^{N}\{ f(\Delta d_i, \sigma_{d_i}, \lambda)\cdot \Delta d_i \cdot \tilde{P}_{ij}^{\prime} \nonumber \\  
    & & + [1-f(\Delta d_i, \sigma_{d_i}, \lambda)]\cdot \Delta d_i \cdot \delta_{ij}\} \ .
\label{func:unfoldingFormula}
\end{eqnarray*}
It is expressed in terms of a correction to the true value of the initial model ($ C t_{j}$).  
The second term is that part of the difference between the initial
model and the data that is to be unfolded using the unfolding matrix
$\tilde{P'}_{ij}$. The third term is the part of the difference
attributed to statistical or systematic fluctuation that is not
unfolded.  This procedure was iterated,
but its output was found to have converged after just one application,
so that first iteration provides the results presented below.
The procedure for unfolding the Doppler smearing is identical except that
the optimal value of $\lambda=1.0$. 

\subsection{Results of the Unfolding}
\label{sec:unfolding_results}

\begin{table}[!b]  
  \addtolength{\extrarowheight}{1.5pt}
 \begin{center}
  \caption{The unfolded \egcmstrue spectrum and its uncertainties in numbers
   of produced events.  The total error is the sum
   in quadrature of the statistical, systematic and model-dependence errors}
  \label{tab:egamma_yields_after_unfolding_resolution}
  \vspace{0.1in}
  \begin{tabular*}{8.6cm}{@{\extracolsep{\fill}}crrrrr} \hline \hline
    \multirow{2}{*}{\egcmstrue (GeV)} & \multirow{2}{*}{Yield} &\multicolumn{4}{c}{Error (events)} \\ \cline{3-6}
          &   &\multicolumn{1}{c}{Stat.}&\multicolumn{1}{c}{Syst.}&\multicolumn{1}{c}{Model} &\multicolumn{1}{c}{Total}    \\  \hline
    1.53 to 1.60  & 24620 & 15193 & 24749 &   657 & 29115  \\   
    1.60 to 1.70  & 51556 & 12190 & 21593 &   365 & 25140  \\   
    1.70 to 1.80  & 4244  & 10631 & 15598 &    42 & 18427  \\  
    1.80 to 1.90  & 22346 &  8999 & 11278 &   208 & 14217  \\  
    1.90 to 2.00  & 22506 &  7252 &  7565 &    94 & 10626  \\  
    2.00 to 2.10  & 22177 &  5705 &  4708 &  1512 &  7461  \\  
    2.10 to 2.20  & 27518 &  4773 &  2865 &   939 &  5406  \\  
    2.20 to 2.30  & 42298 &  4140 &  2037 &  1073 &  4673  \\  
    2.30 to 2.40  & 39193 &  4010 &  1542 &  1256 &  4384  \\  
    2.40 to 2.50  & 43214 &  3755 &  1671 &  1140 &  4164  \\  
    2.50 to 2.60  & 29488 &  3560 &  1065 &   611 &  3789  \\  
    2.60 to 2.70  & 20025 &  2784 &   723 &    75 &  2857  \\  
    2.70 to 2.80  & 610   &  2446 &   179 &   141 &  2467  \\ \hline \hline
  \end{tabular*}
 \end{center}
\end{table}

\begin{figure}[!tb]
 \begin{center}   
  \includegraphics[width=.47\textwidth]{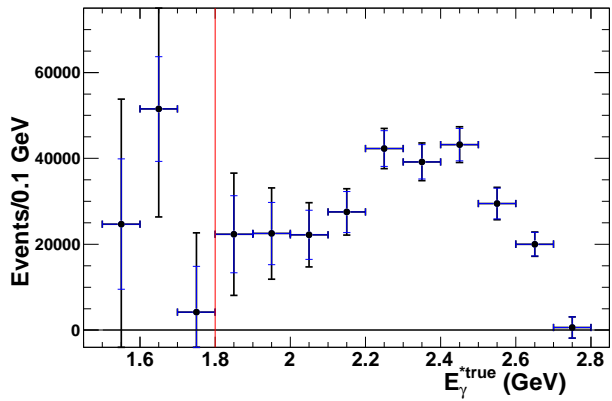}   
 \end{center}
 \vspace{-0.2in}
 \caption{The photon spectrum after unfolding the effects of
 calorimeter resolution and correcting for the selection efficiency and
 detector acceptance. The inner error is statistical only, the outer
 includes statistical, systematic and model-dependence errors added
 in quadrature.  The vertical line shows the boundary between
 the lower control region and the signal region.}
 \label{fig:egamma_after_unfolding_resolution} 
\end{figure}

The measured spectrum shown Fig.~\ref{fig:egamma_dataminusbackground}
and the corresponding yields and uncertainties in
Table~\ref{tab:unblindedYields} are the starting point for the
unfolding.  First the spectrum is corrected for the selection
efficiency, taking into account the additional correlated errors
between the efficiency and the background estimation described in
Sec.~\ref{sec:bf_syst_total}. Then the resolution smearing is unfolded
and the resultant spectrum corrected for detector acceptance to give a 
spectrum in bins of \egcmstrue, presented in 
Table~\ref{tab:egamma_yields_after_unfolding_resolution}.
The estimation of the
statistical, systematic and model-dependence uncertainties is
described in Sec.~\ref{sec:unfolding_errors}.
To provide complete yield uncertainties, the 3.1\% energy-independent
uncertainty on efficiency (see Table~\ref{tab:totalSyst}) is included in 
the systematic uncertainty.  The spectrum is shown in
Fig.~\ref{fig:egamma_after_unfolding_resolution}.

\begin{table}[!b]
  \addtolength{\extrarowheight}{1.5pt}
  \begin{center}
   \caption{Partial branching fraction in bins of \egb obtained from the 
   unfolded spectrum. These values describe the shape of the spectrum
   in the \PB rest frame and provide a crosscheck (see Sec. XI.E) of the
   integrated branching fractions, but are not intended as primary
   branching fraction results.  (The integrated branching fractions
   reported in Table~\ref{tab:bfResultsTable} are more precise and less 
   susceptible to bias, as explained in Sec.~\ref{sec:resOverview}.)  The
   total error is the sum in quadrature of the statistical, systematic and
   model-dependence errors. The model error is relatively large in the bins 
   above 2.4\gev, but anticorrelated between neighboring bins, as 
   discussed in Sec.~\ref{sec:unfolding_errors}.  Hence combined
   200-MeV bins for this region are shown at the bottom of this table
   and in Fig.~\ref{fig:egamma_after_unfolding_resolution_and_doppler}.  }
   \label{tab:egamma_yields_after_unfolding_resolution_and_doppler}
   \vspace{0.1in}
   \begin{tabular*}{8.6cm}{@{\extracolsep{\fill}}ccrrrr} \hline \hline
     \multirow{2}{*}{\egb (GeV)} & $\Delta\BR(\bxsdg)$ &\multicolumn{4}{c}{Error } \\ \cline{3-6}
                &($10^{-5}$) &\multicolumn{1}{c}{Stat}&\multicolumn{1}{c}{Syst}&\multicolumn{1}{c}{Model}&\multicolumn{1}{c}{Total}    \\  \hline  
 1.53 to 1.60 &     2.53 &     1.59 &      2.52 &      0.33 &     2.97 \\
 1.60 to 1.70 &     7.76 &     1.95 &      3.90 &      0.31 &     4.44 \\
 1.70 to 1.80 &     0.25 &     1.53 &      2.07 &      0.06 &     2.48 \\
 1.80 to 1.90 &     2.81 &     1.30 &      1.45 &      0.03 &     1.87 \\
 1.90 to 2.00 &     3.16 &     1.05 &      1.03 &      0.10 &     1.45 \\
 2.00 to 2.10 &     2.67 &     0.83 &      0.65 &      0.28 &     1.06 \\
 2.10 to 2.20 &     3.56 &     0.70 &      0.38 &      0.16 &     0.76 \\
 2.20 to 2.30 &     5.44 &     0.60 &      0.28 &      0.26 &     0.69 \\
 2.30 to 2.40 &     5.37 &     0.58 &      0.23 &      0.16 &     0.62 \\
 2.40 to 2.50 &     5.80 &     0.53 &      0.24 &      0.99 &     1.13 \\
 2.50 to 2.60 &     6.46 &     0.59 &      0.26 &      0.80 &     1.02 \\
 2.60 to 2.70 &     0.00 &     0.11 &      0.01 &      0.12 &     0.16 \\
 2.70 to 2.80 &    -0.12 &     0.21 &      0.01 &      0.10 &     0.23 \\ \hline 								       
 2.40 to 2.60 &    12.25 &     0.79 &      0.47 &      0.19 &     0.92 \\
 2.60 to 2.80 &    -0.12 &     0.24 &      0.01 &      0.22 &     0.32 \\ \hline \hline
   \end{tabular*}
  \end{center}
\end{table}

\begin{figure}[!tb]
 \begin{center}   
  \includegraphics[width=.47\textwidth]{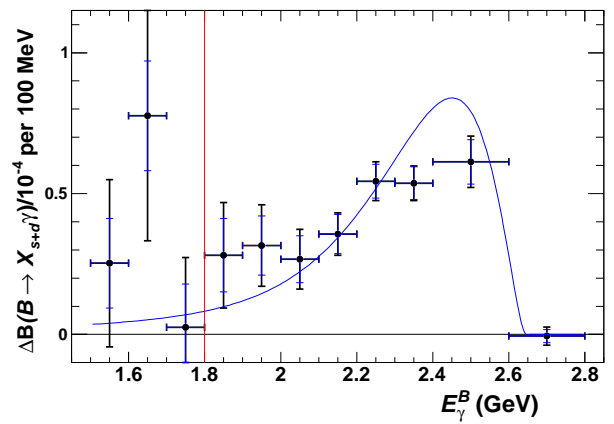}
 \end{center}
 \vspace{-0.2in}
 \caption{The photon spectrum after unfolding resolution and Doppler
  smearing, shown as a partial branching fraction ($\Delta$B, see
  Table~\ref{tab:egamma_yields_after_unfolding_resolution_and_doppler}
  caption).  The inner error is statistical only, the outer includes
  statistical, systematic and model-dependence errors added in
  quadrature. Section~\ref{sec:unfolding_errors} explains why results 
  above 2.4\gev are shown in wider bins.  The vertical line
  shows the boundary between
  the lower control region and the signal region.  The curve shows 
  the spectrum in a kinetic-scheme model (see text), normalized 
  to the data from 1.8 to 2.8\gev.}
 \label{fig:egamma_after_unfolding_resolution_and_doppler} 
\end{figure}

\begin{table*}[t]
\addtolength{\extrarowheight}{1.5pt}
\begin{center}  
\caption{ The correlation matrix for the errors on signal yields shown in
Table~\ref{tab:egamma_yields_after_unfolding_resolution}, in bins 
of \egcmstrue. Statistical, systematic, and model errors are included. Columns
are labeled by the value of \egcmstrue at the lower edge of the bin.} 
\label{tab:EGammaCMSTrueCorrelation}
\vspace{0.1in}
\begin{tabular*}{14cm}{@{\extracolsep{\fill}}cc|cccccccccccccc} \hline \hline 
 \quad \quad & \egcms (GeV) &1.53&1.6&1.7&1.8&1.9&2.0&2.1&2.2&2.3&2.4&2.5&2.6&2.7  & \quad \quad \\ \hline 
 \quad \quad &1.53 to 1.6&  1.00 &  0.65 &  0.63 &  0.60 &  0.55 &  0.44 &  0.29 &  0.16 &  0.11 &  0.13 &  0.07 &  0.03 &  0.04  & \quad \quad\\   
 \quad \quad &1.6 to 1.7 &       &  1.00 &  0.63 &  0.59 &  0.55 &  0.46 &  0.31 &  0.20 &  0.09 &  0.15 &  0.06 &  0.09 &  0.05  & \quad \quad\\   
 \quad \quad &1.7 to 1.8 &       &       &  1.00 &  0.57 &  0.52 &  0.41 &  0.29 &  0.16 &  0.10 &  0.12 &  0.04 &  0.04 &  0.03  & \quad \quad\\   
 \quad \quad &1.8 to 1.9 &       &       &       &  1.00 &  0.50 &  0.39 &  0.26 &  0.15 &  0.09 &  0.17 &  0.05 &  0.09 &  0.01  & \quad \quad\\   
 \quad \quad &1.9 to 2.0 &       &       &       &       &  1.00 &  0.38 &  0.25 &  0.14 &  0.09 &  0.12 &  0.05 &  0.01 &  0.05  & \quad \quad\\   
 \quad \quad &2.0 to 2.1 &       &       &       &       &       &  1.00 &  0.25 &  0.18 &  0.02 &  0.04 &  0.03 &  0.03 & -0.02  & \quad \quad\\   
 \quad \quad &2.1 to 2.2 &       &       &       &       &       &       &  1.00 &  0.16 &  0.00 &  0.07 &  0.06 &  0.03 &  0.01  & \quad \quad\\   
 \quad \quad &2.2 to 2.3 &       &       &       &       &       &       &       &  1.00 &  0.00 &  0.03 &  0.06 &  0.08 & -0.02  & \quad \quad\\   
 \quad \quad &2.3 to 2.4 &       &       &       &       &       &       &       &       &  1.00 &  0.17 &  0.07 &  0.04 &  0.05  & \quad \quad\\   
 \quad \quad &2.4 to 2.5 &       &       &       &       &       &       &       &       &       &  1.00 &  0.11 &  0.05 &  0.05  & \quad \quad\\   
 \quad \quad &2.5 to 2.6 &       &       &       &       &       &       &       &       &       &       &  1.00 &  0.02 & -0.06  & \quad \quad\\   
 \quad \quad &2.6 to 2.7 &       &       &       &       &       &       &       &       &       &       &       &  1.00 &  0.02  & \quad \quad\\   
 \quad \quad &2.7 to 2.8 &       &       &       &       &       &       &       &       &       &       &       &       &  1.00  & \quad \quad\\  
\hline\hline  
\end{tabular*}  
\end{center}
\end{table*} 
 
\begin{table*}[!t]
\addtolength{\extrarowheight}{1.5pt}
\begin{center}  
\caption{ The correlation matrix for the errors on partial branching 
 fractions shown in 
 Table~\ref{tab:egamma_yields_after_unfolding_resolution_and_doppler}, in bins 
 of \egb.  Statistical, systematic, and model errors are included. Columns
 are labeled by the value of \egb at the lower edge of the bin.}  
\label{tab:EGammaBCorrelation}
\vspace{0.1in}
\begin{tabular*}{14cm}{@{\extracolsep{\fill}}cc|cccccccccccc} \hline \hline 
 \quad \quad & \egb (GeV) & 1.53  &  1.6  &   1.7 &  1.8  &  1.9  &  2.0  &  2.1  &  2.2  &  2.3  &  2.4  &  2.6 & \quad \quad \\ \hline 
 \quad \quad & 1.53 to 1.6&  1.00 &  0.51 &  0.53 &  0.53 &  0.50 &  0.36 &  0.20 &  0.05 &  0.07 &  0.13 &  0.18 & \quad \quad\\   
 \quad \quad & 1.6 to 1.7 &       &  1.00 &  0.61 &  0.57 &  0.55 &  0.48 &  0.28 &  0.15 &  0.01 &  0.12 &  0.08 & \quad \quad\\   
 \quad \quad & 1.7 to 1.8 &       &       &  1.00 &  0.54 &  0.48 &  0.41 &  0.24 &  0.11 &  0.03 &  0.07 &  0.10 & \quad \quad\\   
 \quad \quad & 1.8 to 1.9 &       &       &       &  1.00 &  0.48 &  0.35 &  0.21 &  0.08 &  0.01 &  0.15 &  0.11 & \quad \quad\\   
 \quad \quad & 1.9 to 2.0 &       &       &       &       &  1.00 &  0.37 &  0.15 &  0.05 &  0.05 &  0.11 &  0.16 & \quad \quad\\   
 \quad \quad & 2.0 to 2.1 &       &       &       &       &       &  1.00 &  0.27 &  0.12 & -0.07 &  0.04 & -0.12 & \quad \quad\\   
 \quad \quad & 2.1 to 2.2 &       &       &       &       &       &       &  1.00 &  0.19 & -0.14 &  0.07 & -0.09 & \quad \quad\\   
 \quad \quad & 2.2 to 2.3 &       &       &       &       &       &       &       &  1.00 &  0.00 & -0.03 & -0.23 & \quad \quad\\   
 \quad \quad & 2.3 to 2.4 &       &       &       &       &       &       &       &       &  1.00 &  0.11 &  0.20 & \quad \quad\\   
 \quad \quad & 2.4 to 2.6 &       &       &       &       &       &       &       &       &       &  1.00 &  0.16 & \quad \quad\\   
 \quad \quad & 2.6 to 2.8 &       &       &       &       &       &       &       &       &       &       &  1.00 & \quad \quad\\  \hline \hline
\end{tabular*}  

\end{center}
\end{table*}

The Doppler smearing is then unfolded starting from 
Fig.~\ref{fig:egamma_after_unfolding_resolution} and
Table~\ref{tab:egamma_yields_after_unfolding_resolution}.  The 
resulting yields in bins of \egb are converted to partial branching fractions
by dividing by the number of \PB mesons in the on-resonance
data sample, $2N_{\BB}$.  These branching fractions are presented in 
Table~\ref{tab:egamma_yields_after_unfolding_resolution_and_doppler}.
An additional 1.1\% has been included in the systematic
error to account for the uncertainty in $N_{\BB}$.
Figure~\ref{fig:egamma_after_unfolding_resolution_and_doppler} shows
this photon spectrum in the \PB rest frame.  The spectrum is
compared to that for a kinetic-scheme model with parameters 
$\mb = 4.60\gev$ and $\mupsq = 0.45\,\mathrm{GeV}^2$, close to HFAG world
averages.  (The \PKsti has not been substituted for the highest-energy
part of the spectrum, because the unfolded data cannot resolve such a
peak.)  

The correlation matrices corresponding to 
Table~\ref{tab:egamma_yields_after_unfolding_resolution} 
and~\ref{tab:egamma_yields_after_unfolding_resolution_and_doppler} are
given in Tables~\ref{tab:EGammaCMSTrueCorrelation}
and~\ref{tab:EGammaBCorrelation}, respectively.  
These matrices have a complex structure because many effects contribute.
At low energies (the upper left quadrant) they are dominated by the
highly-correlated uncertainties in the \BB backgrounds.  At higher
energies, the uncorrelated statistical uncertainty is relatively
more important, along with the smaller but fully-correlated systematic
uncertainty on efficiency.  The contributions from the unfolding
itself and from model dependence can be negative.  
Hence in the lower-right quadrant, where other correlations are weak,
the net result can be close to zero or negative.

The numbers in
Tables~\ref{tab:egamma_yields_after_unfolding_resolution_and_doppler} 
and~~\ref{tab:EGammaBCorrelation} can be used to fit the measured
spectral shape to any theoretical prediction in the \PB-meson rest frame.

\subsection{Statistical, Systematic and Model-Dependence Uncertainties 
 in the Unfolding}
\label{sec:unfolding_errors}

\begin{table}[!tb] 
\addtolength{\extrarowheight}{1.5pt}
\begin{center}
\caption{ The change in the number of events in each bin of the
unfolded photon spectrum after shifting the photon energy scale by
$\pm 0.3\%$. The absolute value of the largest difference ($+$ or
$-$) is shown after resolution unfolding (\egcmstrue bins) and both
resolution and Doppler smearing unfolding (\egb bins). In both cases
efficiency and acceptance corrections have been applied. These changes
are included in the final systematic errors in
Tables~\ref{tab:egamma_yields_after_unfolding_resolution} and
~\ref{tab:egamma_yields_after_unfolding_resolution_and_doppler}
assuming 100\% correlation between the bins.}
\label{tab:UnfoldingSyst}
\vspace{0.1in}
\begin{tabular*}{8.6cm}{@{\extracolsep{\fill}}ccc} \hline \hline 
  \multirow{2}{*}{Energy Range (GeV)}  & \multicolumn{2}{c}{Change (events)} \\ \cline{2-3}
               &  \egcmstrue Bins & \egb Bins  \\ \hline
  1.53 to 1.60 &    222.1 &    220.2  \\
  1.60 to 1.70 &    190.6 &    191.0  \\
  1.70 to 1.80 &    261.1 &    261.6  \\
  1.80 to 1.90 &    354.4 &    354.8  \\
  1.90 to 2.00 &    493.2 &    492.0  \\
  2.00 to 2.10 &    622.9 &    622.2  \\
  2.10 to 2.20 &    640.3 &    658.5  \\
  2.20 to 2.30 &    428.4 &    461.1  \\
  2.30 to 2.40 &    528.7 &    598.9  \\
  2.40 to 2.50 &   1184.2 &   1292.5  \\
  2.50 to 2.60 &   1080.6 &    967.6  \\
  2.60 to 2.70 &    490.8 &    475.7  \\ \hline \hline
 \end{tabular*}
\end{center}
\end{table}

The dominant uncertainty in the bins of the unfolded spectrum is due to the \BB subtraction described
in Sec.~\ref{sec:yields}. The statistical and systematic errors on the efficiency-corrected yields
 are propagated using the ensemble MC technique described previously. A number of possible uncertainties in the unfolding procedure were considered.
These included  changing the 
regularization parameter $\lambda$ to zero which changes $f$ to $1.0$ in all bins, 
changing the normalization factor $C$ according to the 10\% uncertainty in the measured value of $\BR(\bxsg)$, 
varying the energy scale by $\pm 0.3\%$, and smearing the calorimeter resolution  in the MC simulation
by an additional $1\%$, as determined by data comparisons in Sec.~\ref{sec:bf_syst_photon}. The only 
significant effects are found to be in the photon energy scale shift. 
Table~\ref{tab:UnfoldingSyst} shows the bin-by-bin change in the event yields due to the photon energy shift. 
For each bin, the absolute value of the largest difference ($+$ or $-$) is 
taken as the systematic uncertainty, and 100\% bin-to-bin
correlation is assumed. This error is combined in quadrature with the systematic error propagated
from the measured  \egcms spectrum and is included in 
Tables~\ref{tab:egamma_yields_after_unfolding_resolution}
and~\ref{tab:egamma_yields_after_unfolding_resolution_and_doppler}.

To assess the model dependence, the unfolding is performed with a range of models. 
In each case the same model is used for the entire procedure including efficiency and acceptance corrections,
and unfolding the detector resolution and Doppler smearing. Figure~\ref{fig:dataVsModels} shows
two models that could plausibly describe the data at the one-sigma level. These are a shape function
model with ($\mb=4.40 \gev, \mupsq=0.52 \gev^2$) and a kinetic-scheme model ($\mb=4.60 \gev, \mupsq=0.45 \gev^2, 
\mugsq=0.27 \gev^2$). To set the model-dependence error we unfold the nominal simulated model 
(shape function: $\mb=4.51 \gev ,\mupsq =0.46 \gev^{2}$) with one of these two models. The larger bin-by-bin difference is taken as the model error in the 
unfolded spectrum with 100\% correlation between each bin. The model-dependence
error is generally much smaller than the systematic error except for the unfolding of the Doppler smearing 
close to the kinematic limit ($\egb \approx m_{B}/2$). The steeply falling spectrum at this limit leads to a much greater sensitivity
to the model, which results in a large error that is anti-correlated between the 2.4-2.5\gev and 
2.5-2.6\gev bins. To avoid this edge effect the two bins are summed. This is
also done for the 2.6-2.8\gev range. 

\subsection{Crosscheck of Branching Fraction}
\label{sec:unfolding_bf}

The numbers in 
Table~\ref{tab:egamma_yields_after_unfolding_resolution_and_doppler}
are used to obtain integrated branching fractions \BR(\bxsg) for purposes 
of comparison with the reported results from Sec.~\ref{sec:bf_results}.  
The $\Delta\BR(\bxsdg)$ values are summed over \egb intervals, with the
errors combined including correlations (Table~\ref{tab:EGammaBCorrelation}).
Lastly, a factor of 0.958 is applied to account for the \bxdg contribution.
As explained in Sec.~\ref{sec:resOverview}, the unfolded yields are based
on a different choice of model than that used to extract the \BR(\bxsg)
results for this analysis, and hence are not intended to be used for 
such results.  
This procedure has been carried out for one energy range,
$1.8 < \egb < 2.8\gev$.

There are two contributions to the uncertainty beyond those implied 
by Tables~\ref{tab:egamma_yields_after_unfolding_resolution_and_doppler}
and~\ref{tab:EGammaBCorrelation}.
First, there is the small (1.1\%) uncertainty on $N_{\BB}$.
Second, because the range of models used to estimate model-dependence
uncertainty is data-driven, that uncertainty 
is positively correlated with the systematic uncertainty on the signal 
yield.  This gives a \BR(\bxsg) for $1.8 < \egb < 2.8\gev$ of
$(3.36 \pm 0.19 \pm 0.34 \pm 0.08)\times 10^{-4} = 
(3.36 \pm 0.43)\times 10^{-4}$, where the first set of errors are statistical,
systematic and model, and their combination in the second form
takes the model-systematic correlation into account.

This value may be compared to the reported branching fraction of
$(3.20 \pm 0.15 \pm 0.29 \pm 0.08)\times 10^{-4} = 
(3.20 \pm 0.33)\times 10^{-4}$ from Table~\ref{tab:bfResultsTable};
the three uncertainties (independent in that case) are added in
quadrature.  The difference in the central values is due to the different
choice of the central model; if a data-like model had been used in
Sec.~\ref{sec:bf}, the extracted branching fraction would have
been $3.36 \times 10^{-4}$,
the same value obtained with unfolding.  

The smaller statistical and systematic uncertainties on the branching
fraction from Table~\ref{tab:bfResultsTable} are in large part a 
consequence of applying the efficiency correction to a single wide bin 
of photon energy.  As discussed in Sec.~\ref{sec:resOverview}, 
this de-emphasizes the importance of the uncertainties in the
lowest-energy region, where signal efficiency is lowest and background
uncertainties are largest.  The branching fraction as derived from the 
unfolded spectrum of necessity relies upon efficiency corrections in
100-MeV bins.  In addition, the combined uncertainty on the latter
result is increased by the model-background correlation discussed
above, an effect which does not occur when the model range is chosen
as described in Sec.~\ref{sec:bf_modeldep}.

\subsection{Moments of the Spectrum}
\label{sec:unfolding_moments}

The moments of the spectrum provide information to measure the HQET parameters $\mb$ and $\mupsq$
in the kinetic scheme~\cite{Benson:2004sg}. The first, second and third spectral moments, $E_{1},E_{2},E_{3}$
are defined in Sec.~\ref{sec:intro}, Eq.~(\ref{eq:momentdefs}).
They are measured for three photon energy ranges: 1.8 to 2.8\gev, 
1.9 to 2.8\gev and 2.0 to 2.8\gev.  The
moments are computed directly from the unfolded spectrum in 100-MeV bins given in Tables~~\ref{tab:egamma_yields_after_unfolding_resolution} and ~\ref{tab:egamma_yields_after_unfolding_resolution_and_doppler} using the correlation
matrices given in Tables~\ref{tab:EGammaCMSTrueCorrelation} and ~\ref{tab:EGammaBCorrelation}.

The behavior of the moments for different photon energy ranges  has been studied theoretically in the
kinetic scheme. The spectral moments in \egcmstrue are given in
Table~\ref{tab:unfolded_data_moments_egcms}. The correlations between the moments are given 
in Table~\ref{tab:MomentsCorrelationsEGammaStarTrue} to allow fits to predictions 
of the moments. The \egb spectral moments and correlations between the moments are given in 
Tables~\ref{tab:unfolded_data_moments_egb} and ~\ref{tab:MomentsCorrelationsEGammaB}.

\section{Conclusions}

\begin{table*}[p] \footnotesize
 \begin{center}
 \caption{The \egcmstrue spectral moments and errors ($\pm$ statistical $\pm$ systematic $\pm$ model-dependence).  
Moments are defined by Eq.~(\ref{eq:momentdefs}) in Sec.~\ref{sec:intro}.}
 \label{tab:unfolded_data_moments_egcms}
\vspace{0.1in}
\addtolength{\extrarowheight}{2pt}
  \begin{tabular*}{17.6cm}{@{\extracolsep{\fill}}cccc} \hline \hline
   \egcms Range (GeV)       & $E_1$ (GeV) & $E_2$ ($\gev^2$) & $E_3$ ($\gev^3$)\\ \hline
   1.8 to 2.8 & $2.275 \pm 0.018 \pm 0.032 \pm 0.003$   & $0.0546 \pm 0.0049 \pm 0.0074 \pm 0.0005$  & $-0.0031\pm 0.0011 \pm 0.0013 \pm 0.0004$  \\
   1.9 to 2.8 & $2.314 \pm 0.013 \pm 0.017 \pm 0.004$   & $0.0417 \pm 0.0032 \pm 0.0028 \pm 0.0003$  & $-0.0013 \pm 0.0007 \pm 0.0005 \pm 0.0003$ \\
   2.0 to 2.8 & $2.350 \pm 0.010 \pm 0.008 \pm 0.005$   & $0.0317 \pm 0.0022 \pm 0.0010 \pm 0.0005$  &  $0.0001 \pm 0.0005 \pm 0.0002 \pm 0.0002$ \\ \hline   
    \hline
  \end{tabular*}
\end{center}
\end{table*}

\begin{table*}[p] \footnotesize
 \begin{center}
 \caption{The \egb spectral moments and errors ($\pm$ statistical $\pm$ systematic $\pm$ model-dependence).  
Moments are defined by Eq.~(\ref{eq:momentdefs}) in Sec.~\ref{sec:intro}.}
 \label{tab:unfolded_data_moments_egb}
\vspace{0.1in}
\addtolength{\extrarowheight}{2pt}
  \begin{tabular*}{17.6cm}{@{\extracolsep{\fill}}cccc} \hline \hline
   \egb Range (GeV)    & $E_1$ (GeV) & $E_2$ ($\gev^2$)   & $E_3$ ($\gev^3$) \\ \hline
   1.8 to 2.8   & $2.267\pm 0.019 \pm 0.032 \pm 0.003$    & $0.0484 \pm 0.0053 \pm 0.0077 \pm 0.0005$  & $-0.0048 \pm 0.0011 \pm 0.0011 \pm 0.0004$ \\ 
   1.9 to 2.8   & $2.304\pm 0.014 \pm 0.017 \pm 0.004$    & $0.0362 \pm 0.0033 \pm 0.0033 \pm 0.0005$  & $-0.0029 \pm 0.0007 \pm 0.0004 \pm 0.0002$ \\  
   2.0 to 2.8   & $2.342\pm 0.010 \pm 0.008 \pm 0.005$    & $0.0251 \pm 0.0021 \pm 0.0013 \pm 0.0009$  & $-0.0013 \pm 0.0005 \pm 0.0002 \pm 0.0001$ \\ \hline \hline  \\  
  \end{tabular*}
 \end{center}
\end{table*}

 \begin{table*}[p] 
\addtolength{\extrarowheight}{1.5pt}
  \begin{center}
\caption{ The correlation matrix of the \egcmstrue spectral moments.
  Superscripts denote the lower end of the energy range in GeV.}
\label{tab:MomentsCorrelationsEGammaStarTrue}
\vspace{0.1in} 
\begin{tabular*}{8.6cm}{@{\extracolsep{\fill}}c|rrrrrrrrr} \hline \hline
        & $E_{1}^{1.8}$  &  $E_{2}^{1.8}$   &     $E_{3}^{1.8}$ 
        & $E_{1}^{1.9}$  &  $E_{2}^{1.9}$   &     $E_{3}^{1.9}$ 
        & $E_{1}^{2.0}$  &  $E_{2}^{2.0}$   &     $E_{3}^{2.0}$ \\ \hline 
  $E_{1}^{1.8}$  &   1.00 &  -0.88 &  -0.09 &   0.84 &  -0.68 &  -0.26 &   0.61 &  -0.29 &  -0.30 \\       
  $E_{2}^{1.8}$  &        &   1.00 &  -0.27 &  -0.58 &   0.71 &   0.19 &  -0.25 &   0.43 &   0.24 \\       
  $E_{3}^{1.8}$  &        &        &   1.00 &  -0.29 &   0.19 &   0.55 &  -0.23 &   0.26 &   0.35 \\       
  $E_{1}^{1.9}$  &        &        &        &   1.00 &  -0.75 &  -0.31 &   0.75 &  -0.28 &  -0.34 \\       
  $E_{2}^{1.9}$  &        &        &        &        &   1.00 &   0.12 &  -0.17 &   0.64 &   0.30 \\       
  $E_{3}^{1.9}$  &        &        &        &        &        &   1.00 &  -0.61 &   0.62 &   0.85 \\       
  $E_{1}^{2.0}$  &        &        &        &        &        &        &   1.00 &   0.25 &  -0.41 \\       
  $E_{2}^{2.0}$  &        &        &        &        &        &        &        &   1.00 &  -0.50 \\       
  $E_{3}^{2.0}$  &        &        &        &        &        &        &        &        &   1.00 \\ \hline \hline
\end{tabular*}
\end{center}
\end{table*}

\begin{table*}[p]
\addtolength{\extrarowheight}{1.5pt}
\begin{center}
\caption{ The correlation matrix of the \egb spectral moments.
  Superscripts denote the lower end of the energy range in GeV.}
\label{tab:MomentsCorrelationsEGammaB}
\vspace{0.1in} 
\begin{tabular*}{8.6cm}{@{\extracolsep{\fill}}c|rrrrrrrrr} \hline \hline
        & $E_{1}^{1.8}$  &  $E_{2}^{1.8}$   &     $E_{3}^{1.8}$ 
        & $E_{1}^{1.9}$  &  $E_{2}^{1.9}$   &     $E_{3}^{1.9}$ 
        & $E_{1}^{2.0}$  &  $E_{2}^{2.0}$   &     $E_{3}^{2.0}$ \\ \hline 
  $E_{1}^{1.8}$  &     1.00 &  -0.90 &   0.10 &   0.84 &  -0.73 &  -0.05 &   0.53 &  -0.46 &  -0.09 \\       
  $E_{2}^{1.8}$  &          &   1.00 &  -0.35 &  -0.60 &   0.73 &  -0.07 &  -0.21 &   0.45 &   0.12 \\       
  $E_{3}^{1.8}$  &          &        &   1.00 &  -0.23 &   0.14 &   0.48 &  -0.29 &   0.30 &   0.26 \\       
  $E_{1}^{1.9}$  &          &        &        &   1.00 &  -0.82 &  -0.05 &   0.68 &  -0.50 &  -0.08 \\       
  $E_{2}^{1.9}$  &          &        &        &        &   1.00 &  -0.11 &  -0.27 &   0.59 &   0.14 \\       
  $E_{3}^{1.9}$  &          &        &        &        &        &   1.00 &  -0.50 &   0.52 &   0.59 \\       
  $E_{1}^{2.0}$  &          &        &        &        &        &        &   1.00 &  -0.59 &  -0.06 \\       
  $E_{2}^{2.0}$  &          &        &        &        &        &        &        &   1.00 &   0.20 \\       
  $E_{3}^{2.0}$  &          &        &        &        &        &        &        &        &   1.00 \\ \hline \hline
\end{tabular*}
\end{center}
\end{table*}

In summary, the \bxsdg photon energy spectrum in the CM frame has been
measured in \onlumi of data taken with the \babar\ experiment. It is
used to extract measurements of the direct \CP asymmetry for the
sum of \bxsg and \bxdg, the branching fraction for \bxsg, and the
spectral shape and its energy moments in the \PB-meson rest frame.  
The result for \CP asymmetry is
\begin{equation*}
  \acp  =   0.057 \pm 0.060 (\mathrm{stat}) \pm 0.018 (\mathrm{syst}) \ .
\end{equation*}
The branching fraction and moments are presented for three ranges of the 
photon energy in the \PB-meson rest frame, 1.8, 1.9 and 2.0 to 2.8\gev 
(Tables~\ref{tab:bfResultsTable} and~\ref{tab:unfolded_data_moments_egb}).
For example, in the 1.8 to 2.8\gev range:
\begin{eqnarray*}
 &&  \BR(\bxsg) = (3.21 \pm 0.15 \pm 0.29 \pm 0.08)\times 10^{-4}\ ,  \\
 &&  E_1 = (2.267\pm 0.019 \pm 0.032 \pm 0.003)\gev \ \ \mathrm{and} \\
 &&  E_2 = (0.0484 \pm 0.0053 \pm 0.0077 \pm 0.0005) \gev^2 \ ,  
\end{eqnarray*}
where the errors are from ststistics, systematics and model dependence,
respectively, and the moments are defined in Eq.~\ref{eq:momentdefs}.

\begin{figure}[bt]
 \begin{center}
  \includegraphics[width=0.5\textwidth]{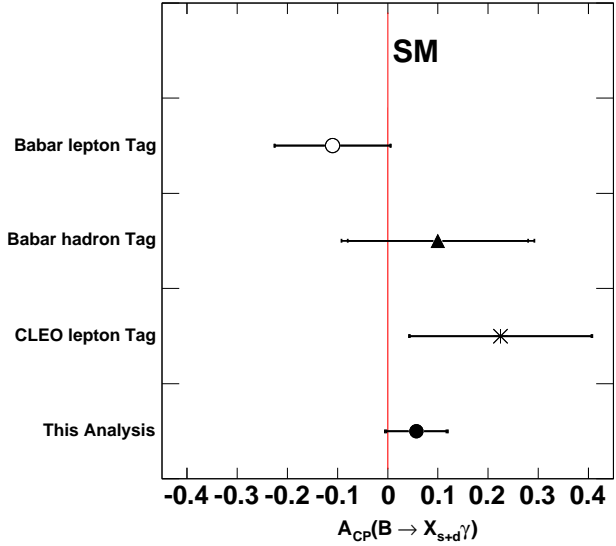}
  \vspace{-0.1in}
  \caption{Measurements of \acp(\bxsdg), with statistical and systematic
   errors.  The three published results, top to
   bottom, are from references~\cite{Aubert:2006gg}, \cite{Aubert:2007my}
   and~\cite{Coan:2000pu}, respectively.  The uppermost
   result is based on a subset of the data used in the current analysis.}
  \label{fig:ACPcompare}
 \end{center}
\end{figure}

Figure~\ref{fig:ACPcompare} compares the measured \acp(\bxsdg) to
previous measurements and to the SM prediction. No
asymmetry is observed, consistent with SM expectation. The current
measurement is the most precise to date.

\begin{figure}[hbtp]
 \begin{center}
  \includegraphics[width=0.5\textwidth]{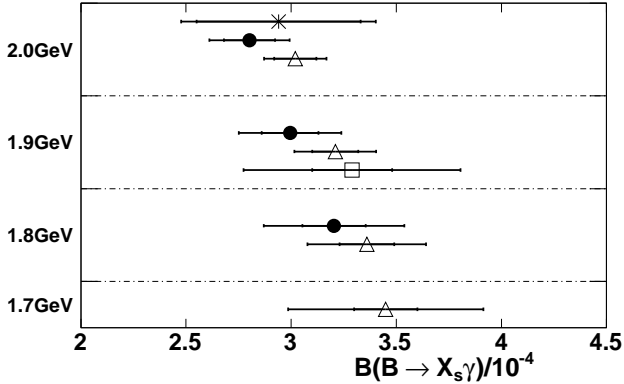}
  \vspace{-0.1in}
  \caption{ The measured branching fraction for this analysis
    ($\CIRCLE$) compared to previous measurements for different \eg
    ranges (minimum energies \egb given on the left axis).  The
    previous measurements are from CLEO ($\ast$)~\cite{Chen:2001fja}, Belle
    ($\bigtriangleup$)~\cite{Limosani:2009qg}, and \babar\ using the
    semi-inclusive technique ($\square$)~\cite{Aubert:2005cua}.  Error
    bars show total uncertainties.}
  \label{fig:BFcompare}
 \end{center}
\end{figure}

\begin{figure}[!b]
 \begin{center}
  \includegraphics[width=0.4\textwidth]{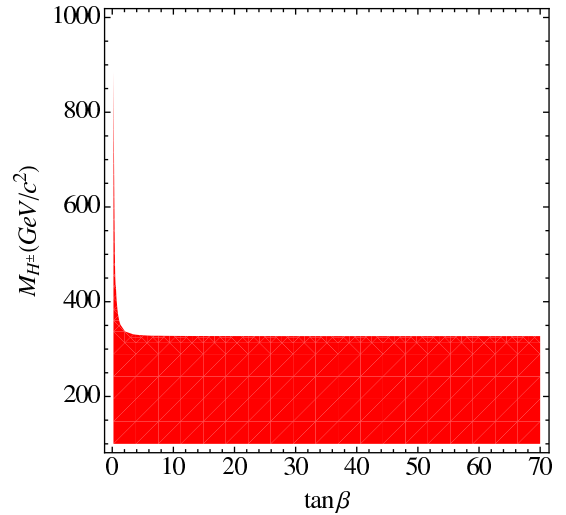}
  \vspace{-0.1in}
  \caption{The shaded area shows the excluded region 
   (at the 95\% confidence level) in charged
   Higgs mass \vs $\tan{\beta}$ for a type-II two-Higgs-doublet model,
   using the measured value of $\BR(\bxsg) = (3.31 \pm 0.16 \pm 0.30 \pm
   0.10) \times 10^{-4}$\,($\egb > 1.6\gev$) from this analysis. 
   This plot is based on predictions in 
   references~\cite{Misiak:2006zs} and~\cite{Haisch:2008ar}.}
  \label{fig:THDM}
 \end{center}
\end{figure}

Figure~\ref{fig:BFcompare} compares the measured branching fraction to
previous measurements performed for different \eg ranges. This
measurement supersedes the previous fully inclusive measurement from
\babar. It is consistent with previous measurements and
of comparable precision to the recent Belle
measurement~\cite{Limosani:2009qg}. In order to compare with
theoretical predictions the measurement for $\egb > 1.8 \gev$ can be
extrapolated down to 1.6\gev using a factor provided by the
HFAG collaboration~\cite{TheHeavyFlavorAveragingGroup:2010qj}. 
They fit results from previous measurements of \bxsg and
\bxclnu to predictions in the kinetic scheme to yield average values of 
\mb and \mupsq.  These parameters are then used to generate
a \bxsg model in the kinetic scheme which gives an extrapolation
factor of $1/(0.968 \pm 0.006)$. When applied to the present
result this gives $\BR(\bxsg) = (3.31 \pm 0.16 \pm 0.30 \pm 0.10)
\times 10^{-4}$\,($\egb > 1.6\gev$) which is in excellent agreement
with the SM prediction $\BR(\bxsg) = (3.15 \pm 0.23) \times 10^{-4}
(\eg>1.6 \gev)$ ~\cite{Misiak:2006zs} and can be used to provide
stringent constraints on new physics. An example is shown in
Figure~\ref{fig:THDM}. The effects of a type-II two-Higgs-doublet model
(THDM) on $\BR(\bxsg)$ at next-to-leading order are presented in
Refs.~\cite{Misiak:2006zs,Haisch:2008ar}.  Software provided by
the author of Ref.~\cite{Haisch:2008ar} computes an excluded region,
following a procedure described in Ref.~\cite{Ciuchini:1997xe}.
The branching fraction, including both the SM and the THDM contributions, is
calculated for each point in the $M_{H^{\pm}}$ \vs $\tan{\beta}$ plane.
The various theoretical uncertainties are assumed to have
Gaussian distributions, and are combined in quadrature.
A point is then excluded if the negative $1\sigma$
deviation of the prediction lies above the 95\% confidence-level upper
limit of the measured branching fraction extrapolated to 1.6\gev.
The region $M_{H^{\pm}}<327\gev$ 
is excluded at the 95\% confidence level, independent of $\tan{\beta}$.

\begin{figure}[hbtp]
\begin{center}   
\includegraphics[width=0.4\textwidth]{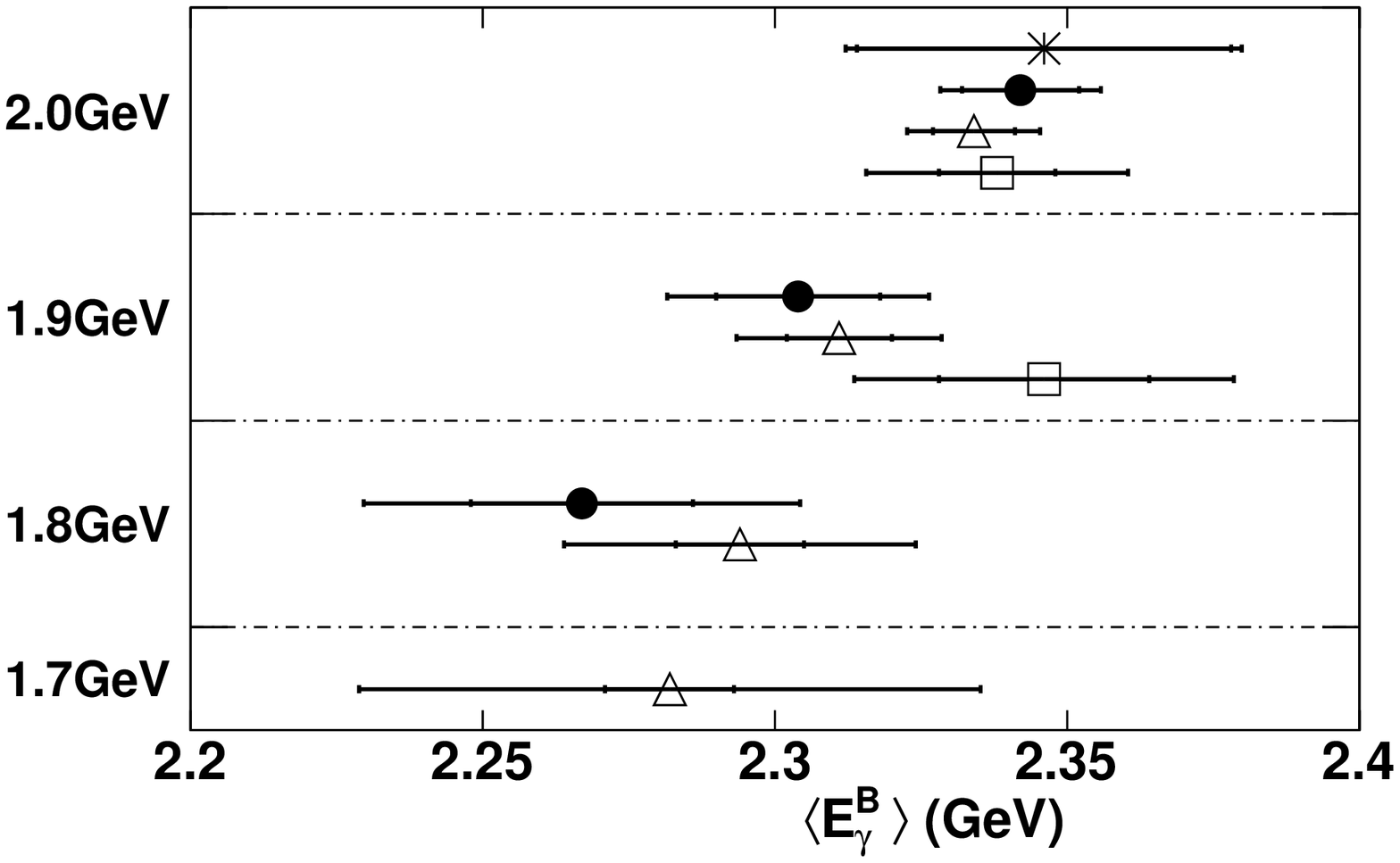}
\vspace{0.1in}
\includegraphics[width=0.4\textwidth]{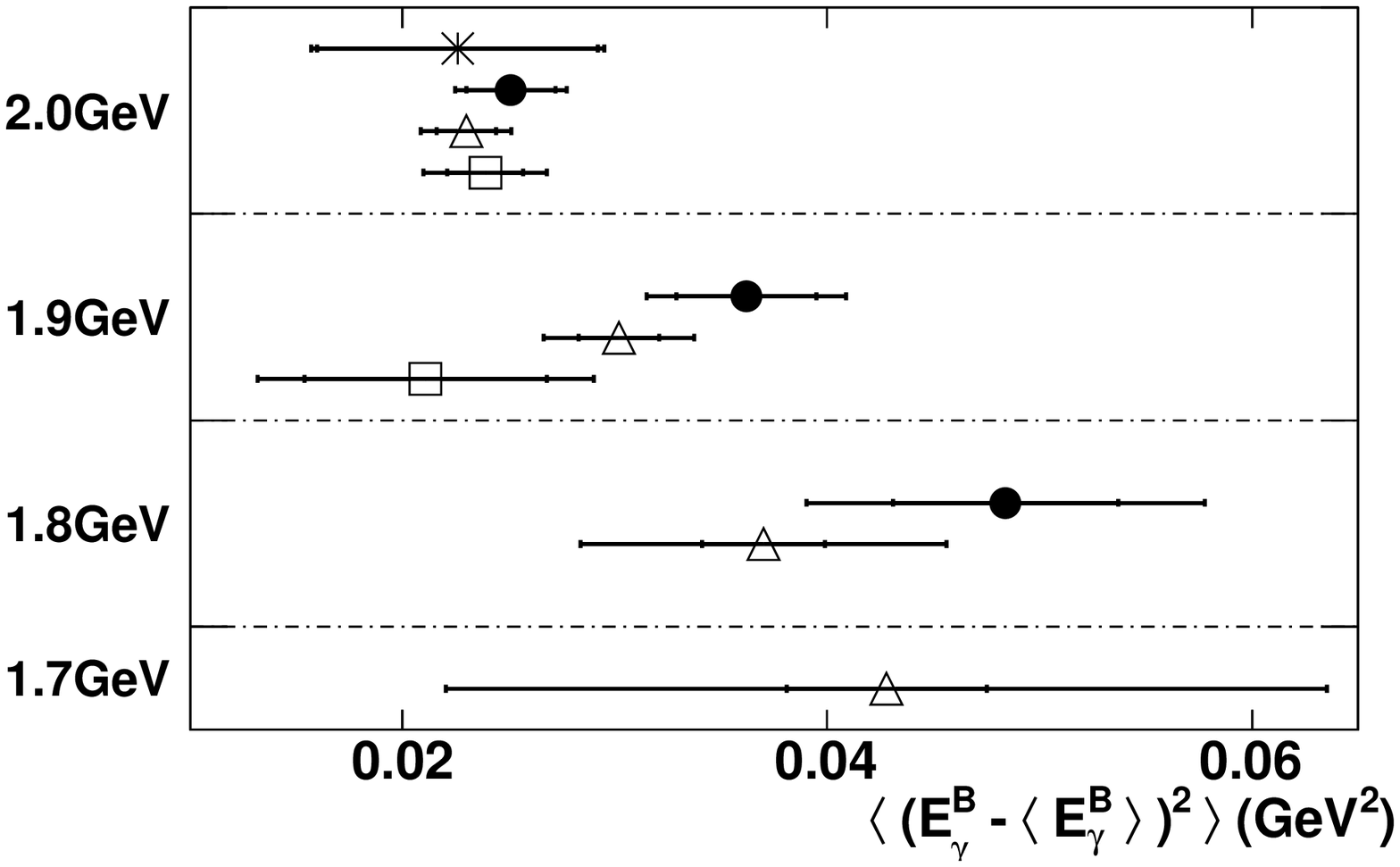}
\vspace{0.1in}
\caption{The measured first (top) and second (bottom) moments from this
  analysis ($\CIRCLE$) compared with the previous measurement for
  different \eg ranges (minimum energies given on the left
  axis). These previous measurements are CLEO
  ($\ast$)~\cite{Chen:2001fja}, \babar\ semi-inclusive
  ($\square$)~\cite{Aubert:2005cua}, and Belle
  ($\bigtriangleup$)~\cite{Limosani:2009qg}.  Error bars show total
  uncertainties.}
\label{fig:MomentsComparison} 
\end{center}

\end{figure}

The effects of detector resolution and Doppler smearing are unfolded to present the photon spectrum in the 
\PB-meson rest frame for the first time in Fig.~\ref{fig:egamma_after_unfolding_resolution_and_doppler}. 
This spectrum may be used to extract information on HQET parameters in two ways. First, the full covariance
matrix is provided to allow any theoretical model to be fit to the entire spectrum. Secondly the moments have 
been extracted and can be compared to predictions for difference energy ranges. Figure~\ref{fig:MomentsComparison} 
compares the measured moments to previous measurements.


\section*{ACKNOWLEDGMENTS}
We are grateful for the 
extraordinary contributions of our \pep2\ colleagues in
achieving the excellent luminosity and machine conditions
that have made this work possible.
The success of this project also relies critically on the 
expertise and dedication of the computing organizations that 
support \babar.
The collaborating institutions wish to thank 
SLAC for its support and the kind hospitality extended to them. 
This work is supported by the
US Department of Energy
and National Science Foundation, the
Natural Sciences and Engineering Research Council (Canada),
the Commissariat \`a l'Energie Atomique and
Institut National de Physique Nucl\'eaire et de Physique des Particules
(France), the
Bundesministerium f\"ur Bildung und Forschung and
Deutsche Forschungsgemeinschaft
(Germany), the
Istituto Nazionale di Fisica Nucleare (Italy),
the Foundation for Fundamental Research on Matter (The Netherlands),
the Research Council of Norway, the
Ministry of Education and Science of the Russian Federation, 
Ministerio de Ciencia e Innovaci\'on (Spain), and the
Science and Technology Facilities Council (United Kingdom).
Individuals have received support from 
the Marie-Curie IEF program (European Union) and the A. P. Sloan Foundation (USA).

\appendix
\section{Parameterization of Branching Fraction Factors}
\label{sec:app}

\begin{table*}[tb]
\begin{center}
 \caption{Coefficients in fits to signal efficiency (\sigeff) and adjustment
   factor ($\alpha$) as a function of \mb and \mupsq, using the functional
   form of Eq.~(\ref{eq:mb_mupsq}).}
 \label{tab:efficiencyFits}
 \vspace{0.1in}
 \addtolength{\extrarowheight}{2pt}
 \begin{tabular*}{14cm}{@{\extracolsep{\fill}}ccccccccc}
  \hline\hline
  & Energy Range  & Quantity &  $f_0$  &  $f_1$  &  $f_2$  &  $f_3$  &  $f_4$   &\\
  \hline                                                                        
  & 1.8 to 2.8\gev&$\sigeff$ & 0.025823& 0.004638&-0.000802&-0.011207&-0.008734 &\\
  &               &$\alpha$  & 1.02189 &-0.01859 & 0.01699 & 0.02896 & 0.06966  &\\
  & 1.9 to 2.8\gev&$\sigeff$ & 0.026099& 0.004380&-0.000501&-0.011818&-0.008565 &\\
  &               &$\alpha$  & 1.03440 &-0.04529 & 0.012169& 0.09818 & 0.11165  &\\
  & 2.0 to 2.8\gev&$\sigeff$ & 0.026463& 0.003876&-0.000371&-0.012715&-0.008907 &\\
  &               &$\alpha$  & 1.06264 &-0.04162 & 0.03924 & 0.35858 & 0.29559  &\\
  \hline\hline
 \end{tabular*}
\end{center}
\end{table*}

The central values of the partial branching fractions reported in
Section~\ref{sec:bf_results} depend on the signal efficiency \sigeff and
adjustment factor $\alpha$ computed for a kinetic-scheme model with
parameters \mb and \mupsq set to current HFAG world-average 
values~\cite{TheHeavyFlavorAveragingGroup:2010qj}
(4.591\gevcc and $0.454\,\mathrm{(GeV/c)}^2$, respectively).  This
Appendix provides functional forms for the dependence of \sigeff
and $\alpha$ on these HQET parameters.  
In the event of possible future changes in the HFAG values, the
information presented here would allow for a corresponding adjustment
of the branching fraction central values.
Since each partial branching fraction in the \PB rest frame is proportional
to $\alpha/\sigeff$, the adjustment would be made by dividing out
that combination computed for the current values of HQET parameters
and multiplying by the same combination computed for the new values.

These functions have no physical significance.  They result from fits to
the \sigeff and $\alpha$ values computed by MC simulation 
for a wide range of parameters.  For a grid of models spanning 
$4.5 \le \mb \le 4.7\gevcc $ and $0.3 \le \mupsq \le 0.7\,\mathrm{(GeV/c)}^2$,
these fits have fractional accuracy of better than 0.2\% for \sigeff and 
0.1\% for $\alpha$.
The functional form used for both quantities is
\begin{eqnarray} 
  f(\mb,\mupsq) &=& f_0 \nonumber \\ 
  & +& f_1(\mb - 4.6\gevcc) + f_2(\sqrt{\mupsq}-0.6\gevc) \nonumber \\
  & +& f_3(\mb - 4.6\gevcc)(\sqrt{\mupsq}-0.6\gevc) \nonumber \\
  & +& f_4(\mb - 4.6\gevcc)^2\ , 
 \label{eq:mb_mupsq}
\end{eqnarray}
where the coefficients $f_0$ through $f_4$ have appropriate units to make
each term dimensionless.  
Table~\ref{tab:efficiencyFits} gives the values of these coefficients 
for \sigeff and $\alpha$ for each of the three photon-energy ranges in 
which branching fractions are reported.

\clearpage

%
%
%

\bibliography{note2347}
\bibliographystyle{apsrev}

\end{document}